\documentclass[11pt,a4paper]{article}
\usepackage[T1]{fontenc}
\usepackage{setspace}

\usepackage{amsmath}
\usepackage{amsfonts}
\usepackage{amssymb}
\usepackage[small]{caption}
\usepackage{color}
\usepackage{fullpage,epsfig,graphics,amsbsy,url,epstopdf}
\usepackage{epstopdf}
\usepackage{cancel}
\usepackage{psfrag}
\usepackage{verbatim}

\newcommand{\bea}{\begin{eqnarray}}
\newcommand{\eea}{\end{eqnarray}}
\newcommand{\eq}[1]{\begin{align}#1\end{align}}
\newcommand{\bal}{\begin{aligned}}
\newcommand{\eal}{\end{aligned}}
\newcommand{\beq}{\begin{equation}}
\newcommand{\eeq}{\end{equation}}

\newcommand{\dt}{\! \cdot \!}
\newcommand{\hth}{\ensuremath{\hat{\theta}_}}
\newcommand{\VEV}[1]{\langle#1\rangle}
\newcommand{\nn}{\nonumber}
\newcommand{\ee}[2]{\epsilon_{#1} \!\cdot\! \epsilon_{#2}}

\newcommand{\ap}{\alpha^{\prime}}
\long\def\symbolfootnote[#1]#2{\begingroup
\def\thefootnote{\fnsymbol{footnote}}\footnote[#1]{#2}\endgroup}

\newcommand{\fermi}[2]{\langle H_{#1} H_{#2}\rangle}
\newcommand{\eqe}{\!\!\! &=& \!\!\!}
\newcommand{\sme}{\!\! &\simeq& \!\!}

\newcommand{\bt}[3]{\frac{\Gamma(#1)\Gamma(#2)}{\Gamma(#3)}}
\newcommand{\bfs}{\boldsymbol}
\newcommand{\sdot}{\hspace{-3pt}\cdot\hspace{-3pt}}

\begin{document}
\setlength{\captionmargin}{20pt}
\begin{titlepage}
\begin{flushright}
\phantom{UFIFT-HEP-10-}
\end{flushright}

\vskip 2.5cm

\begin{center}
\begin{Large}
{\bf On Type 0 Open String Amplitudes and the Tensionless Limit\symbolfootnote[2]{Supported 
in part by the Department
of Energy under Grant No. DE-FG02-97ER-41029 and FAPESP grant 2012/05451-8}
}
\end{Large}
\vskip 2cm
{\large 
Francisco Rojas\,\symbolfootnote[1]{{\tt frojasf@ift.unesp.br}}
}
\vskip0.20cm
\centerline{{\it  
Instituto de F\'{i}sica Te\'{o}rica, UNESP-Universidade Estadual Paulista}} \centerline{{\it 
R. Dr. Bento T. Ferraz 271, Bl. II, S\~ao Paulo 01140-070, SP, Brasil}}
\vskip0.20cm
%\vskip12pt(\today)
\vskip 1.0cm
\end{center}

\begin{abstract}\noindent
The sum over planar multi-loop diagrams in the NS+ sector of type 0 open strings in flat spacetime has been proposed by Thorn as a candidate to 
resolve non-perturbative 
issues of gauge theories in the large $N$ limit. With $SU (N)$ Chan-Paton factors, the sum over planar open string multi-loop diagrams describes the 't Hooft limit $N\to \infty$ with $Ng_s^2$ held fixed. By including only planar diagrams in the sum the usual mechanism for the cancellation of loop divergences (which occurs, for example, among the planar and M\"obius strip diagrams by choosing a specific gauge group) is not available and a renormalization procedure is needed. In this article the renormalization is 
achieved by suspending total momentum conservation by an amount $p\equiv \sum_i^n k_i\neq 0$ at the level of the integrands
in the integrals over the moduli and analytically continuing them to 
$p=0$ at the very end. This procedure has been successfully tested for the 2 and 3 gluon planar loop amplitudes by Thorn. Gauge invariance is respected and the correct running of the coupling in the limiting gauge field theory was also correctly obtained. 
In this article we extend those results in two directions. First, we generalize the renormalization 
method to an arbitrary $n$-gluon planar loop amplitude giving full details for the 
4-point case. 
One of our main results is to provide a fully renormalized amplitude which is free of both UV and the usual spurious divergences leaving only the physical singularities in it. Second, using the complete renormalized amplitude, we extract the high-energy scattering regime at fixed angle (tensionless limit). Apart from obtaining the usual exponential falloff at high energies, 
we compute the full dependence on the scattering angle which shows the existence of a smooth connection between the Regge and hard scattering regimes. 

%With the open strings attached to a stack of $N$ coincident D$p$-branes
%with $SU(N)$ Chan-Paton factors, the low energy limit of these amplitudes
%become the amplitudes for the gluons in $N$ pure Yang-Mills theory 
%which allows us to compute the high-energy limit at fixed angle for 
%gluon scattering. 
\end{abstract}
\vfill
\end{titlepage}

%\tableofcontents 

\section{Introduction}
Ever since 't Hooft's original suggestion that the large $N$ limit of gauge theories should possess a 
dual string  description \cite{'tHooft:1973jz} there has been an enormous amount of efforts to find the 
corresponding dual description for large $N$ QCD. With the advent of the AdS/CFT 
correspondence \cite{Maldacena:1997re,Witten:1998qj,Gubser:1998bc} much has been learned about 
the nonperturbative regime of gauge field theories, however, the precise string picture dual to QCD in 
the large $N$ limit still remains undelivered.  

A different approach for resolving nonperturbative issues such as confinement in gauge theory has been put 
forward by Thorn \cite{Thorn:Summing,Thorn:Digital} 
where the strategy is to perform 
the summation of \emph{open string} multi-loop diagrams instead of \emph{field theoretic} multi-loop diagrams, delaying
the $\alpha' \to 0$ limit for only after computing the sum\footnote{More recent 
developments in this program have been reported 
in \cite{Papathanasiou:2012mi,Papathanasiou:2012fn,Papathanasiou:2013nta}. For other 
directions on the connections between string amplitudes and field theory Feynman 
diagrams see \cite{Magnea:2013lna}.}. It is important to recall that 't Hooft's limit corresponds to summing all the 
planar Feynman diagrams of the field theory, and that these diagrams are the $\alpha' \to 0$ limit of 
the planar open string multi-loop diagrams order by order in the perturbative expansion.
The main idea is that, since the perturbative expansion 
in string theory has far fewer diagrams than the field theory one, the multi-loop sum could be 
more tractable for string diagrams rather than field theory diagrams. 

In \cite{Thorn:Summing,Thorn:Digital} this program was initiated using type 0 strings mainly for two reasons: (i) the spectrum 
of type 0 strings is purely bosonic and the one of large $N$ QCD is straightforward to obtain from the low energy limit of the 
open sector of the type 0 theory, and (ii) the presence of a tachyon in its closed string spectrum could produce the desired instability 
to drive its perturbative vacuum to the true (large $N$) QCD vacuum \cite{Klebanov:1998yya, Klebanov:1998yy, Klebanov:1999ch, Minahan:1999yr}. If the stabilization indeed occurs, it should manifest after 
the multi-loop summation is performed.

The first tests of type 0 open string theory as a viable model for the multi-loop diagram summation 
of \cite{Thorn:Summing,Thorn:Digital} were performed in 
\cite{Thorn:Subcritical} where it was obtained the correct running coupling behavior of the limiting 
gauge theory by studying the planar 2 and 3-gluon amplitudes at one-loop. At this point 
is it important to stress a crucial fact: since only planar diagrams participate in the multi-loop sum 
of \cite{Thorn:Summing,Thorn:Digital} the usual cancellation of loop UV divergences, 
which occurs for example among the planar and Moebius strip diagrams by choosing a 
specific gauge group for the Chan-Paton factors\footnote{For the type I superstring for example, 
the Chan-Paton gauge group is SO(32).}, no longer takes place and a renormalization 
procedure is necessary to manage these infinities.
For the 2 and 3 gluon cases the renormalization was achieved by an analytic continuation which consists in suspending 
total momentum conservation by an amount $\boldsymbol{p}$ \emph{before} performing the integrals over the 
moduli (called GNS regularization in 
\cite{Thorn:Subcritical}), i.e., one takes $\sum_i^n k_i = \boldsymbol{p} \neq 0$ at the level of  the integrands, 
where $k_i$ are the 
momenta of the $n$ external gluons. Only after performing the integrations one analytically continues 
the answer to $\boldsymbol{p}=0$. 

As an example of how this procedure works, 
consider the planar 1-loop amplitude for two external gluons in bosonic string theory. The 
amplitude is proportional to
\beq\bal\label{intro2gluon}
\mathcal{M}_2^{\rm Bose} &=\int_0^{\pi} d\theta \left[\sin \theta \right]
^{2\alpha^{\prime} k_1 \cdot k_2-2},
\eal\eeq
where $k_1$ and $k_2$ are the momenta of the external gluons. From here we see that, since for 
gluons we have $k_i^2=0$, the exponent above is $2\alpha^{\prime} k_1 \sdot k_2=\alpha'(k_1+k_2)^2=0$ due to total momentum conservation $k_1+k_2=0$. We therefore have
\eq{
\mathcal{M}_2^{\rm Bose} &=\int_0^{\pi} d\theta \left[\sin \theta \right]
^{-2},
}
which diverges due to the the infinite contributions to the integral coming 
from the regions where $\theta\sim 0$ and $\theta\sim \pi$. These are spurious divergences that typically occur 
in open string loop diagrams that usually come as integral representations outside their domain of 
convergence. Therefore, one needs to analytically continue the integrals in order to get rid of the spurious infinities leaving only 
the physical ones such as infrared and collinear divergences.
The usual way would be to integrate by parts in \eqref{intro2gluon} to extend its domain of convergence 
as a function of the complex variable $k_1\cdot k_2$. This is indeed not hard to do here, but it 
is impractical for higher point amplitudes that involve multi-dimensional integrals over many $\theta$ 
variables. Already for the 3-gluon amplitude this gets very intricate.

Our method is to suspend momentum conservation at the level of the integrands by taking 
$\sum_i^n k_i=p\neq 0$, and then to analytically continue the result to $p=0$ at the end. This 
way we now have  $2\alpha^{\prime} k_1 \cdot k_2=\alpha'(k_1+k_2)^2=\alpha'p^2$ instead of 
zero. The amplitude \eqref{intro2gluon} now reads
\eq{
\mathcal{M}_2^{\rm Bose} &=\int_0^{\pi} d\theta \left[\sin \theta \right]
^{\alpha'p^2-2}= 
\frac{\Gamma(1/2)\Gamma(\alpha^{\prime} p^2/2-1/2)}{\Gamma(\alpha^{\prime} p^2/2)} =  -\frac{\pi \alpha^{\prime} p^2}{4} + \mathcal{O}(p^4)
\label{eulerian},
}
where, for Re$(\alpha'p^2)>1$, we recognize it in the second equal sign as the integral representation for the the Euler beta function that has a smooth $p\to 0$ 
limit as shown. Even better, from the power series 
expansion in $p$, we see that the continuation to $p\to 0$ gives $\mathcal{M}_2^{\rm Bose}=0$, which is very welcome for the 2-gluon 
amplitude at 1-loop since gauge invariance must also hold 
order by order in perturbative string theory, this is, the gluon mass must not receive loop corrections. 

For the 3-gluon amplitude for the planar one loop the procedure also 
works but it is considerably more complicated than the 2 gluon case (see section 4.2 in   
\cite{Thorn:Subcritical}). Based on these results, it does not seem obvious that the procedure continues to work for higher point amplitudes.

One of the main results of the present article is that we show that the analytic continuation procedure does extend to an arbitrary number 
of external gluons (planar loop $n$-gluon amplitude) and we also give the
full details of the computation for 4 gluons, providing a novel renormalized 
expression for the amplitude which is completely free of spurious and UV divergences. As a result, the 
renormalized amplitude we give contains physical divergences only and, for example, is ready to provide the correct field theory limit by taking 
$\alpha'\to 0$ without worrying about the known spurious infinities that arise from the usual 
integral representations of stringy loop amplitudes\footnote{At the level of an $n$-point amplitude the spurious divergences are the ones that arise from the integration regions where all or all but one vertex operators get 
arbitrarily close to each other in moduli space. See section 
9.5  in J.~Polchinski's, ``String theory. Vol. 1: An introduction to the bosonic string,'' \cite{Polchinski:1998rq} for a more detailed 
discussion.}. We also show that the UV divergences and all the spurious ones can be regulated altogether by means of single counterterm 
using the regulator $p\equiv \sum_i k_i\neq0$. After this is done, we analytically continue the amplitude to $p=0$ and 
arrive at the final renormalized expression. 

The second part of this article concerns the high energy limit for the scattering of type 0 open 
strings. Here we extend our analysis of \cite{ThornRojas} by studying the high energy regime of the 
planar one-loop amplitude for 4 gluons at fixed scattering angle 
(hard scattering). 
One of the main results of this second part is that we explicitly show that the hard-scattering and Regge regimes 
are smoothly connected since there is an overlapping region in the moduli 
where the two approaches 
yield the same results. Note also that, since 
all the Mandelstam variables come multiplied with a factor of $\alpha'$, the hard scattering 
regime is exactly equivalent to the tensionless limit ($\alpha'\to \infty$) with external particles 
held at fixed momenta.

By carefully analyzing all dominant regions we extract the leading behavior 
of the amplitude providing its complete kinematic dependence. This includes the exact dependence 
on the scattering angle that multiplies the usual exponentially decaying factor. 
Although in order to compare our results with those of \cite{ThornRojas} we focus on the particular 
polarization structure $\epsilon_1 \cdot  \epsilon_4 \, \epsilon_2\cdot \epsilon_3$ (the dominant one in the Regge regime), 
our results are general and can be straightforwardly extended to all the other polarization structures. 

For the planar loop amplitude the leading behavior we obtain for large $\alpha'|s|$ with fixed $\lambda\equiv -t/s$ (fixed angle)
is
\eq{
\mathcal{M}\sim  F(\lambda)\, e^{-\alpha'
|s| f(\lambda)} 
\left(\frac{1}{\ln\alpha' |s|}\right)^{\gamma-1}(-\alpha' s)^{3/2}
\label{Mpfinal2}
}
where $f(\lambda) \equiv \lambda \ln(-\lambda)+(1-\lambda)\ln(1-\lambda)$ and the function 
$F(\lambda)$ is 
given by
\eq{
F(\lambda)\equiv \int_0^{\pi}d\theta\int_0^{\infty}dr \frac{r\sin^2\theta\, (r^2+2r\cos\theta+1)^{-1}}
{r^2(1-\lambda)^2+2r(1-\lambda)\cos\theta+1}}

The usual case occurs when $\gamma=1$ which corresponds to a space-time filling D-brane, but 
for smaller dimensional D-branes the behavior gets soften (since $\gamma>1$) by the logarithmic factor above.
The $\lambda \sim 0$ analysis of $F(\lambda)$ allows to see that there exists a smooth 
connection between the hard scattering and Regge regimes; to our knowledge this is also a new result 
and it is explained in detail in section (4.3).

\begin{comment}
This regularization procedure respects gauge invariance as shown in 
\cite{Thorn:Subcritical, neveuscherkrenorm,ThornRojas} and Minahan \cite{Minahan:1987ha} showed that 
both conformal and modular invariances are not violated either. We also provide the systematic procedure to obtain the renormalized 
expression for an arbitrary $n$-gluon planar loop diagram in Section 3. 
\end{comment}

In \cite{ThornRojas} we studied the one-loop correction to the open string Regge trajectory 
$\alpha(t)=1+\alpha't + g^2\Sigma(t)$ and also extracted its field theory limit in order to deepen our understanding 
of the suitability of type 0 open strings  as an `uplifted' tensionful model ($\alpha'\neq 0$) of Yang-Mills theory. 
In \cite{ThornRojas}, by using the regulator $p=\sum_i^4 k_i$ for the $4-$gluon amplitude 
and carefully taking the $\alpha' \to 0$ limit in the renormalized expression we obtained for $\Sigma(t)$ using 
the analytic continuation procedure previously mentioned, we were able to recover the known answer for the 
one-loop gluon Regge trajectory in dimensionally regularized Yang-Mills theory \cite{kunsztst,chakrabartiqt,chakrabartiqt2,thornresir}. 
%
\begin{comment}
In order to obtain pure Yang-Mills theory in the 
$\alpha'\to 0$ limit, we also projected out the massless scalars from the loop using the method 
proposed by Thorn in \cite{Thorn:Nonabelian}. Since we were interested in planar diagrams only, and due to 
the absence of
supersymmetry in type 0 models (there are no spacetime fermions),
ultraviolet divergences do arise in loop amplitudes and it is necessary to regulate them
and perform an analytic continuation in order to obtain the correct field theory 
limit\footnote{For instance, for type I superstrings in flat spacetime, 
the absence of UV divergences at one loop 
occurs only through a cancellation of infinities among the planar and Moebius strip diagrams.}. 
\end{comment}
%

The high energy behavior of one-loop open string amplitudes has been studied 
since the very early days of string theory 
\cite{Alessandrini:1972jy,Dorn:1974er,DornKaiser,Otto:1976pu}, and more recently in \cite{Moeller:2005ez}. In 
\cite{Alessandrini:1972jy} Alessandrini, Amati, and Morel studied the 
high energy limit at fixed angle (hard scattering) for the one-loop non-planar amplitude 
of four open string tachyons.  The same high energy regime for an arbitrary number loops 
in the bosonic open string was studied in the late 1980's by Gross and Ma\~nes \cite{GrossManes}. 
One of their main conclusions was 
that, similarly to the study of closed strings in \cite{GrossMende1}, the amplitude for four external 
open string tachyons had a dominant saddle point at all genus, implying that the leading behavior can be obtained by 
analyzing the contribution to the amplitude around these saddle points. Moreover, extending the closed string semi-classical 
analysis of \cite{GrossMende1} to the open string case, the authors of \cite{GrossManes} found that the open string planar 
amplitude does not possess saddle points in the interior of moduli space. Therefore, the only regions that could 
potentially give the dominant behavior at high energies and fixed angle are the boundaries of the moduli space. This conclusion extends 
immediately to type 0 open strings because the relevant dependence on the external 
momenta is identical in both, the bosonic and the type 0 string models.

The organization of this paper is as follows. In section 2 we familiarize the reader on the computation of the annulus amplitude for 
external gluons in %the even G-parity sector of the original Neveu-Schwarz model 
type 0 open string theory, and we introduce 
the analytic continuation procedure to regulates both UV and spurious divergences. We show the calculation of the 2-gluon amplitude 
\cite{Thorn:Subcritical} as a simple example of the method, and give the full details of our procedure for the 4-gluon case. 
In section 3 we give the systematics of the generalization for an arbitrary number of external gluons. 
Once having obtained the full renormalized expression for the 4-gluon amplitude, in section 4 we 
compute its tensionless limit, i.e., the high energy regime at fixed-angle (hard scattering). By taking 
$s\gg t$ and comparing this with the $\alpha'\to \infty$ limit of the Regge behavior of the 4-gluon amplitude found in 
\cite{ThornRojas}, we find perfect agreement with our results, thus, explicitly 
showing the smooth connection between the hard and Regge regimes. In appendix A we show a different procedure to 
project out massless scalars circulating in the loop based on an orbifold projection \cite{ThornRojas}, and in appendix B we provide 
the explicit form of the counterterms needed in the 4-gluon case.

\begin{comment}
However, it has been pointed out that their results 
may not be valid for external string states other than tachyons  
\cite{Moeller:2005ez}. A detailed one-loop analysis is also provided in 
reference \cite{Moeller:2005ez}.
\end{comment}

\section{One loop planar amplitude and renormalization} \label{chap:GNSrenorm}
We start with a very brief discussion about the basic elements of type 0 theories. These are ten dimensional string theories 
that are obtained by the GSO projection
\bea
\frac{1}{2}(1+(-1)^F)\label{GSO}
\eea
on the open string sector, and
\bea
\frac{1}{2}(1+(-1)^{F+\tilde{F}})
\eea
on the closed string sector, where $F$ is the world-sheet fermion number. The closed string spectrum is
\begin{align*}
\mbox{type 0A : } & (NS-,NS-)\oplus(NS+,NS+)\oplus(R+,R-)\oplus(R-,R+) \\
\mbox{type 0B : } & (NS-,NS-)\oplus(NS+,NS+)\oplus(R+,R+)\oplus(R-,R-)
\end{align*}
Although there are no fermions in the spectrum, these projections produce modular invariant partition 
functions \cite{thornsantafe,Dixon:1986iz,Seiberg:1986by}. Note also that, although the GSO 
projection eliminates the open string tachyon from the spectrum, there remains a closed string tachyon 
from the $(NS-,NS-)$ sector. However, the doubling of R-R fields has an stabilizing effect 
on the closed string tachyon by giving its mass-squared a positive shift \cite{Klebanov:1998yya}.
The approach proposed by Thorn suggests 
that this instability could also resolved by the planar multi-loop summation of type 0 open string diagrams 
\cite{Thorn:Subcritical,Thorn:Nonabelian,Thorn:Summing,Thorn:Digital}. 

In this article we are mainly interested in the open string sector of the type 0 model. Its free spectrum, 
after the GSO projection \eqref{GSO} is $\alpha' M^2 = 0,1,2,\dots$. The lowest mass state is 
$\epsilon \cdot b_{-1/2}|0,k\rangle$ with $k^2=0$ and 
$k\cdot \epsilon=0$. This massless gauge state will be called the ``gluon'' in the rest of this article.

By projecting out the states with odd fermion worldsheet number, the tachyon of the NS sector 
is removed and the low energy excitations of a D$p$-brane correspond to massless gauge fields 
and scalars only \cite{Klebanov:1998yya}. This result also holds if one considers a stack of $N$ parallel 
like-charged D$p$-branes. Thus, the world-volume theory of this configuration of D$p$-branes in type 
0 theories describes a pure glue $U(N)$ gauge theory in $p$ spacetime dimensions coupled to $(9-p)$ 
massless adjoint scalars \cite{Klebanov:1998yya,Bergman:1997rf}. If one is only interested in 
pure Yang-Mills theory, these scalars can be removed by using orbifold projections or by using 
the nonabelian D-branes procedure of \cite{Thorn:Nonabelian}.

\subsection{Analytic continuation}
With the metric signature $\{-++\cdots \}$ the Mandelstam variables are conventionally defined as $s=-(k_1+k_2)^2$, $t=-(k_2+k_3)^2$, and $u=-(k_2+k_4)^2$. The integral 
expression for the $M$-point planar one-loop amplitude is plagued with  divergences in various ``corners'' of the integration region. We will examine these in detail in sections \ref{lineardiv} and \ref{logdiv}. These infinities simply arise from the use of an integral 
representation outside its domain of convergence \cite{neveuscherkrenorm}. The point we would like to 
stress here is that, since these divergences are a direct consequence of momentum 
conservation, if we allow for $\sum_{i=1}^M k_i \equiv p \neq 0$, we can regulate 
and track the effects of all of these divergences. Finally, we analytically continue the 
integrals to $p=0$ at the very end of our calculations. We will see that 
this technique leads to physically meaningful consequences such as gauge invariance 
because it allows to prove that massless vector bosons remain massless 
at one loop \cite{Minahan:1987ha,Thorn:Subcritical,ThornRojas}. In \cite{Minahan:1987ha} Minahan shows 
that such prescription does not violate conformal nor modular invariance. 
It will also prove to be important when we study the high energy regime at fixed scattering angle 
in section \ref{sec:hard}. This technique
was proposed and used long ago by Peter Goddard \cite{goddardreg} and Andr\'e Neveu and Joel Scherk 
\cite{neveuscherkrenorm} in the early days of string theory. 
The variable $p$ that represents the temporary 'suspension' of 
momentum conservation is referred to, in this article, as the Goddard-Neveu-Scherk or GNS 
regulator for short \cite{Thorn:Subcritical}.

We will first begin by writing the full type 0 open string planar one-loop amplitude for 
the scattering of $M$ ``gluons'' \cite{Thorn:Subcritical}. The open string coupling $g$ is normalized 
so that in the $\alpha' \to 0$ limit it is related to the QCD strong coupling
$g_s$ by $\alpha_s N=g_s^2N/4\pi=g^2/2\pi$. Thus $g$ is held fixed in the
large $N$ limit. 
We should also clarify that an overall group theory factor of 
$tr(T^{a_1}T^{a_2}T^{a_3}T^{a_4})$, coming from the $SU(N)$ Chan-Paton factors, is implicit in 
all of our expressions for the planar amplitudes. Having said this, the properly normalized $M$-gluon amplitude is
$(g\sqrt{2\alpha^\prime})^M$ times 
\bea
{\cal M}_M=\frac{1}{2}({\cal M}_M^+-{\cal M}_M^-)\label{M+-M-}
\eea
where $\mathcal{M}^+$ and $\mathcal{M}^-$ come from the $1$ and $(-1)^F$ respective parts 
of the GSO projection in \eqref{GSO}. The difference 
between these two expressions realizes the projection onto states with even fermion worldsheet number.

The complete 
expressions for $\mathcal{M}^{\pm}$ are
\bea\label{mgluononeloop}
{\cal M}_M^\pm
&=&\int \frac{dw}{w}  
\prod_{i=2}^M \frac{dy_i}{y_i}w^{-1/2}
\left({-1\over4\pi\alpha^\prime\ln w}\right)^{D/2}
\exp\left\{\alpha^\prime
\sum_{i< j}k_i\cdot k_j{\ln^2{y_i/y_j}\over\ln w}\right\}
\nonumber\\
&&%(1\mp w^{1/2})^{10-D-S}
\VEV{{\hat{\cal P}}(y_1)\cdots{\hat{\cal P}}(y_M)}^\pm{\prod_r(1\pm w^r)^8
\over\prod_n(1-w^n)^8}
\prod_{i<j}\left[2i{\theta_1\left(-i\ln\sqrt{y_i/ y_j},\sqrt{w}\right)
\over
\theta_1^\prime(0,\sqrt{w})}
\right]^{2\alpha^\prime k_i\cdot k_j}.
\label{empee}
\eea
Here, $n=1,2,\cdots$, $r=1/2,3/2,\cdots$. We use the notation and conventions
of \cite{Thorn:Subcritical}. The  
Koba-Nielsen variables $y_i$ are integrated over the range:
\bea
0<w<y_M<y_{M-1}<\cdots <y_2<y_1=1\;.
\eea
The presence of the factor $\left({-1\over4\pi\alpha^\prime\ln w}\right)^{D/2}$ 
in \eqref{mgluononeloop} comes from the fact that we are allowing the open 
string ends to be attached to a stack of $N$ coincident D$p$-branes for $p=D-1$. 
In the planar one-loop calculation, 
this amounts to integrating over only
the first $D$ components of the loop momentum and setting the
remaining components to zero. 

If we take the $\alpha^{\prime}\to 0$ limit 
at this point, we will not obtain the $M$-gluon amplitude in pure Yang-Mills theory, but Yang-Mills coupled 
to $10-D$ adjoint massless scalars \cite{Klebanov:1998yya}. The scalar excitations arise from the vibrations of the string in the 
directions perpendicular to the D-brane. In order to have just gluons circulating 
in the loop we need to project out these scalars. There is not a unique way to 
achieve this and the procedure we use here is the projection 
proposed in \cite{Thorn:Nonabelian}. A different procedure to eliminate the scalars from the loops 
is by introducing an orbifold projection as explained in \cite{ThornRojas}. We briefly show in 
appendix \ref{app:orb} that the orbifold projection produces the 
same answer in the field theory limit (i.e., $\alpha'\to 0$) as one we use here, but their effects differ as $\alpha'$ departs from zero. The projection \cite{Thorn:Nonabelian} 
produces an extra factor of $(1\mp w^{1/2})^{10-D-S}$ in the integrand above, where $S$ is the number of scalars remaining after the projection, which we also need to include. 
If one is interested large $N$ QCD, there are certainly 
no adjoint massless scalars in the spectrum, so we would need $S=0$. However, we will 
leave $S$ arbitrary in order to make our expressions more general. 

The factors in \eqref{empee} that contain the Jacobi $\theta_1$
function can be expressed in terms of an infinite product representation as
\bea
\prod_{i<j}y_j^{2\alpha^\prime k_i\cdot k_j}
\prod_{i<j}\left[2i{\theta_1\left(-i\ln\sqrt{y_i/y_j},\sqrt{w}\right)
\over
\theta_1^\prime(0,\sqrt{w})}
\right]^{2\alpha^\prime k_i\cdot k_j}&=&\nonumber\\
&&\hskip-1.5in\prod_{i<j}\left[\left(1-{y_j\over y_i}\right)\prod_n
{\left(1-w^n{y_i/y_j}\right)\left(1-w^n{y_j/y_i}\right)
\over(1-w^n)^2}\right]^{2\alpha^\prime k_i\cdot k_j}\;.
\eea
Following \cite{Thorn:Subcritical}, the gluon vertex operator is $V=e^{i k\cdot x}(\epsilon\cdot{\cal P}
+\sqrt{2\alpha^\prime}k\cdot H\epsilon\cdot H)\equiv e^{i k\cdot x}
{\hat{\cal P}}$.
The $\VEV{\cdots}$ correlator involves a finite number of
${\cal P}$ (bosonic) and $H$ (fermionic) worldsheet fields and it's determined by its Wick expansion
with the following contraction rules:
\bea
\VEV{{\cal P}(y_l)}&=&\sqrt{2\alpha^\prime}\sum_i k_i\left[
-{\ln(y_i/y_l)\over\ln w}+{1\over2}{y_i+y_l\over y_l-y_i}
+\sum_{n=1}^\infty\left({y_iw^n\over y_l-y_iw^n}-{y_lw^n\over y_i-y_lw^n}
\right)\right]\nonumber\\
\VEV{{\cal P}^\mu(y_i){\cal P}^\nu(y_l)}&=&\VEV{{\cal P}^\mu(y_i)}
\VEV{{\cal P}^\nu(y_l)}
+\eta^{\mu\nu}\left[
-{1\over \ln w}+{y_i y_l\over(y_i-y_l)^2}
\right.\nonumber\\
&&\hspace{140pt}\left.+\sum_{n=1}^\infty\left({y_i y_lw^n\over(y_l-y_iw^n)^2}
+{y_i y_lw^n\over(y_i-y_lw^n)^2}\right)\right]\nonumber\\
\VEV{H^\mu(y_i)H^\nu(y_j)}^+&=&\eta^{\mu\nu}
\sum_r{(y_j/y_i)^r+(wy_i/y_j)^r\over
1+w^r}\nonumber\\
\VEV{H^\mu(y_i)H^\nu(y_j)}^-&=&\eta^{\mu\nu}
\sum_r{(y_j/y_i)^r-(wy_i/y_j)^r\over
1-w^r} .
\eea
The two types of traces over the $b_r$ oscillators are distinguished with the $\pm$ superscript: for $+$ odd
and even G-parity states contribute with the same sign, whereas
$-$ denotes the contributions with opposite signs. In the ${\cal F}_2$
picture, the difference of the two traces, which amounts to taking $\mathcal{M}^+-\mathcal{M}^-$ 
\eqref{M+-M-}, projects out all the odd G-parity states; the open string tachyon being one of them. 
 
\begin{figure}[ht!]
\centering
\includegraphics[scale=0.65]{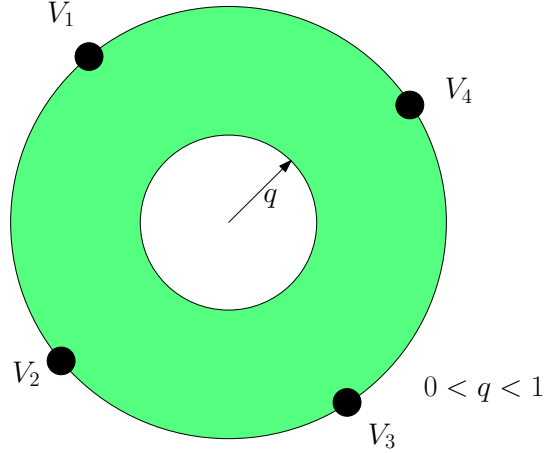}
\caption{The one-loop planar diagram with four external states. Notice that all states are located at only 
one of the two boundaries of this topology, namely, the outer boundary. 
A non-planar diagram would have particles attached to both, outer and inner boundaries}
\label{fig:annulus}
\end{figure}

In the cylinder variables $\theta_i=\pi\ln y_i/\ln w$
and $\ln q=2\pi^2/\ln w$, the planar one-loop amplitude is 
\bea
{\cal M}_M^+&=&
2^M\left({1\over8\pi^2\alpha^\prime}\right)^{D/2}
\int \prod_{k=2}^M {d\theta_k}\int_0^1{dq\over q}
\nonumber\\&&\qquad
%{\prod_r(1+q^{2r})^{8}\over\prod_n(1-q^{2n})^{8}}
\left({-\pi\over\ln q}\right)^{(10-D)/2}
P_+(q)\prod_{l<m}\left[\psi(\theta_m-\theta_l,q)
\right]^{2\alpha^\prime k_l\cdot k_m}
\VEV{{\hat{\cal P}}_1{\hat{\cal P}}_2\cdots{\hat{\cal P}}_M}^+\nonumber\\
\label{cylmgluon+}\\
{\cal M}_M^-&=&
2^M\left({1\over8\pi^2\alpha^\prime}\right)^{D/2}
\int \prod_{k=2}^M {d\theta_k}\int_0^1{dq\over q}
\nonumber\\&&\qquad
\left({-\pi\over\ln q}\right)^{(10-D)/2}P_-(q)
\prod_{l<m}\left[\psi(\theta_m-\theta_l,q)
\right]^{2\alpha^\prime k_l\cdot k_m}
\VEV{{\hat{\cal P}}_1{\hat{\cal P}}_2\cdots{\hat{\cal P}}_M}^-
\label{cylmgluon-}
\eea
where
\bea
P_+(q)&\equiv&q^{-1}(1-w^{1/2})^{10-D-S}
{\prod_r(1+q^{2r})^{8}\over\prod_n(1-q^{2n})^{8}}\label{P+}\\
P_-(q)&\equiv&2^{4}(1+w^{1/2})^{10-D-S}
{\prod_n(1+q^{2n})^{8}\over\prod_n(1-q^{2n})^{8}}\label{P-}\\
\psi(\theta,q)&=&\sin{\theta}\prod_n{(1-q^{2n}e^{2i\theta})
(1-q^{2n}e^{-2i\theta})\over(1-q^{2n})^2}\label{qpsi}\\
{\hat{\cal P}}&=& \epsilon\cdot
{\cal P}+\sqrt{2\alpha^\prime}k\cdot H\epsilon\cdot H ,
\eea
Figure \ref{fig:annulus} shows the one-loop planar diagram (annulus) for the $M=4$ case.
The expressions for $P_{\pm}(q)$ above include the aforementioned factor of $(1\mp w^{1/2})^{10-D-S}$ 
that accounts for the projection that leaves $S$ massless scalars circulating in the loop. As an example, 
consider the 
more familiar case with D3-branes and 6 adjoint massless scalars. In this case $D=4, S=6$, gives 
$(1\mp w^{1/2})^{10-D-S}=1$ yielding the usual partition function. The average $\VEV{\cdots}$ is evaluated 
with contractions:
\bea
\VEV{{\cal P}_l}&=&\sqrt{2\alpha^\prime}\sum_i k_i
\left[{1\over2}\cot{\theta_{il}}
+\sum_{n=1}^\infty{2q^{2n}\over1-q^{2n}}
\sin 2n\theta_{il}\right]\\
\VEV{{\cal P}_i{\cal P}_l}
-\VEV{{\cal P}_i}\VEV{{\cal P}_l}
&=&{1\over4}\csc^2{\theta_{il}}
-\sum_{n=1}^\infty n{2q^{2n}\over1-q^{2n}}
\cos 2n\theta_{il}\label{PP}\\
\VEV{H_iH_j}^+&\equiv&\chi_+(\theta_{ji})
={1\over2\sin\theta_{ji}}
-2\sum_r{q^{2r}\sin 2r\theta_{ji}\over
1+q^{2r}}={1\over2}\theta_2(0)\theta_4(0){\theta_3(\theta_{ji})\over
\theta_1(\theta_{ji})}\label{chi+def}\\
\VEV{H_iH_j}^-&\equiv&\chi_-(\theta_{ji})
={\cos\theta_{ji}\over2\sin\theta_{ji}}
-2\sum_n{q^{2n}\sin 2n\theta_{ji}\over1+q^{2n}} 
={1\over2}\theta_3(0)\theta_4(0){\theta_2(\theta_{ji})\over
\theta_1(\theta_{ji})}\;.\label{chi-def}\label{HHm}
\eea
We have abbreviated $\theta_{ji}=\theta_j-\theta_i$
and space-time indices were suppressed.
Finally  the range of integration is
\bea
0=\theta_1<\theta_2<\cdots<\theta_N<\pi .
\eea 

To see the GNS regulator at work, consider the one-loop 2-gluon function studied 
in \cite{Thorn:Subcritical} which controls the mass shifts of the gluon in perturbation theory. For the coefficient of 
$\epsilon_1 \dt \epsilon_2$, the bosonic part of the string amplitude 
is\footnote{We are omitting here all constant pre-factors in the amplitude for convenience.}:
\bea\label{2gluon}
\mathcal{M}_2^{Bose} &=& \int_0^1 [dq]^{\pm} \int_0^{\pi} d\theta \left[\sin \theta \prod_{n=1}^{\infty} \frac{1-2q^{2n} \cos 2\theta 
+ q^{4n}}{(1-q^{2n})^2}\right]^{2\alpha^{\prime} k_1 \cdot k_2}\times\nn\\
&&\left[\frac{1}{4}\csc^2 \theta - \sum_{n=1}^{\infty} n \frac{2q^{2n}}{1-q^{2n}} \cos 2 n \theta \right]
\eea
where we use $[dq]^{\pm}$ as a short-hand for 
$\frac{dq}{q}\left(\frac{-\pi}{\ln q}\right)^{(10-D)/2}P_{\pm}(q)$ 
since this factor is not relevant for the discussion below. 

\noindent Momentum conservation implies $2 k_1 \dt k_2 = (k_1+k_2)^2 =0$, thus
\bea
\mathcal{M}_2^{Bose} &=& \int_0^1 [dq] \int_0^{\pi} d\theta \left[\frac{1}{4}\csc^2 \theta 
-\sum_{n=1}^{\infty} n \frac{2q^{2n}}{1-q^{2n}} \cos 2 n \theta \right]
\eea
from where we see that the first term is clearly divergent in the 
$\theta\sim 0,\pi$ regions. However, by using the GNS regulator we will show that 
this is a spurious 
divergence due to an integral representation outside its domain of convergence.
In order to analytically continue the amplitude, 
we suspend momentum conservation in the intermediate steps by using the regulator $p=\sum_i k_i$, 
so that now we have $2 k_1 \dt k_2 = p^2$ instead of $2 k_1 \dt k_2 =0$. This makes integral 
perfectly convergent for ${\rm Re} (\alpha^{\prime} p^2)>1$. We then analytically continue 
to $p \to 0$ at the end. Notice that there is only one angular integration in the two 
gluon function. This will allow us to perform the the analytic continuation to $p=0$ rather 
straightforwardly as we shall now see. This is in contrast with four and higher point functions where 
the angular integrals becomes multi-dimensional and technically more complicated. Writing \eqref{2gluon} again, 
but this time with the $p$ regulator turned on, reads
\bea
\mathcal{M}_2^{Bose} &=& \int_0^1 [dq] \int_0^{\pi} d\theta [\sin \theta]^{\alpha^{\prime} p^2} 
\left[ \prod_{n=1}^{\infty} \frac{1-2q^{2n} \cos 2\theta + q^{4n}}{(1-q^2n)^2}\right]^{\alpha^{\prime} p^2}\nonumber\\ 
&&\hspace{1.8cm}\times \left[\frac{1}{4}\csc^2 \theta - \sum_{n=1}^{\infty} n \frac{2q^{2n}}{1-q^{2n}} \cos 2 n \theta \right]
\eea
Expanding the infinite product up to first order in $p^2$ is enough for our purposes. 
Doing this and performing a resummation yields
\bea
\left[\prod_{n=1}^{\infty} \frac{1-2q^{2n} \cos 2\theta + q^{4n}}{(1-q^{2n})^2}\right]^{\alpha^{\prime} p^2} = 1+
\alpha^{\prime} p^2\sum_{m=1}^{\infty} \frac{1}{m} \frac{2q^{2m}}{1-q^{2m}} (1-\cos 2 m \theta) + \mathcal{O}(p^2)
\label{infprodexp} 
\eea
therefore
\bea
\mathcal{M}_2^{Bose} &=& \int_0^1 [dq] \left[\frac{1}{4}\int_0^{\pi} d\theta \, [\sin \theta]^{\alpha^{\prime} p^2-2} +\right.\nn\\
&& -\sum_{n=1}^{\infty} n  \frac{2q^{2n}}{1-q^{2n}}  \int_0^{\pi} d\theta 
\,[\sin \theta]^{\alpha^{\prime} p^2}  \cos 2 n \theta + \nn\\
&& + \alpha^{\prime} p^2 \sum_{m=1}^{\infty} \frac{1}{m} \frac{2q^{2m}}{1-q^{2m}}  
\frac{1}{4}\int_0^{\pi} d\theta \, [\sin \theta]^{\alpha^{\prime} p^2-2} (1-\cos 2 m \theta)+\nn\\
&& \left.-  \alpha^{\prime} p^2 \sum_{m,n=1}^{\infty} \frac{1}{m} \frac{2q^{2m}}{1-q^{2m}}  
 \frac{n2q^{2n}}{1-q^{2n}} \int_0^{\pi} d\theta \, 
[\sin \theta]^{\alpha^{\prime} p^2}  \cos 2 n \theta (1-\cos 2 m \theta)\right]
\eea
Without the regulator, the only problematic term here is the first one, since putting $p^2=0$ in the integrand shows a linear divergence 
in the $\theta$ integration. However if we assume that ${\rm Re} (\alpha'p^2)>1$ we have
\bea
\frac{1}{4}\int_0^{\pi} d\theta \, [\sin \theta]^{\alpha^{\prime} p^2-2} = \frac{1}{4}
\frac{\Gamma(1/2)\Gamma(\alpha^{\prime} p^2/2-1/2)}{\Gamma(\alpha^{\prime} p^2/2)} =  -\frac{\pi \alpha^{\prime} p^2}{4} + \mathcal{O}(p^4)
\label{eulerian}
\eea
Thus, taking the right hand side to be the analytic continuation of the 
left-hand side as $p\to 0$, we have a convergent expression.
The rest of the integrals are completely convergent even if we set $p^2=0$ directly in their integrands. Thus, we now have
a new expression which we take it to be the analytic continuation of 
\eqref{2gluon} to $p\to 0$, that reads
\bea
\mathcal{M}_2^{Bose} &=&  \int_0^1 [dq] \left[-\frac{\pi \alpha^{\prime} p^2}{4} + \alpha^{\prime} p^2 \sum_{n=1}^{\infty} \frac{2q^{2n}}{1-q^{2n}} n \frac{\pi}{2n} + \alpha^{\prime} p^2 \frac{1}{4}\sum_{m=1}^{\infty} \frac{1}{m} \frac{2q^{2m}}{1-q^{2m}}2\pi m +\right.\nn\\
&& \left.+  \alpha^{\prime} p^2 \sum_{m=1}^{\infty} \frac{1}{m} \frac{2q^{2m}}{1-q^{2m}}  
\sum_{n=1}^{\infty} n \frac{2q^{2n}}{1-q^{2n}} \frac{\pi}{2} \delta_{n,m}\right]\nn\\
&=& \pi \alpha^{\prime} p^2  \int_0^1 [dq]\left[-\frac{1}{4} + \sum_{n=1}^{\infty} \frac{q^{2n}}{1-q^{2n}}  + 
\sum_{m=1}^{\infty} \frac{q^{2m}}{1-q^{2m}} +  \sum_{n=1}^{\infty} \frac{2q^{4n}}{(1-q^{2n})^2} \right]\nn\\
&=& \pi \alpha^{\prime} p^2  \int_0^1 [dq]\left[-\frac{1}{4} + \sum_{n=1}^{\infty} \frac{2q^{2n}}{(1-q^{2n})^2} \right]
\label{2gluonresult}
\eea
which shows that, not only the limit $p \to 0$ is finite, but that it is actually zero. This is very welcome here since 
the vanishing of the two-gluon function guarantees that the gluon remains massless in perturbation theory, which is a consequence of gauge invariance. \\
The complete two-gluon amplitude is \cite{Thorn:Subcritical} given by
\bea
\mathcal{M}_2^{+} &\sim& \pi \alpha^{\prime} p^2 \int [dq]^+ \left[-\frac{1}{2}+4\sum_{n=1}^{\infty} \frac{q^{2n}}{(1-q^{2n})^2}+
4\sum_{r=1/2}^{\infty} \frac{q^{2r}}{(1+q^{2r})^2}\right]\\
\mathcal{M}_2^{-} &\sim& \pi \alpha^{\prime} p^2 \int [dq]^- \left[4\sum_{n=1}^{\infty} \frac{q^{2n}}{(1-q^{2n})^2}+
4\sum_{n=1}^{\infty} \frac{q^{2n}}{(1+q^{2n})^2}\right]
\eea
which shows that the analytically continued result to $p \to 0$ for the full two-gluon function is indeed zero.

To motivate the general result for the $M$-gluon amplitude for the planar loop, let us consider 
the four gluon function. In order to be able to compare the calculations we do in this article 
with the results of \cite{ThornRojas} 
we will focus on a particular polarization structure, namely the coefficient 
of $\epsilon_1 \cdot \epsilon_4 \epsilon_2 \cdot \epsilon_3$. The main reason to 
do this is that the coefficient of this factor is the one that dominates in the Regge limit 
($s\to-\infty$ with $t$ fixed) at tree and one-loop levels. 
At tree level, the 4-gluon amplitude for the type 0 string (for the polarization above and omitting numerical coefficients) is
\bea
M_4^{tree} = g^2 \frac{\Gamma(1-\alpha^{\prime} s)\Gamma(-\alpha^{\prime} t)}{\Gamma(-\alpha^{\prime} s -\alpha^{\prime} t)}
\label{nstree}
\eea
At one-loop the general form of the 4-gluon amplitude is given by  
\bea
\mathcal{M}_4 = \frac{1}{2} \left(\mathcal{M}_4^+-\mathcal{M}_4^-\right)
\eea
with
\bea
\mathcal{M}_4^+ &=& 2^4\left(\frac{1}{8 \pi \alpha^{\prime}}\right)^{D/2} \int_0^1 \frac{dq}{q} 
\left(\frac{-\pi}{\ln q}\right)^{(10-D)/2} q^{-1}(1-w^{1/2})^{10-D-S} \frac{\prod_{r}^{\infty} (1+q^{2r})^8}{\prod_{n}^{\infty} (1-q^{2n})^8\nonumber}\\
&& \int \prod_{k=2}^4 d\theta_k \prod_{i<j} [\psi (\theta_{ji})]^{2\alpha^{\prime} k_i \cdot k_j} 
\langle \hat{\mathcal{P}}_1 \hat{\mathcal{P}}_2 \hat{\mathcal{P}}_3 \hat{\mathcal{P}}_4 \rangle^+ \label{cylmgluon+2}\\
\mathcal{M}_4^- &=& 2^4\left(\frac{1}{8 \pi \alpha^{\prime}}\right)^{D/2} \int_0^1 \frac{dq}{q} 
\left(\frac{-\pi}{\ln q}\right)^{(10-D)/2} 2^4 (1+w^{1/2})^{10-D-S} \frac{\prod_{r}^{\infty} (1+q^{2n})^8}{\prod_{n}^{\infty} (1-q^{2n})^8}\nonumber\\
&& \int \prod_{k=2}^4 d\theta_k \prod_{i<j} [\psi (\theta_{ji})]^{2\alpha^{\prime} k_i \cdot k_j} 
\langle \hat{\mathcal{P}}_1 \hat{\mathcal{P}}_2 \hat{\mathcal{P}}_3 \hat{\mathcal{P}}_4 \rangle^- \label{cylmgluon-2}
\eea
Picking out the combination that multiplies $\epsilon_1 \cdot \epsilon_4 \epsilon_2 \cdot \epsilon_3$ from the corrrelator gives
\bea
\VEV{{\hat{\cal P}}_1{\hat{\cal P}}_2{\hat{\cal P}}_3{\hat{\cal P}}_4}
&\to&\epsilon_2\cdot\epsilon_3 \epsilon_1\cdot\epsilon_4\bigg(
\VEV{{{\cal P}}_2{{\cal P}}_3}\VEV{{{\cal P}}_1{{\cal P}}_4}-\VEV{{{\cal P}}_2{{\cal P}}_3}\VEV{H_1H_4}^22\alpha^\prime k_1\cdot k_4\nonumber\\
&&\hskip-1in
-\VEV{{{\cal P}}_1{{\cal P}}_4}\VEV{H_2H_3}^22\alpha^\prime k_2\cdot k_3
+4\alpha^{\prime2}\VEV{H_2H_3}\VEV{H_1H_4}
\VEV{k_1\cdot H_1 k_2\cdot H_2 k_3\cdot H_3 k_4\cdot H_4}\bigg)
\nonumber\\
&\to&\epsilon_2\cdot\epsilon_3 \epsilon_1\cdot\epsilon_4\bigg(
\VEV{{{\cal P}}_2{{\cal P}}_3}\VEV{{{\cal P}}_1{{\cal P}}_4}-\VEV{{{\cal P}}_2{{\cal P}}_3}\VEV{H_1H_4}^22\alpha^\prime k_1\cdot k_4\nonumber\\
&&\hskip-1in
-\VEV{{{\cal P}}_1{{\cal P}}_4}\VEV{H_2H_3}^22\alpha^\prime k_2\cdot k_3
+4\alpha^{\prime2}\VEV{H_2H_3}\VEV{H_1H_4}
(k_1\cdot k_2k_3\cdot k_4\VEV{H_1 H_2}\VEV{H_3H_4}\nonumber\\
&&-k_1\cdot k_3k_2\cdot k_4\VEV{H_1 H_3}\VEV{H_2H_4}
+k_1\cdot k_4k_2\cdot k_3\VEV{H_2 H_3}\VEV{H_1H_4}\bigg)
\nonumber\\
&\to&\epsilon_2\cdot\epsilon_3 \epsilon_1\cdot\epsilon_4\bigg(
(\VEV{{{\cal P}}_2{{\cal P}}_3}+\alpha^\prime t\VEV{H_2H_3}^2)(
\VEV{{{\cal P}}_1{{\cal P}}_4}+\alpha^\prime t\VEV{H_1H_4}^2)\nonumber\\
&&\hskip-1in
+ \VEV{H_2H_3}\VEV{H_1H_4}
(\alpha^{\prime2}s^2\VEV{H_1 H_2}\VEV{H_3H_4}
-\alpha^{\prime2}(s+t)^2\VEV{H_1 H_3}\VEV{H_2H_4})\bigg)
\eea
We will call this combination of contractions $\VEV{T}$, hence
\bea
\VEV{T}&\equiv& 
\left(\VEV{{{\cal P}}_2{{\cal P}}_3}+\alpha^\prime t\VEV{H_2H_3}^2\right)\left(
\VEV{{{\cal P}}_1{{\cal P}}_4}+\alpha^\prime t\VEV{H_1H_4}^2\right)\nonumber\\
&& + \VEV{H_2H_3}\VEV{H_1H_4}
\left(\alpha^{\prime2}s^2\VEV{H_1 H_2}\VEV{H_3H_4}
-\alpha^{\prime2}(s+t)^2\VEV{H_1 H_3}\VEV{H_2H_4}\right)
\eea
thus, the correlator becomes
\bea
\VEV{{\hat{\cal P}}_1{\hat{\cal P}}_2{\hat{\cal P}}_3{\hat{\cal P}}_4} \to \epsilon_1 \dt \epsilon_4 
\, \epsilon_2 \dt \epsilon_3 \, \VEV{T}
\eea
As pointed out before, the expressions \eqref{cylmgluon+2} and \eqref{cylmgluon-2} 
diverge in various corners of integration region over the $\theta_k$ variables. We already encountered a 
divergence of the linear type in the 2-gluon amplitude due to the behavior of $(\csc \theta)^2 $ near 
the end points $\theta\sim 0,\pi$. We showed that this divergence was spurious and it was healed by 
suspending momentum conservation in $\sum_{i=1}^M p_i = p \neq 0$ temporarily. After that, we were able to identify the integral in 
\eqref{eulerian} as the 
Euler Beta function which allowed us to analytically continue the left hand side to the complete 
complex $p$-plane. Undoubtedly, for the three and higher point amplitudes a closed form 
is practically impossible to obtain. However, our approach to the problem will not be to attempt this, but to extract the divergent contributions from the singular regions and track the consequences 
of these seemingly divergent terms. What we will find is that the analytic continuation to $p=0$ of 
the linearly divergent terms precisely combine and give the tree amplitude following the steps of 
\cite{neveuscherkrenorm}. Although the coefficient of this term is an infinite number (which can also be viewed due to the 
presence of the closed string tachyon which introduces a singularity in the $q\sim 0$ region), 
the fact that it is proportional to the tree amplitude allows us re-interpret it as a renormalization of the string coupling constant. We will then find that the logarithmically divergent corners, when continued to $p=0$ also produce terms proportional to the tree amplitude, although in this case the coefficient in front of it is a finite number and these corners will simply correct the coupling by a finite amount. We will now make these statements more explicit with the following calculations.
\subsection{Linear Divergences}\label{lineardiv}
We will now extract the leading divergences in the $\theta_k$ integrations at fixed $q$ and show that they are linear divergences in the relevant angular 
variables. We construct the necessary counterterms to cancel these infinities and 
show that after analytic continuation, the limit $p \to 0$ of the angular integrals is finite\footnote{By angular integrals we mean the integration over 
all the $\theta_k$ variables, or in other words, everything except 
the integration over $q$}. 
We will follow closely the analysis done by Neveu and Scherk \cite{neveuscherkrenorm} adapted for our case, 
open string massless vector external states ('gluons') in the type 0 model, 
and show that not only the limit is finite, but also that its continuation to $p \to 0$ gives precisely the tree amplitude. This allows us to absorb the corresponding 
counterterms into the open string coupling.

\begin{figure}[ht!]
\centering
\includegraphics[scale=0.75]{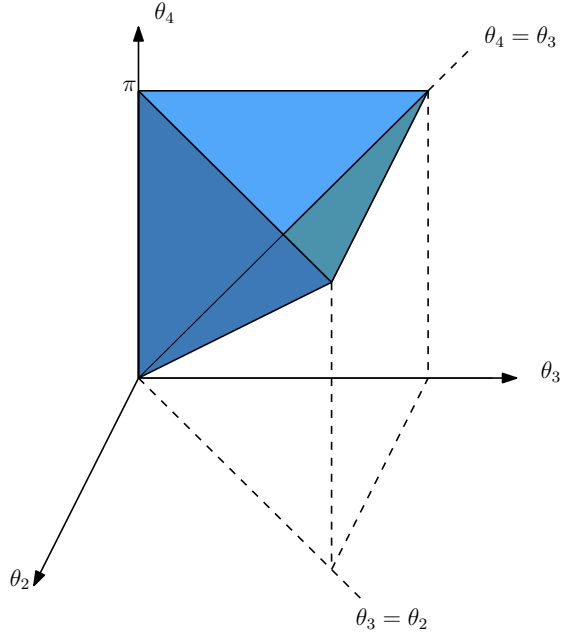}
\caption{The 3-simplex above shows the region of integration at fixed $q$ 
for the 4-point amplitude. Edges and vertices correspond to the places where 
spurious and real divergences can occur.}
\label{fig:simplex}
\end{figure}

For the $M$-point planar one-loop amplitude, 
the integration region in $\int\prod_k d\theta_k$ is given by  
$0<\theta_2<\theta_3<\cdots<\theta_M<\pi$, which is an $(M-1)$-simplex that 
has $M$ vertices and $M!/2!(M-2)!$ edges. For example, the integration region over 
the $\theta_k$ variables for the 4-point amplitude is shown 
in figure \ref{fig:simplex}.
The leading divergences are linear and arise from each of the $M$ 
vertices in the $M$-gluon amplitude as we will show next.

\begin{figure}[ht!]
\centering
\includegraphics[scale=0.65]{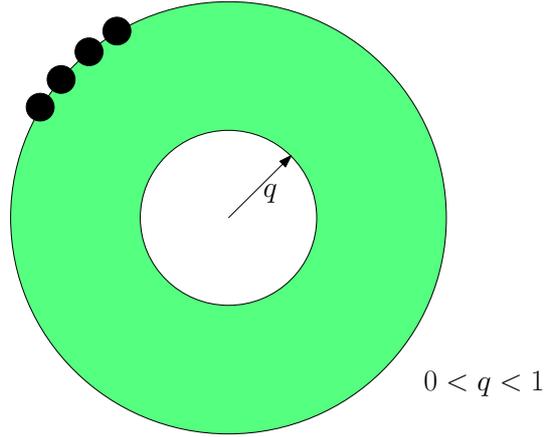}
\caption{Configuration corresponding to the moduli region where all vertex operators come arbitrarily close to each other.}
\label{fig:annulus coincident}
\end{figure}

We can study the vertices of the ($M-1$)-simplex by remembering that they correspond to the configuration in parameter 
space where all the vertex operators coincide (see figure \ref{fig:annulus coincident} which shows the 4-point case). 
For instance, we can examine the one where 
$\theta_M\sim\theta_{M-1}\sim\cdots\sim\theta_2\sim0$ by studying the $\theta_M \sim 0$ limit and performing the changes
\bea
\theta_{j-1}&\equiv& \theta_j \hat{\theta}_{j-1} \quad j=3,\cdots M
\eea
For the 4-gluon amplitude, and keeping only the most divergent terms in the 
$\theta_k$ integrations, we have
\bea
\prod_{i<j} \psi(\theta_{ji})^{2\alpha^{\prime} k_i \cdot k_j} &\simeq& 
\theta_4^{\alpha^{\prime} p^2} 
\hat{\theta}_3^{2\alpha^{\prime} k_4 \cdot p}(1-\hth3)^{2\alpha^{\prime} k_4 \cdot k_3} (1-\hth3\hth2)^{2\alpha^{\prime} k_4 \cdot k_2} 
\hth2^{2\alpha^{\prime} k_2 \cdot k_1}(1-\hth2)^{2\alpha^{\prime} k_3 \cdot k_2}\nn\\{}
\eea
where we have also only kept the leading terms in $p$ in the exponents. Also
\bea
\VEV{T}^+ &\simeq& \frac{1}{4\theta_4^2}(1+\alpha^{\prime} t)\frac{1}{4 \theta_{32}}(1+\alpha^{\prime} t) + 
\frac{1}{2\theta_4}\frac{1}{2\theta_{32}}\left[(\alpha^{\prime} s)^2\frac{1}{2\theta_2}\frac{1}{2\theta_{43}}
-\alpha^{\prime 2}(s+t)^2\frac{1}{2\theta_3}\frac{1}{2\theta_{42}} \right]\nn\\
 &\simeq&\frac{1}{16 \, \theta_4^4 \hth3^2(1-\hth2)}\left[\frac{(1+\alpha^{\prime} t)^2}{1-\hth2}+\frac{(\alpha^{\prime} s)^2}{\hth2 (1-\hth3)}-\frac{\alpha^{\prime 2}(s+t)^2}{1-\hth3\hth2}\right]
\eea
thus
\bea
&&\int_0^{\epsilon} d\theta_4\int_0^{\theta_4}d\theta_3 \int_0^{\theta_3} d\theta_2 \prod_{i<j} \psi (\theta_{ji})^{2\alpha^{\prime} k_i \cdot k_j} 
\VEV{T}^+ \simeq  \nn\\
&\simeq& \frac{1}{16} \int_0^{\epsilon} d\theta_4 \theta_4^{\alpha^{\prime} p^2-2} \int_0^1 d\hth3 \hth3^{2\alpha^{\prime} k_4 \cdot p-1} 
(1-\hth3)^{2\alpha^{\prime} k_4 \cdot k_3} \int_0^1 d\hth2 \hth2^{2\alpha^{\prime} k_2 \cdot k_1} (1-\hth2)^{2\alpha^{\prime} k_3 \cdot k_2-1} \nn\\
&&\times(1-\hth3\hth2)^{2\alpha^{\prime} k_4 \cdot k_2}\left[\frac{(1+\alpha^{\prime} t)^2}{1-\hth2}+\frac{(\alpha^{\prime} s)^2}{\hth2(1-\hth3)}-\frac{\alpha^{\prime 2}(s+t)^2}{1-\hth3\hth2}\right]
\eea
from which we see that, if we put $p=0$ directly in the integrand, the leading divergence near $\theta_4=0$ is linear. It is worth noticing that in this corner of the integration region the integral factorizes and shows a pole at $\alpha^{\prime} p^2=1$ which corresponds to the propagation of a closed string tachyon disappearing into 
the vacuum. In contrast to superstring theories where this kind of divergences are absent due to supersymmetry, 
the planar one-loop diagram in the type 0 model is not divergence free, but it is renormalizable \cite{Neveu:1970iq}. 
The cancellation of these divergences is achieved with the introduction of counter-terms just as in the early days of the dual resonance models. 
We now proceed to cancel this and all of the other linear divergences which come from all the vertices of the 
simplex\footnote{The edge-type divergences will be taken care of in the next section when we deal with logarithmic divergences.} 
with one single counter-term. We subtract and add back the following counter-term:
\bea\label{C+ct}
C_4^+ \equiv 2^4 \left(\frac{1}{8\pi^2 \alpha^{\prime}}\right)^{D/2} \int_0^1 [dq]^+ \int \prod_{k=2}^4 d\theta_k 
\prod_{i<j}[\sin \theta_{ji}]^{2\alpha^{\prime} k_i \cdot k_j} \VEV{T}^+_C
\eea
where $\VEV{T}^+_C$ is simply $\VEV{T}^+$ evaluated at $q=0$. Following Neveu and Scherk \cite{neveuscherkrenorm}, 
we will now prove that the analytic continuation to $p=0$ of $C_4^+$ goes to the tree amplitude \eqref{nstree}. Making the change 
of integration 
variables
\bea
r(\theta_3)= \frac{\sin \theta_{43}}{\sin \theta_3}\, \qquad x(\theta_2)=\frac{\sin \theta_2 \sin \theta_{43}}{\sin \theta_3 \sin \theta_{42}}
\eea
and solving for the various sine functions we need in the integrand, yields
\begin{align}
\sin \theta_{43} &= \frac{r \sin \theta_4}{\sqrt{r^2+2r \cos \theta_4 +1}} \quad &  
\sin \theta_{42} &= \frac{r/x \sin \theta_4}{\sqrt{(r/x)^2+2r/x \cos \theta_4 +1}}\nn\\
\sin \theta_{32} &= \frac{ r/x(1-x) \sin \theta_4}{\sqrt{(r/x)^2+2r/x \cos \theta_4 +1}\sqrt{r^2+2r\cos \theta_4 +1}} \quad &
\sin \theta_{3} &= \frac{\sin \theta_4}{\sqrt{r^2+2r \cos \theta_4 +1}}\nn\\
\sin \theta_{2} &= \frac{\sin \theta_4}{\sqrt{(r/x)^2+2r/x \cos \theta_4 +1}}
\end{align}
Thus,
\bea
\prod_{i<j}[\sin \theta_{ji}]^{2\alpha^{\prime} k_i \cdot k_j} &=& r^{2\alpha^{\prime} k_1 \cdot p +\alpha^{\prime} p^2}[\sin \theta_4]^{\alpha^{\prime} p^2} 
\left(r^2+2r\cos\theta_4+1\right)^{\alpha^{\prime} k_3 \cdot p} \left(\frac{r^2}{x^2}+\frac{2r}{x}\cos\theta_4+1\right)^{\alpha^{\prime} k_2 \cdot p}\nn\\
&& \times \, x^{-\alpha^{\prime} s + 2\alpha^{\prime} k_2 \cdot p} (1-x)^{-\alpha^{\prime} t}\\
d\theta_3 d\theta_2 &=& r \, [\sin\theta_4]^2 (r^2+2r\cos\theta_4+1)^{-1} \left(\frac{r^2}{x^2}+\frac{2r}{x}\cos\theta_4+1\right)^{-1} x^{-2} dr \, dx\\
\VEV{T}_C^+ &=& \frac{1}{16}\csc^2\theta_4 \csc^2\theta_{32}(1+\alpha^{\prime} t)^2 + \nn\\
&&+\frac{1}{4} \csc\theta_4\csc\theta_{32} \left[\frac{(\alpha^{\prime} s)^2}{4}\csc\theta_{2}\csc\theta_{43}
-\frac{\alpha^{\prime 2}(s+t)^2}{4}\csc\theta_3 \csc\theta_{42}\right]\nn\\
&=& \frac{1}{16}r^{-2}[\sin\theta_4]^{-4}(r^2+2r\cos\theta_4+1) \left(\frac{r^2}{x^2}+\frac{2r}{x}\cos\theta_4+1\right) \nn\\
&&\times \left[(1+\alpha^{\prime} t)^2\frac{x^2}{(1-x)^2}+(\alpha^{\prime} s)^2\frac{x}{1-x}-\alpha^{\prime 2}(s+t)^2\frac{x^2}{1-x}\right]
\eea
Therefore,
\bea
&&\int \prod_{k=2}^4 d\theta_k 
\prod_{i<j}[\sin \theta_{ji}]^{2\alpha^{\prime} k_i \cdot k_j} \VEV{T}^+_C = \nn\\
&=&\frac{1}{16} \int_0^1 \!\!dx\int_0^{\infty} \!\!dr\int_0^{\pi} \!\!d\theta_4 \,  r^{2\alpha^{\prime} k_1 \cdot p +\alpha^{\prime} p^2-1}[\sin \theta_4]^{\alpha^{\prime} p^2-2} 
 x^{-\alpha^{\prime} s} (1-x)^{-\alpha^{\prime} t}(r^2+2r\cos\theta_4+1)^{\alpha^{\prime} k_3 \cdot p} \nn\\
&&\times \left(\frac{r^2}{x^2}+\frac{2r}{x}\cos\theta_4+1\right)^{\alpha^{\prime} k_2 \cdot p}\left[(1+\alpha^{\prime} t)^2\frac{x^2}{(1-x)^2}+(\alpha^{\prime} s)^2\frac{x}{1-x}-\alpha^{\prime 2}(s+t)^2\frac{x^2}{1-x}\right]
\eea
 The strategy is to do the integrals 
in the following order: first we do the integration over $\theta_4$, then the one over $r$ and at the end, 
after the analytic continuation to $p\to 0$ has been achieved, we perform the integral over $x$. It is because of this 
that we have used $-\alpha^{\prime} s+2\alpha^{\prime} k_2 \cdot p \to -\alpha^{\prime} s$ since we can always choose $-\alpha^{\prime} s$ to be 
positive enough such that the integral over $x$ in convergent. Let us now focus on the integrals over $r$ and $\theta_4$. For this purpose, define
\bea
I&\equiv& \int_0^{\infty} dr \,r^{2\alpha^{\prime} k_1 \cdot p +\alpha^{\prime} p^2-1}\int_0^{\pi} d\theta_4 [\sin \theta_4]^{\alpha^{\prime} p^2-2} (r^2+2r\cos\theta_4+1)^{\alpha^{\prime} k_3 \cdot p}\nn\\
&&\times \left(\frac{r^2}{x^2}+\frac{2r}{x}\cos\theta_4+1\right)^{\alpha^{\prime} k_2 \cdot p}
\eea
As $p\to 0$, the only non-zero contributions to the integral come from only two corners \cite{neveuscherkrenorm}: 
$\theta_4\sim \pi$ and $r$ is near either $r=1$ or $r=x$. Each corner gives the same answer which is $\pi$, therefore:
\bea
I\to 2\pi \quad \mbox{as $p\to 0$}
\eea  
Therefore,
\bea
&&\int \prod_{k=2}^4 d\theta_k 
\prod_{i<j}[\sin \theta_{ji}]^{2\alpha^{\prime} k_i \cdot k_j} \VEV{T}^+_C = \nn\\
&=&\frac{2\pi}{16} \int_0^1 dx \, x^{-\alpha^{\prime} s} (1-x)^{-\alpha^{\prime} t}
\left[(1+\alpha^{\prime} t)^2\frac{x^2}{(1-x)^2}+(\alpha^{\prime} s)^2\frac{x}{1-x}-\alpha^{\prime 2}(s+t)^2\frac{x^2}{1-x}\right]\nn\\
&=&-\frac{\pi}{8} \bt{1-\alpha^{\prime} s}{-\alpha^{\prime} t}{-\alpha^{\prime} s -\alpha^{\prime} t}
\eea
from where we see that this is precisely proportional to the tree amplitude 
\eqref{nstree}. Therefore, after 
analytic continuation to $p=0$, the counter-term $C_4^+$ becomes:
\beq\bal
C_4^+ \!\!\!&=& \!\!\!-\frac{\pi}{4}  \underbrace{\bt{1-\alpha^{\prime} s}{-\alpha^{\prime} t}{-\alpha^{\prime} s -\alpha^{\prime} t}}_\text{Tree} 
\left(\frac{1}{8 \pi \alpha^{\prime}}\right)^{D/2}\hspace{-6pt} \int_0^1 \frac{dq}{q} 
\left(\frac{-\pi}{\ln q}\right)^{(10-D)/2} \!\!\!\!\!\!\!q^{-1}(1-w^{1/2})^{10-D-S} 
\frac{\prod_{r}^{\infty} (1+q^{2r})^8}{\prod_{n}^{\infty} (1-q^{2n})^8\nonumber}
\eal\eeq 
As mentioned before, the counter-term is the product of the tree amplitude and a divergent factor. This infinity 
comes from the divergent region $q=0$ in the expression above\footnote{There is also a divergence from 
the $q \sim 1$ region. However, this will get explicitly canceled by the $M_4^-$ part of the full one-loop amplitude. 
This is simply a consequence of the projection onto even G-parity states.} which signals 
the presence of the tachyon in the closed string sector. This 
counter-term was originally introduced in \cite{neveuscherkrenorm} and \cite{nsetal} to precisely cancel this type of 
divergence, and the fact that it is proportional 
to the tree amplitude here allows us to absorb this divergence into a coupling constant renormalization.  The remarkable feature of this counter-term is that it allows to cancel both, the $q=0$ 
singularity, and the spurious linear divergences of the $\theta_k$ integrations at the same time. 
This is a consequence of the functional form of the correlator $\VEV{\hat{\mathcal{P}}_1\cdots \hat{\mathcal{P}}_M}$ 
since the $\theta^{-2}$ divergent terms that arise from the  $\csc^2\theta$ functions only come from the $q=0$ part 
of Wick expansion of $\VEV{\hat{\mathcal{P}}_1\cdots \hat{\mathcal{P}}_M}$. We thus now have a new expression free 
of both, the spurious linear divergences\footnote{There are still more divergent 
regions (edges of the  3-simplex) which need to be taken care of. Their removal is the focus of the following section.} in the $\theta_k$ variables, and the UV one coming from $q=0$. Therefore, our expressions for the + 
part of the amplitude need the replacement
\bea
\mathcal{M}_4^+ \to \mathcal{M}_4^+-C_4^+
\eea
Now we need to address the $\mathcal{M}_4^-$ part of the amplitude. Notice in 
\eqref{cylmgluon-} that the presence of 
D-branes, which brings the extra logarithmic factor $\left(\frac{-\pi}{\ln q}\right)^{(10-D)/2}$, makes the $q$-integration completely finite near $q=0$ as long as $D<8$ and hence there is no need for a counterterm for the $\mathcal{M}_4^-$ part of the amplitude\footnote{For $D\!\!=\!\!8$ and $D\!\!=\!\!9$ however, these subleading divergences are still present, but they can be taken care of by a renormalization of $\alpha'$}. However, we still need to deal with the same linear and logarithmic divergences in the $\theta_k$ 
integration as in the $\mathcal{M}_4^+$ case. For the leading divergences, the natural choice would be
the same one we used for the $\mathcal{M}_4^+$ case, but now with $\VEV{\hat{\mathcal{P}}_1\cdots \hat{\mathcal{P}}_M}^+$ replaced by 
$\VEV{\hat{\mathcal{P}}_1\cdots \hat{\mathcal{P}}_M}^-$. However, we have not been able to obtain the analytic 
continuation to $p=0$ for such an expression. The main difficulty comes from the 
fact that the $\VEV{\hat{\mathcal{P}}_1\cdots \hat{\mathcal{P}}_M}^-$ correlators 
involve $\cot\theta_{ji}$ functions which change sign in the integration region 
$0<\theta_2<\cdots<\theta_M<\pi$. This 
did not happen for the + correlators since they contain $\sin\theta_{ji}$ functions 
instead.

However, since we only need to cancel the linear divergences, we 
simply choose the same correlator as before, i.e., $\VEV{\hat{\mathcal{P}}_1\cdots \hat{\mathcal{P}}_M}^+$. Thus, 
we only need to 
adapt the counter-term for the $\mathcal{M}^-$ part of the amplitude by integrating with the $[dq]^-$ measure. This means that we choose:
\bea\label{C-ct}
C_4^- \equiv 2^4 \left(\frac{1}{8\pi^2 \alpha^{\prime}}\right)^{D/2} \int_0^1 [dq]^- \int \prod_{k=2}^4 d\theta_k 
\prod_{i<j}[\sin \theta_{ji}]^{2\alpha^{\prime} k_i \cdot k_j} \VEV{T}^+_C
\eea
Summarizing, we now write $\mathcal{M}_4^{\pm}$ as:
\bea
\mathcal{M}_4^{\pm} = \mathcal{M}_4^{\pm}- \mathcal{C}_4^{\pm} + \mathcal{C}_4^{\pm}
\eea
The last term will be discarded later on since we are going to absorb it into a coupling constant 
renormalization. The full expression for the first two terms is then:
\beq\bal
\mathcal{M}_4^{\pm}- \mathcal{C}_4^{\pm} &=& 2^4 \left(\frac{1}{8\pi^2 \alpha^{\prime}}\right)^{D/2} \!\!\!\int_0^1 [dq]^{\pm} 
\int \prod_{k=2}^4 d\theta_k \prod_{i<j} \left[\psi (\theta_{ji})^{2\alpha^{\prime} k_i \cdot k_j} \VEV{T}^{\pm}
-[\sin \theta_{ji}]^{2\alpha^{\prime} k_i \cdot k_j} \VEV{T}^{\pm}_C\right]
\label{linren}
\eal\eeq
The expression above is completely free of both, the spurious linear divergences in the $\theta_k$ integrations and the UV divergence from the $q\sim0$ region. However, it still has logarithmic divergences in the $\theta_k$ integration which we take care of in the next section. We will show that they are also spurious divergences and can 
be also cancelled with the introduction of suitable counterterms. Moreover, after analytic 
continuation to $p=0$, we will show that the corresponding counterterms are again
proportional to the tree level amplitude which amounts to a finite renormalization 
of the coupling.
\subsection{Logarithmic Divergences}\label{logdiv}
The expression \eqref{linren} is the starting point to continue our treatment of the divergences of the 
original `bare' amplitude. Our task now is to study the last type of divergences in the  $\theta_k$ integrals left in \eqref{linren}, which are 
logarithmic. 

Logarithmic divergences in the angular integrations come from the regions where all vertex operators 
\emph{but one} come together in parameter space. %(Figure \ref{fig:selfenergy}). 
It is a well-known fact that these divergences correspond to loop corrections to the mass of the external 
states\footnote{See, for instance, subsection 9.5 in \cite{Polchinski:1998rq} for a more detailed discussion.}. 
Since we are dealing with massless string states, 
we expect these divergences to be completely absent after continuation to $p=0$ because the 
massless vector string states must remain 
massless in perturbation theory due to gauge invariance. We will indeed find this result for the $4$-gluon amplitude. The mechanics of 
the procedure is very well illustrated by the 4-gluon amplitude and will allow us to see how to extend it 
for an arbitrary number of gluons. 

Recall that for the $M$-gluon amplitude the integration region over $\theta_k$ is an $(M-1)$-simplex, which has 
$M!/(2!(M-2)!)$ edges (see figure \ref{fig:simplex} for the 4-gluon amplitude in which case there are 6 edges). %(see appendix). 
Each of these edges correspond to processes where an open string 
loop is inserted between two string states. If one of these states correspond to one of the $M$ external states, then we have the situation where an internal propagator gets evaluated on-shell, producing an infinity. Before proceeding
with the analysis of these infinities, let us do some counting first. We see that there are
\bea
\frac{M!}{2!(M-2)!} -M = \frac{M}{2}(M-3) \label{internalchannels}
\eea
edges left which do not correspond to radiative corrections to the external legs. Therefore, the number of edges 
that correspond to a loop insertion in the internal channels of the $M$-gluon amplitude has to be given by equation 
\eqref{internalchannels}. On the other hand, we know that the number of planar channels in an $M$-point string 
amplitude is $M/2(M-3)$ 
which precisely matches the number above.

\begin{figure}[ht!]
\centering
\includegraphics[scale=0.75]{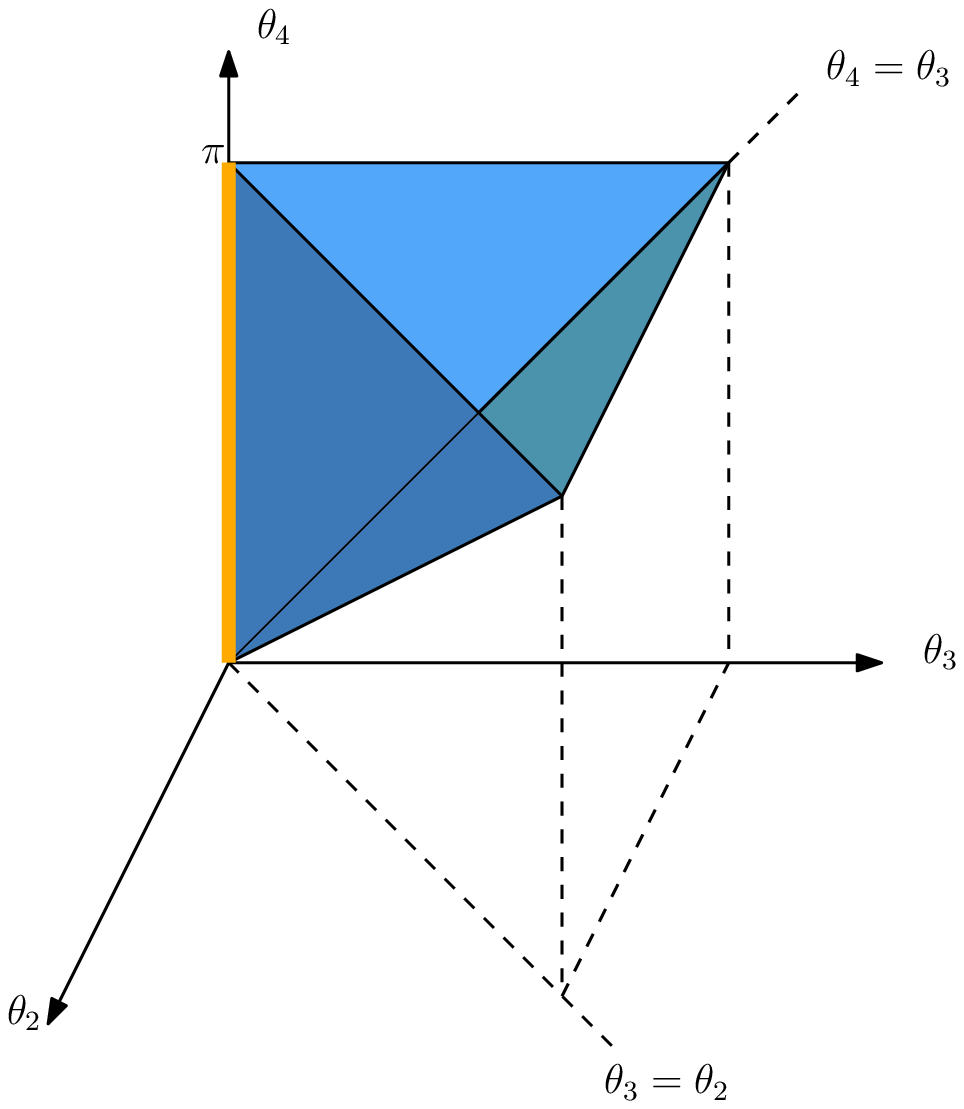}
\caption{The edge corresponding to $\theta_3\sim\theta_2\sim0$ is 
shown as the highlighted line in the figure. This region corresponds to 
a loop insertion in one of the external states which forces the propagator for state number 4 to be evaluated on-shell producing a divergence}
\label{fig:highlighted}
\end{figure}

Let us now focus on the 4-gluon amplitude. This has four (out of six) edges that should correspond to an open 
string loop inserted for each external leg\footnote{The other two edges evidently correspond to loop insertions in 
each of the two planar channels: $s$ and $t$. The $t$-channel is the relevant one in the Regge limit as studied in 
\cite{ThornRojas}.}. 
We now study one of them, namely the edge $\theta_3\sim\theta_2\sim0$ which is 
highlighted in figure \ref{fig:highlighted}.
This corresponds to the region where the vertex operators 
associated with external states 1, 2 and 3 get close together in parameter 
space and it reflects a radiative correction 
to the mass of the external leg 4. %as shown in Figure \ref{fig:bubble}.
To analyze this region, it is convenient to make the change  
$\theta_2 \equiv \theta_3 \hat{\theta}_2$ and study the small $\theta_3$ behavior, namely
\bea
\prod_{i<j} \psi(\theta_{ji})^{2\alpha^{\prime} k_i \cdot k_j} &\simeq& \psi(\theta_4)^{-2\alpha^{\prime} k_4 \cdot p} \theta_3^{2\alpha^{\prime} k_4 \cdot p +\alpha^{\prime} p^2} 
\hat{\theta}_2^{-\alpha^{\prime} s} (1-\hat{\theta}_2)^{-\alpha^{\prime} t}\label{prodedge1} 
\eea 
and also
\bea
\VEV{T} &\simeq& \left(\mathcal{P}_{14}+\alpha^{\prime} t \fermi{1}{4}^2\right)(1+\alpha^{\prime} t) \frac{1}{4\theta_{32}^2} 
+ \fermi{1}{4}^2 \frac{1}{2\theta_{32}} \left[\frac{(\alpha^{\prime} s)^2}{2\theta_2}-\frac{\alpha^{\prime 2} (s+t)^2}{2\theta_3}\right]\nonumber\\
&=& \frac{1}{4\theta_3^2} \left[\mathcal{P}_{14}\frac{(1+\alpha^{\prime} t)}{(1-\hat{\theta}_2)^2} + 
\fermi{1}{4}^2 \left(\frac{\alpha^{\prime} t(1+\alpha^{\prime} t)}{(1-\hat{\theta}_2)^2} + \frac{(\alpha^{\prime} s)^2}{\hat{\theta}_2(1-\hat{\theta}_2)}-\frac{\alpha^{\prime 2}(s+t)^2}{(1-\hat{\theta}_2)}\right) \right]\nonumber\\
&&{}\label{corredge1}
\eea
From $\prod_{k=2}^4 d\theta_k = d\theta_3 \theta_3 d\hat{\theta}_2$ and equations \eqref{prodedge1} and 
\eqref{corredge1}, we see that the leading behavior of the integral over the three angles separates 
into three independent integrals. The integration over the $\theta_k$ variables in \eqref{linren} becomes
\bea
&& \int \prod_{k=2}^4 d\theta_k\left[ \prod_{i<j} \psi (\theta_{ji})^{2\alpha^{\prime} k_i \cdot k_j} \VEV{T}^{\pm}- \prod_{i<j} [\sin \theta_{ji}]^{2\alpha^{\prime} k_i \cdot k_j} \VEV{T}^{\pm}_C\right] =\nonumber\\
&\simeq& \frac{1}{4}\int_0^\pi d\theta_4 \int_0^{\epsilon} d\theta_3 \int_0^1 d\hat{\theta}_2 \, 
\psi(\theta_4)^{-2\alpha^{\prime} k_4 \cdot p} \, \theta_3^{2\alpha^{\prime} k_4 \cdot p +\alpha^{\prime} p^2-1} \,
\hat{\theta}_2^{-\alpha^{\prime} s} (1-\hat{\theta}_2)^{-\alpha^{\prime} t}  \nn \\
&& \times \left[(\mathcal{P}_{14}-\mathcal{P}_{14C})\frac{(1+\alpha^{\prime} t)}{(1-\hat{\theta}_2)^2} \right.\nn\\
&&\hspace{50pt} \left.+ 
(\fermi{1}{4}^2-\fermi{1}{4}_C^2) \left(\frac{\alpha^{\prime} t(1+\alpha^{\prime} t)}{(1-\hat{\theta}_2)^2} + \frac{(\alpha^{\prime} s)^2}{\hat{\theta}_2(1-\hat{\theta}_2)}-\frac{\alpha^{\prime 2}(s+t)^2}{(1-\hat{\theta}_2)}\right) \right]\nn\\
\label{intedge1}
\eea
If we take $p=0$ in the integrand, i.e., if we go back to the original calculation before the introduction 
of the GNS regulator, we see the logarithmic divergence
\bea
\int_0^{\epsilon} d\theta_3 \, \theta_3^{-1}
\eea
Notice that 
\bea
\mathcal{P}_{14}-\mathcal{P}_{14C} =\mathcal{O}(1) \quad \text{and} \quad \fermi{1}{4}^2-\fermi{1}{4}_C^2 =\mathcal{O}(1)
\eea
as $\theta_4 \to 0,\pi$, thus there are no linear divergences  near this edge either, which is simply a 
consequence of the subtraction made in the previous subsection that was introduced precisely to get rid of 
this type of divergences.
Therefore, we need to subtract \eqref{intedge1}, evaluated at $p=0$, from \eqref{linren} and 
we will have a new expression which is free from all linear and the one logarithmic divergence that arises 
from the $\theta_3\sim\theta_2\sim0$ edge\footnote{We will also take care of the other three edges, but we will 
see that the treatment is exactly the same.}. Let us write this new expression in terms of the original $\theta$ variables as
\bea
&&\int_0^\pi d\theta_4 \int_0^{\theta_4} d\theta_3 \int_0^{\theta_3} d\theta_2 \left[ \prod_{i<j} \psi (\theta_{ji})^{2\alpha^{\prime} k_i \cdot k_j} \VEV{T}^{\pm}
- \prod_{i<j} [\sin \theta_{ji}]^{2\alpha^{\prime} k_i \cdot k_j} \VEV{T}^{\pm}_C - B_4\right]\nn\\{}
\eea
where $B_4$ denotes the integrand corresponding to the loop insertion on leg 4 which, in the new variables, is
\bea
B_4 \!\!\!\!&=& \!\!\!\!\frac{1}{4} \,\theta_3^{2\alpha^{\prime} k_4 \cdot p +\alpha^{\prime} p^2-1} \,
\hat{\theta}_2^{-\alpha^{\prime} s} (1-\hat{\theta}_2)^{-\alpha^{\prime} t} \left[\left(\mathcal{P}(\theta_4)-\mathcal{P}(\theta_4)_C\right)\frac{(1+\alpha^{\prime} t)}{(1-\hat{\theta}_2)^2} + \right.\nn\\ 
&&\left.+\left(\chi_+^2(\theta_4)-\chi_+^2(\theta_4)_C\right) \left(\frac{\alpha^{\prime} t(1+\alpha^{\prime} t)}{(1-\hat{\theta}_2)^2} + 
\frac{(\alpha^{\prime} s)^2}{\hat{\theta}_2(1-\hat{\theta}_2)}-\frac{\alpha^{\prime 2}(s+t)^2}{(1-\hat{\theta}_2)}\right) \right]\nn\\
&&{}
\eea
Now that we have taken care of the divergence by subtracting the counterterm $B_4$, let us see what is the result of the 
analytic continuation to $p=0$ when we add this term back. With the GNS regulator put back on, we now need to compute
\bea
\int_0^\pi d\theta_4 \int_0^{\theta_4} d\theta_3 \int_0^{1} d\hat{\theta}_2 \, 
\psi(\theta_4)^{-2\alpha^{\prime} k_4 \cdot p}\,B_4(\theta_4.\theta_3,\hat{\theta}_2)
\label{bubble4ct}
\eea
and then we need to perform on this expression the analytic continuation to $p=0$. The integral over $\theta_3$ is
\bea\label{ppole}
\int_0^{\theta_4} d\theta_3 \, \theta_3^{2\alpha^{\prime} k_4 \cdot p + \alpha^{\prime} p^2 -1} = \frac{\theta_4^{2\alpha^{\prime} k_4 \cdot p + \alpha^{\prime} p^2}}{2\alpha^{\prime} k_4 \cdot p + \alpha^{\prime} p^2} \to \frac{\theta_4^{2\alpha^{\prime} k_4 \cdot p }}{2\alpha^{\prime} k_4 \cdot p }
\eea
as $p\to 0$.
We will now solve for the rest of the integrals. We will find that 
the integral over $\theta_4$ precisely vanishes as $\mathcal{O}(\alpha^{\prime} k_4 \cdot p)$, cancelling the $\frac{1}{2\alpha^{\prime}k_4 \cdot p}$ pole in 
\eqref{ppole}, thus giving a finite result which is exactly what we desire. Proceeding this way, we have
\begin{equation}
\begin{aligned}
\label{bubble4ct2}
&\frac{1}{8\alpha^{\prime} k_4 \cdot p }\int_0^\pi d\theta_4  \int_0^1 d\hat{\theta}_2 \, 
\psi(\theta_4)^{-2\alpha^{\prime} k_4 \cdot p} \,\theta_4^{2\alpha^{\prime} k_4 \cdot p }
\hat{\theta}_2^{-\alpha^{\prime} s} (1-\hat{\theta}_2)^{-\alpha^{\prime} t}  \\
&\left[\left(\mathcal{P}(\theta_4)-\mathcal{P}(\theta_4)_C\right)\frac{(1+\alpha^{\prime} t)}{(1-\hat{\theta}_2)^2} 
\left(\chi_+^2(\theta_4)-\chi_+^2(\theta_4)_C\right) \left(\frac{\alpha^{\prime} t(1+\alpha^{\prime} t)}{(1-\hat{\theta}_2)^2}  + 
\frac{(\alpha^{\prime} s)^2}{\hat{\theta}_2(1-\hat{\theta}_2)}-\frac{\alpha^{\prime 2}(s+t)^2}{(1-\hat{\theta}_2)}\right) \right]
\end{aligned}
\end{equation}
We start with computing first the integral over $\theta_4$ for the $\mathcal{P}(\theta_4)-\mathcal{P}(\theta_4)_C$ contribution:
\bea
I_{\mathcal{P}} \equiv \int_0^{\pi} d\theta_4 \, \psi(\theta_4)^{-2\alpha^{\prime} k_4 \cdot p} 
\left[-\sum_{n=1}^{\infty} \frac{2q^{2n}}{1-q^{2n}} n \cos 2n\theta_4\right] \, 
\theta_4^{2\alpha^{\prime} k_4 \cdot p}
\eea
If we set $p=0$ in the integrand we see that the integral is convergent and it is actually zero. However, we take the opportunity 
here to remind ourselves that we need to know the precisely way on how it goes to zero as a function of $p$, since we have a factor of $\mathcal{O}(p^{-1})$ multiplying this quantity. Expanding the factor $\theta_4^{2\alpha^{\prime} k_4 \cdot p}$ in powers of $p$ and using the small 
$p$ expansion \eqref{infprodexp} in the integrand we have
\bea
I_{\mathcal{P}} &=& -\sum_{n=1}^{\infty} \frac{2q^{2n}}{1-q^{2n}} n \int_0^{\pi} d\theta_4 \, \psi(\theta_4)^{-2\alpha^{\prime} k_4 \cdot p} \cos 2n\theta_4 \, \theta_4^{2\alpha^{\prime} k_4 \cdot p}\nn\\
&=& -\sum_{n=1}^{\infty} \frac{2q^{2n}}{1-q^{2n}} n \int_0^{\pi} d\theta [\sin\theta]^{-2\alpha^{\prime} k_4 \cdot p} 
\left[1+2\alpha^{\prime} k_4\cdot p \ln \theta+\mathcal{O}(p^2)\right]\cos 2n\theta_4\times\nn\\
&&\hspace{3cm}\left[1-2\alpha^{\prime} k_4 \cdot p \sum_{m=1}^{\infty} \frac{1}{m} \frac{2q^{2m}}{1-q^{2m}}(1-\cos 2m\theta) + 
\mathcal{O}(p^2)\right]
\eea
The $\mathcal{O}(1)$ term in $p$ can be analytically continued to $p=0$ by integrating by parts as
\bea
\int_0^{\pi} \sin^z\theta \cos 2n\theta &=& -\frac{z}{2n}\int_0^{\pi} [\sin\theta]^{z-1} 
\cos 2n\theta \sin 2n\theta \nn\\
&\simeq&  -\frac{z}{2n}\int_0^{\pi} [\sin\theta]^{-1} 
\cos 2n\theta \sin 2n\theta \nn \quad \text{as $z\sim 0$, therefore}\\
&\simeq& -\frac{z\pi}{2n}
\eea
The rest of the terms already have an explicit factor of $p$ in front, so we can simply put $p=0$ in their 
integrands obtaining:
\beq\bal
I_{\mathcal{P}} \sme & -\sum_{n=1}^{\infty} \frac{2nq^{2n}}{1-q^{2n}} 
\left[\frac{\pi}{2n}2\alpha^{\prime} k_4 \sdot p + 2\alpha^{\prime} k_4 \sdot p \int_0^{\pi}d\theta \cos 2n\theta\left[\ln\theta-\sum_{m=1}^{\infty} \frac{1}{m} \frac{2q^{2m}}{1-q^{2m}}(1-\cos 2m\theta)\right]\right]\nn\\
\sme& -\sum_{n=1}^{\infty} \frac{2q^{2n}}{1-q^{2n}} \,n 
\,2\alpha^{\prime} k_4\cdot p\left[\frac{\pi}{2n}-\frac{{\rm Si}(2\pi n)}{2n}+\sum_{m=1}^{\infty} \frac{1}{m} \frac{2q^{2m}}{1-q^{2m}}\frac{\pi}{2}\delta_{n,m}\right]\nn\\
\sme& 2\alpha^{\prime} k_4\cdot p\sum_{n=1}^{\infty} \frac{q^{2n}}{1-q^{2n}} 
\left[-\pi+{\rm Si}(2\pi n)-\frac{2q^{2n}}{1-q^{2n}}\pi\right]\nn
\eal\eeq
The new term in the sum, ${\rm Si}(2\pi n)$, where ${\rm Si}(z)\equiv \int_0^z \sin(t)/t\,dt$ is the sine integral, makes the sum converge rather fast at fixed $q$ so there is nothing potentially dangerous coming from this term. Hence, 
the small $p$ behavior of the $\mathcal{P}(\theta_4)-\mathcal{P}(\theta_4)_C$ contribution is
\bea
\eqe\frac{1}{8\alpha^{\prime} k_4 \sdot p }(1+\alpha^{\prime} t)2\alpha^{\prime} k_4\cdot p\sum_{n=1}^{\infty} \frac{q^{2n}}{1-q^{2n}} 
\left[-\pi+{\rm Si}(2\pi n)-\frac{2q^{2n}}{1-q^{2n}}\pi\right]\int_0^1 d\hat{\theta}_2 \, 
\hat{\theta}_2^{-\alpha^{\prime} s} (1-\hat{\theta}_2)^{-\alpha^{\prime} t-2} \nn\\
\eqe \frac{1}{4}(1+\alpha^{\prime} t)\sum_{n=1}^{\infty} \frac{q^{2n}}{1-q^{2n}} 
\left[-\pi+{\rm Si}(2\pi n)-\frac{2q^{2n}}{1-q^{2n}}\pi\right]\frac{\Gamma(1-\alpha^{\prime} s)\Gamma(-1-\alpha^{\prime} t)}{\Gamma(-\alpha^{\prime} s -\alpha^{\prime} t)} \nn\\
\eqe \frac{\pi}{4}\sum_{n=1}^{\infty} \frac{q^{2n}}{1-q^{2n}} 
\left[1-\frac{{\rm Si}(2\pi n)}{\pi}+\frac{2q^{2n}}{1-q^{2n}}\right]\underbrace{\frac{\Gamma(1-\alpha^{\prime} s)\Gamma(-\alpha^{\prime} t)}{\Gamma(-\alpha^{\prime} s -\alpha^{\prime} t)}}_{{\rm Tree}}
\eea
from where we see that this counterterm is also proportional to the tree amplitude.

Before going on and compute the integral over $\theta_4$ for the $\chi(\theta_4)^2-\chi(\theta_4)^2_C$ term in \eqref{bubble4ct}, let us first calculate the integral over $\hat{\theta}_2$ that multiplies it. This is
\bea
&&\int_0^1 d\hat{\theta}_2 \, \hat{\theta}_2^{-\alpha^{\prime} s} (1-\hat{\theta}_2)^{-\alpha^{\prime} t} \left(\frac{\alpha^{\prime} t(1+\alpha^{\prime} t)}{(1-\hat{\theta}_2)^2} + \frac{(\alpha^{\prime} s)^2}{\hat{\theta}_2(1-\hat{\theta}_2)}-\frac{\alpha^{\prime 2}(s+t)^2}{(1-\hat{\theta}_2)}\right)\nn\\
&&=\alpha^{\prime} \frac{\Gamma(1-\alpha^{\prime} s)\Gamma(-\alpha^{\prime} t)}{\Gamma(-\alpha^{\prime} s -\alpha^{\prime} t)} \left[-t-s+s+t\right]
\label{cancellation}\\
&&=0
\eea
Thus the integral over $\theta_4$ of the $\fermi{1}{4}^{\pm}$ term in \eqref{intedge1} does not need to be computed 
since its factor in front vanishes identically! The immediate question is whether we would have obtained the same 
result if we had kept $p\neq 0$ when we performed the Wick contractions on $\VEV{T}$. The answer is yes, although it 
is not totally obvious since if this factor vanishes as $\mathcal{O}(p)$, then we {\emph do} have a non-vanishing contribution 
from this term due to the $\mathcal{O}(p^{-1})$ pole coming from the $\theta_3$ integration (see equation \eqref{ppole}). Luckily, it is easy to show that the cancellation that occurs 
in \eqref{cancellation} is of order $\mathcal{O}(p^2)$. This fact ensures that the fermionic part of 
logarithmic counterterms really vanishes after analytic continuation to $p=0$.  
Had the expression in \eqref{cancellation} been $\mathcal{O}(p)$ instead, the 
$1/p$ factor in \eqref{ppole} would have rendered a nonzero contribution, which 
would have probably spoiled the use of the GNS regulator as an useful renormalization scheme.

We start by writing all the kinematical invariants in terms of the Mandelstam 
variables $s$ and $t$ when total momentum conservation is \emph{not} satisfied, 
but instead we have $\sum_i k_i=p$, this is
\bea\label{kinematicsGNS}
2k_3\cdot k_4 &=& -s +2p (k_1+k_2)+p^2\nn\\
%2 k_1 \cdot k_4 &=& -t +2p\cdot(k_2+k_3)+p^2%\nn\\
2 k_2 \cdot k_4 &=& s+t-2k_2 \cdot p%\nn\\
%2k_1\cdot k_3 &=& s+t+2k_4\cdot p+p^2
\eea 
with similar expressions for $2 k_1 \cdot k_4$ and $2k_1\cdot k_3$. 
Using \eqref{kinematicsGNS} we have that \eqref{cancellation}, after some 
algebra becomes
\bea
\alpha' s\frac{\Gamma(-\alpha' s)\Gamma(-\alpha' t)}{\Gamma(-\alpha' s-\alpha' t)}\left[\frac{2\alpha' k_4 \dt p \, 2\alpha' k_2 \dt p}{\alpha' s+\alpha' t}-\alpha' p^2 
\frac{2\alpha' k_2\dt p}{\alpha' s+\alpha' t}\right]
\eea
which is indeed of order $\mathcal{O}(p^2)$ as required.

After all these intermediate calculations, we can finally write the continuation of \eqref{bubble4ct} to $p=0$, which is
\bea
\eqe\frac{\pi}{4}\sum_{n=1}^{\infty} \frac{q^{2n}}{1-q^{2n}} 
\left[1-\frac{{\rm Si}(2\pi n)}{\pi}+\frac{2q^{2n}}{1-q^{2n}}\right]\frac{\Gamma(1-\alpha^{\prime} s)\Gamma(-\alpha^{\prime} t)}{\Gamma(-\alpha^{\prime} s -\alpha^{\prime} t)} \quad \text{i.e.,}\nn\\
&\propto& \frac{\pi}{4}\sum_{n=1}^{\infty} \frac{q^{2n}}{1-q^{2n}} 
\left[1-\frac{{\rm Si}(2\pi n)}{\pi}+\frac{2q^{2n}}{1-q^{2n}}\right] \times \{{\rm Tree}\}
\label{bubble}
\eea
thus, its complete kinematic dependence is exactly the same as the 
tree amplitude. Therefore, this counterterm can also be absorbed into a (finite) 
coupling renormalization.

We are now ready to write the complete finite expression for the planar one-loop 
amplitude, where momentum conservation is exact. This reads
\beq\bal
&&\int_0^\pi \prod_{i=2}^4 \Theta(\theta_{i+1}-\theta_i) d\theta_i \left[ \prod_{i<j} \psi (\theta_{ji})^{2\alpha^{\prime} k_i \cdot k_j} \VEV{T}^{\pm}
- \prod_{i<j} [\sin \theta_{ji}]^{2\alpha^{\prime} k_i \cdot k_j} \VEV{T}^{\pm}_C - \sum_{k=1}^4 B_k(\theta_i)\right]
\eal\eeq
The $B_k(\theta_i)$ counterterms are listed in appendix \ref{app:Bc}.

\begin{comment}
In the appendix we show that the expression
\bea
 \int_0^\pi \prod_{i=2}^4 \Theta(\theta_{i+1}-\theta_i) d\theta_i \, B_k \equiv \mathcal{B}_k
\label{bubblecont}
\eea
gives the same answer for all the legs, i.e. $\forall i$, the result in \eqref{bubble}. Since it is proportional to the tree amplitude we can absorb it in to a redefinition of the coupling constant.
\end{comment}

We close this section by pointing out that these divergences in the angular integration at fixed $q$ do not always occur when computing one-loop string amplitudes. Take for example the planar one-loop amplitude for $M$ gluons in the type I superstring:
\beq\bal
A_{\rm loop} = 16 \pi^3 g^4 K \int_0^1 \frac{dq}{q} \int_0^1\!\!\!d\nu_i \prod_{i=1}^{M-1} \theta(\nu_{i+1}-\nu_i) \prod_{i<j}\left( \sin \pi\nu_{ji} \prod_{n=1}^{\infty} \frac{1-2q^{2n}\cos2\pi\nu_{ji}+q^{4n}}{(1-q^{2n})^2} \right)^{2\alpha^{\prime} 
k_i \cdot k_j}
\eal\eeq
where $K=K(k_i,\epsilon_j)$ is the kinematical coefficient that depends on the external 
momenta $k_i$ and polarizations $\epsilon_i$ only and it can be found, for example, in \cite{GSW2}.  
 
For this expression, we can clearly see that there are no singular regions in the angular integrals as opposed to the amplitudes we studied above.
\section{Renormalized $M$-gluon amplitude} 

Summarizing our results from the previous section, the complete renormalized expression for the one-loop 4-gluon amplitude which is free of spurious divergences is
\bea\label{mgluonren}
\mathcal{M}_4^{ren} \equiv \int_0^1 \frac{dq}{q} \left(\frac{-\pi}{\ln q}\right)^{5-D/2}\left[\Delta I(q) - \Delta C(q) -\Delta B(q)\right]
\eea
where
\bea
\Delta I&\equiv& 
\int \prod_{k=2}^4 d\theta_k \prod_{i<j} \psi (\theta_{ji})^{2\alpha^{\prime} k_i \cdot k_j} 
\left(P_+\VEV{T}^+-P_-\VEV{T}^-\right)\label{deltaI}\\
\Delta C &\equiv& 
(P_+-P_-)\int \prod_{k=2}^4 d\theta_k \prod_{i<j} [\sin \theta_{ji}]^{2\alpha^{\prime} k_i \cdot k_j} 
\VEV{T}^+_C\label{deltaC}\\
\Delta B &\equiv&  (P_+-P_-) \sum_k \int \prod_{k=2}^4 d\theta_k \, B_k
\label{deltaB}
\eea
with $P_{\pm}$ given in \eqref{P+} and \eqref{P-}. The counterterm integrands $B_k$ 
are given in appendix \ref{app:Bc}.

Also, in all of the expressions above, the GNS regulator $p=\sum_{i=1}^M p_i$ can be already removed, 
i.e., momentum conservation is exact at this stage meaning $p=0$. This is precisely 
what we were after. In particular, with $p=0$, we have
\bea
 \prod_{i<j} \psi (\theta_{ji})^{2\alpha^{\prime} k_i \cdot k_j} = \left[\frac{\psi(\theta_{43})\psi(\theta_{21})}{\psi(\theta_{42})\psi(\theta_{31})}\right]^{-\alpha^{\prime} s} \left[\frac{\psi(\theta_{41})\psi(\theta_{32})}{\psi(\theta_{42})\psi(\theta_{31})}\right]^{-\alpha^{\prime} t}
\eea
The expression for $\Delta B$ is more cumbersome because it is the sum of four terms which correspond to the four different edges that contribute with logarithmic divergences in the $\theta$ integrals. We list them in the 
appendix in equations \eqref{Bct}.

Notice that both $\Delta C$ and $\Delta B$ are directly proportional to $(P_+-P_-)$, which is itself 
independent of the angular integrals since it only depends on the $q$ variable. This is a nice feature because it allows to see explicitly the cancellation of the open string tachyon in all these expressions through the GSO projection, i.e., the `abstruse identity' in this case.

Note also that because of the form of these counter-term integrands, none of them are singular in the $\theta_4\sim\pi$, $\theta_2\sim\theta_3$ region which is the dominant region as $s\to -\infty$ with 
$t$ fixed. Thus, it was this reason why it was not necessary to deal with these divergences in 
\cite{ThornRojas} where the planar one-loop correction to the leading Regge trajectory was obtained.  
The fact that they are also non-singular in the remaining edge, namely $\theta_2\sim0$, $\theta_3 \sim \theta_4$ suggests that they do not contribute either 
to the regime where $t$ is large and $s$ is held fixed.

%
\begin{comment}
Now we would like to extend these results to a general $M$-gluon amplitude which we do in the remaining of this section. As it is shown in the appendix, the analytic continuation of the $\Delta C^{\pm}$ counter-terms \eqref{deltaC} for the general $M$-gluon amplitude is
\bea
\Delta C^{\pm} \to (P_+-P_-) (M-2)\pi \times \{\rm Tree\} \,\,\,\, \text{as $p\to 0$}
\eea
therefore this part is automatic and we can safely absorb the worst divergence for $q\sim 0$ into the coupling constant.\footnote{We recall here that the projection onto even G-parity states gets rid of the $q\sim 1$ (or $w\sim 0$) divergence which is clearly taken care of in the expression for $\Delta C$ by the $(P_+-P_-)$ factor}. The extension of the $B_i^{\pm}$ counter-terms to the $M$-gluon case is not explicitly staightforward but we do the following: 
\end{comment}
%
Inspecting equations \eqref{mgluonren} through \eqref{deltaB}, it is natural 
to conjecture that this structure will remain valid for an arbitrary number 
of external gluons. The analytic continuation to $p=0$ of the $\Delta C$ counterterm was proven that it successfully cancels 
the leading divergences for the scattering of an arbitrary number of external tachyons in 
\cite{neveuscherkrenorm}. It is thus plausible to believe that, since it worked for the 2, 3 and 4 gluon amplitudes, it will continue to do so for an arbitrary number of external gluons\footnote{For an evaluation for the 3-point case, see reference \cite{Thorn:Subcritical}.}. It would be interesting to show this explicitly for the 5-point case.
 
Also, the fact that there is a match between the number of edges and the number of loop insertions in internal channels plus the number of external legs (see equation \eqref{internalchannels}), suggests that the $\Delta B$ 
counterterms can be constructed in the same systematic way we used here for the 
4-point case. 

As mentioned in the introduction, the expression that contains all the 
relevant information in the high energy regime in terms of the external momenta, 
is the factor
\bea
\prod_{i<j} \psi (\theta_{ji})^{2\ap  k_i \cdot k_j} = 
\left[\frac{\psi(\theta_{43})\psi(\theta_{2})}{\psi(\theta_{42})
\psi(\theta_{3})}\right]^{-\ap  s}
 \left[\frac{\psi(\theta_{4})\psi(\theta_{32})}{\psi(\theta_{42})
 \psi(\theta_{3})}\right]^{-\ap  t}
\label{psiprod}
\eea
It will be convenient to write this as
\bea
\prod_{i<j} \psi (\theta_{ji})^{2\ap  k_i \cdot k_j} = 
e^{-\ap  s\left(V_s - \lambda V_t\right)}= e^{\ap  |s| \, V_{\lambda}}
\label{psiprod2}
\eea
where $\lambda \equiv -t/s$, and $V_{\lambda} \equiv V_s-\lambda V_t$ with
\bea
V_s &\equiv&  \ln\frac{\psi(\theta_{43})\psi(\theta_{2})}{\psi(\theta_{42})
\psi(\theta_{3})}\nn\\
V_t &\equiv& \ln \frac{\psi(\theta_{4})\psi(\theta_{32})}{\psi(\theta_{42})
\psi(\theta_{3})}
\label{VsVt}
\eea
Thus, the hard scattering limit $s \to -\infty$ with $\lambda \equiv -t/s$  held fixed corresponds to 
the regions where $V_{\lambda}$ is maximized.

%%%%%%%%%%%%%%%%%%%%%%%%%%%%% Tensionless Section       %%%%%%%%%%%%%
\section{The tensionless limit}\label{sec:hard}
Note that since 
all the Mandelstam variables in the string amplitude come multiplied with a factor of $\alpha'$, the tensionless limit ($\alpha'\to \infty$) 
with $s$ and $t$ held fixed is exactly equivalent to the hard scattering limit (high energy at fixed angle), namely, 
$s,t\gg \alpha'^{-1}$ with the ratio $s/t$ held fixed. 

Recall that the amplitude above has physical resonances in both the $s$ and $t$ channels, i.e., the integral representation \eqref{mgluonren} has open-string poles whenever $\alpha' s = 0,1,2,\dots$ [and also
when $\alpha't =0,1,2,\dots$]. Thus, in order to avoid these poles for computing the high energy limit, 
we take $\alpha's$ and $\alpha't$ both to $\to -\infty$. Note that a similar situation also appears at tree level string 
scattering. For example, if we take the hard scattering limit in
the Veneziano amplitude
\bea
A(s,t)=\frac{\Gamma(-\alpha's -1)
\Gamma(-\alpha't-1)}{\Gamma(-\alpha's-\alpha't-2)},\label{Veneziano}
\eea
we would also `hit' all the poles at $\alpha's=n$ and $\alpha't=n$ for large values of the positive integer $n$ as we take $\alpha's$ and $\alpha't$ to infinity. Note also that, although the usual integral representation of the Veneziano amplitude 
\bea
A(s,t)=\int_0^1 x^{-\alpha's-2}(1-x)^{-\alpha't -2}\label{VenezianoInt}
\eea
only converges for ${\rm Re}(\alpha's)<-1$ (and ${\rm Re}(\alpha't)<-1$), the hard scattering limit obtained by evaluating this integral for $\alpha's \to -\infty$ with $\lambda$ fixed gives the same answer as the one computed from \eqref{Veneziano} which defines the analytic continuation of \eqref{VenezianoInt} to the full 
complex plane.

\subsection{Hard scattering limit through one loop}

We now focus on the hard scattering limit of the 4-gluon amplitude for type 0 strings. We start by writing the fully 
renormalized amplitude for NS+ spin structure, $\mathcal{M}_4^+$. This reads
\beq\bal
\mathcal{M}_{4,\rm ren}^+ &= 2\left(\frac{1}{8 \pi \ap }\right)^{D/2} 
\int_0^1 \frac{dq}{q} \left(\frac{-\pi}{\ln q}\right)^{(10-D)/2} 
P_+(q)\nn\\
&\hspace{30pt}\int \prod_{k=2}^4 d\theta_k \, \left[e^{-\ap s V_{\lambda}}
\langle \hat{\mathcal{P}}_1 
\hat{\mathcal{P}}_2 \hat{\mathcal{P}}_3 \hat{\mathcal{P}}_4 \rangle^+ 
-e^{-\ap s V_{\lambda}^0}
\langle \hat{\mathcal{C}}_1 \hat{\mathcal{C}}_2 \hat{\mathcal{C}}_3 
\hat{\mathcal{C}}_4\rangle- B^+\right]
\label{MpwithC}
\eal\eeq
where $V_{\lambda}^0$ is by definition $V_{\lambda}$ in equation 
\eqref{VsVt} with all the Jacobi theta functions evaluated at $q=0$, i.e.
\bea
V_{\lambda}^0 = \ln\left[\frac{\sin\theta_{43}\sin\theta_{2}}
{\sin\theta_{42}\sin\theta_{3}}\right]- \lambda
\ln \left[\frac{\sin\theta_{4}\sin\theta_{32}}
{\sin\theta_{42}\sin\theta_{3}}\right]
\eea
The counterterm $B^+$ is the sum of the + terms in \eqref{Bct} (appendix \ref{app:Bc}). 

The $s\to-\infty$ limit with $\lambda$ fixed can now be 
extracted by finding the regions where $V_{\lambda} \equiv V_s - 
\lambda V_t$ is a maximum and integrating $V_{\lambda}$ 
around these dominant regions. As it was first observed by Gross and Manes 
for the open superstring in flat space \cite{GrossManes}, all the dominant 
critical points 
for the one-loop planar amplitude lie on the boundary of the integration 
region. Since the exponential dependence on the external momenta in the type 0 model  
is the same as for the superstring, this also holds true here. 
Thus, we will study all possible boundary regions that 
produce a contribution which are not exponentially suppressed. We will see that there are many 
regions that are not exponentially suppressed, therefore, we need to  
compare all the relevant contributions and extract the leading one that 
dominates at high energies.
%It will also be convenient to perform the changes
%\bea
%\alpha =\frac{1}{2\pi}(\theta_2-\theta_{43})\nn\\
%\beta = \frac{1}{2\pi}(\theta_2+\theta_{43})\nn\\
%\gamma=\frac{1}{2\pi}(\theta_{42}+\theta_3)
%\eea
%now
%\bea
%V_s &\equiv&  \ln\frac{\psi(\beta-\alpha)\psi(\beta+\alpha)}
%{\psi(\gamma-\alpha)\psi(\gamma+\alpha)}\nn\\
%V_t &\equiv& \ln \frac{\psi(\gamma+\beta)\psi(\gamma-\beta)}
%{\psi(\gamma-\alpha)\psi(\gamma+\alpha)}
%\eea

An important point is that $V_{\lambda}\leq 0$ throughout the entire integration region $0<q<1$, 
$0 < \theta_2 < \theta_3 < \theta_4< \pi$. 
Therefore, the dominant regions as $|s|\to\infty$ at fixed $\lambda$ (hard scattering) are the ones 
where $V_{\lambda} \sim 0$. Although we do not provide an analytic proof here that $V_{\lambda}\leq 0$ 
everywhere, we have strong numerical evidence that  this is indeed the case.

In order to study the dominant regions better, we note that
\bea
\ln \psi(\theta) = \ln \sin\theta + 2\sum_{n=1}^{\infty} \frac{1}{n} 
\frac{q^{2n}}{1-q^{2n}}(1-\cos2n\theta)
\eea
and defining
\bea
x\equiv \frac{\sin\theta_{43}\sin \theta_{2}}{\sin\theta_{42}
\sin\theta_{3}} %= \frac{\sin\pi(\beta-\alpha)\sin\pi(\beta+\alpha)}
%{\sin\pi(\gamma-\alpha)\sin\pi(\gamma+\alpha)}
\label{NSx} 
\eea
we can write $V_{\lambda}$ as
\bea
V_{\lambda} = \ln x -\lambda \ln(1-x) + 2\sum_{n=1}^{\infty} \frac{1}{n} 
\frac{q^{2n}}{1-q^{2n}}(S_n-\lambda T_n)
\label{newexp}
\eea
where
\bea
S_n &\equiv& 2\cos n(\theta_2-\theta_{43})  \left[\cos n(\theta_{42}
+\theta_3) -\cos n (\theta_2+\theta_{43}) \right]\nn\\
T_n &\equiv& 2\cos n(\theta_{42}+\theta_{3})  \left[\cos n(\theta_{2}
-\theta_{43}) -\cos n (\theta_2+\theta_{43}) \right]
\eea
From \eqref{newexp} we immediately recognize that, at $q=0$, one recovers the tree level 
factor, namely
\beq\bal
\int_0^1 dx \, e^{-\ap s V_{\lambda} } = \int_0^1 dx \, 
e^{-\ap s( \ln x -\lambda \ln(1-x)) } 
 =\int_0^1 dx \, x^{-\ap s} (1-x)^{-\ap t} = 
\frac{\Gamma(1-\alpha's)\Gamma(1-\alpha't)}{\Gamma(2-\alpha's-\alpha't)}
\eal\eeq
Because of this fact, and motivated by the analysis in 
\cite{neveuscherkrenorm}, the integrals are more easily analyzed by going to 
the following variables:\footnote{This change of variables was first used by 
Neveu and Scherk \cite{neveuscherkrenorm} when they were studying the 
one-loop planar amplitude for ``mesons'' in the original dual 
resonance models. This allowed them to prove that the leading divergence at 
one-loop was proportional to the Born term (tree amplitude), thus providing 
evidence of renormalizability in those models. Since the counterterm used 
in \cite{neveuscherkrenorm} arises from the divergence at $q\sim 0$, and 
proved to be proportional to the tree amplitude, it was very likely that 
these set of variables was also useful in our calculations for the type 0 
string.}
\bea
r(\theta_3)=\frac{\sin\theta_{43}}{\sin\theta_3} \quad , 
\quad x(\theta_2)= \frac{\sin\theta_{43}\sin \theta_{2}}
{\sin\theta_{42}\sin\theta_{3}}
\label{nsvariables}
\eea
The variable $x$ allows us to see that $\prod_{i<j}\psi(\theta_{ji})^
{2\ap k_i \cdot k_j}$ 
has a critical point of the second kind at the boundary surface $q=0$ and 
along the plane defined by
\bea
\frac{\sin\theta_{43}\sin \theta_{2}}{\sin\theta_{42}\sin\theta_{3}}
=\frac{1}{1-\lambda}\equiv x_c 
\eea
since
\bea
\frac{\partial V_{\lambda}}{\partial \theta_i}\Big|_{q=0,\,x=x_c} = 
\frac{\partial x}{\partial \theta_i}\frac{\partial V_{\lambda}}{\partial x} 
\Big|_{q=0,\,x=x_c} =0\quad i=2,3,4.
\eea
which is obtained from
\bea
\frac{\partial V_{\lambda}}{\partial x}\bigg|_{x=x_c,\,q=0} &=& 
\left[\frac{1}{x} + 
\frac{\lambda}{1-x} + 2\sum_{m=1}^{\infty} \frac{1}{m} 
\frac{q^{2m}}{1-q^{2m}}(\frac{\partial S_m}{\partial x}-
\lambda \frac{\partial T_m}{\partial x})\right]_{x=x_c,\,q=0}\nn\\
 &=& 2\sum_{n=1}^{\infty} \frac{1}{n} 
\frac{q^{2n}}{1-q^{2n}}\left(\frac{\partial S_n}{\partial \theta_i}
\frac{\partial \theta_i}{\partial x}
-\lambda \frac{\partial T_n}{\partial \theta_i}\frac{\partial \theta_i}
{\partial x}\right)\bigg|_{q=0}\nn\\
&=& 0
\label{partialV}
\eea
From equations \eqref{newexp} and \eqref{partialV} we see that 
we have found a stationary point at $q=0$ since as $q\to 0$ the function 
$V_{\lambda}$ becomes independent of $q$. Expanding $V_{\lambda}$ about 
$(x,q)=(x_c,0)$ gives
\bea
V_{\lambda}(x,q) \simeq -\lambda \ln(-\lambda)-(1-\lambda)\ln(1-\lambda) + 
\frac{(1-\lambda)^3}{2\lambda}(x-x_c)^2+2q^2(S_1-\lambda T_1)
\label{newexpapprox}
\eea
Thus, as $s\to -\infty$ the integral over $q$ is dominated by the region 
$q\sim 0$ provided that $S_n-\lambda T_n$ is not too close to zero. Since 
the expression $(S_n-\lambda T_n)$ depends on the angular variables 
$\theta_i$ which are integrated over the range $0<\theta_i<\pi$, this factor 
could get arbitrarily close to zero in certain regions, even for large 
$|s|$. Then, the small $q$ approximation ceases to be valid and one has 
to integrate over the whole range $0<q<1$ in order to obtain the correct 
leading behavior. We will study these regions separately and show that 
they produce subleading behavior, so we can 
simply avoid those regions for now.\\ 
The first two terms in \eqref{newexpapprox} are independent of the 
integration variables $\theta_k$ and $q$, so we can take them out of the 
integrals as
%Notice that the original integral is only defined for $\ap s<1$, 
%therefore we need to 
%define it by analytic continuation ($s\to-i s$). This allows us to extract 
%the factor 
%({\color{red}{more details needed here}})
\bea
e^{-\ap s V_{\lambda}} &\approx& %e^{-\ap s[-\lambda \ln(-\lambda)-
%(1-\lambda)\ln(1-\lambda)]}e^{-\ap s[\frac{(1-\lambda)^3}{2\lambda}
%(x-x_c)^2+
%2q^2(S_1-\lambda T_1)]}\nn\\
%&=&
e^{\ap s[\lambda \ln(-\lambda)+(1-\lambda)\ln(1-\lambda)]}
e^{-\ap s[\frac{(1-\lambda)^3}{2\lambda}(x-x_c)^2+
2q^2(S_1-\lambda T_1)]}%\nn\\
%&=&e^{-\ap |s|[\lambda \ln(-\lambda)+(1-\lambda)\ln(1-\lambda)]}
%e^{-\ap s[\frac{(1-\lambda)^3}{2\lambda}(x-x_c)^2+2q^2(S_1-\lambda T_1)]}
\label{expapprox}
\eea
Since $\lambda<0$, the term inside the square brackets of the first 
exponential is positive definite giving the overall exponential suppression 
$\exp\{-\ap |s|f(\lambda)\}$ where $f(\lambda)=\lambda \ln(-\lambda)+
(1-\lambda)\ln(1-\lambda)$ for the amplitude as $\ap s\to -\infty$. 
This is the well known exponential falloff characteristic of stringy 
amplitudes in the hard scattering limit. Moreover, it is identical to the 
tree level behavior. The reason is that, as $q\to 0$, the hole of the 
annulus shrinks to a point thus making it indistinguishable from the 
disk amplitude. We now re-write \eqref{expapprox} as
\bea
e^{-\ap s V_{\lambda}} &\approx& e^{-\alpha^{\prime} |s| f(\lambda)}\, 
e^{-\ap s[\frac{(1-\lambda)^3}{2\lambda}(x-x_c)^2+2q^2(S_1-\lambda T_1)]}
\label{expapprox2}
\eea
where 
\bea
f(\lambda) &\equiv & \lambda \ln(-\lambda)+(1-\lambda)
\ln(1-\lambda)\\
S_1-\lambda T_1 &=& 2(\sin^2\theta_2+\sin^2\theta_{43})
-2\lambda\left(\sin^2\theta_4+\sin^2\theta_{32}\right)\nn\\
&&-2\left(1-\lambda\right)\left(\sin^2\theta_{42}+\sin^2\theta_3\right)
\label{S1T1}
\eea
It is also important to stress that, at leading order, the 
combination 
$S_1-\lambda T_1$ must be evaluated at the value where the cross ratio $x$ 
extremizes $V_{\lambda}$ i.e.: at $x=x_c=\frac{s}{s+t}=(1-\lambda)^{-1}$. 
Therefore, we can simplify \eqref{S1T1} using \eqref{NSx} with the 
replacement
\bea
\lambda \to -\frac{\sin\theta_4\sin\theta_{32}}{\sin\theta_{43}\sin\theta_2}
\eea
which yields
\bea
(S_1-\lambda T_1)_{x=x_c} &=& -8 \sin\theta_{32}\sin\theta_3
\sin\theta_{42}\sin\theta_4 
\label{S1T1crit}
\eea
From the fact that for the planar amplitude the $\theta_i$ variables are 
ordered, i.e. $0\leq \theta_2\leq\theta_3\leq\theta_4\leq\pi$, we see that 
$(S_1-\lambda T_1)_{x=x_c}$ is a negative number. We have mentioned earlier 
that we only have numerical evidence that $V_{\lambda}$ negative-definite 
in the integration region. However, from \eqref{S1T1crit} and 
\eqref{newexpapprox} we see analytically that this is true at least along 
the surface $x=\frac{\sin\theta_2\sin\theta_{43}}
{\sin\theta_{42}\sin\theta_3}=x_c$ which will dominate at the end.  
After writing the integrals in the new set of variables given in 
\eqref{nsvariables} we can make this more explicit as we will show next. 

Now 
we go ahead and estimate the leading behavior of  \eqref{MpwithC} that comes from the 
$x\sim x_c,\, q\sim 0$ saddle point. 
 We re-write \eqref{MpwithC} here for convenience,
\bea
\mathcal{M}_{4,\rm ren}^+&=& 2\left(\frac{1}{8 \pi \ap }\right)
^{D/2} \int_0^1 \frac{dq}{q} 
\left(\frac{-\pi}{\ln q}\right)^{(10-D)/2} P_+(q)\nn\\
&&\times\int \prod_{k=2}^4 d\theta_k \, \left[e^{-\ap s V_{\lambda}}
\langle \hat{\mathcal{P}}_1 \hat{\mathcal{P}}_2 \hat{\mathcal{P}}_3 
\hat{\mathcal{P}}_4 \rangle^+ -e^{-\ap s V_{\lambda}^0}\langle 
\hat{\mathcal{C}}_1 \hat{\mathcal{C}}_2 \hat{\mathcal{C}}_3 
\hat{\mathcal{C}}_4 \rangle - B^+\right]
\label{MpwithC2}
\eea
%In the $w$ variables we have
%\bea
%\psi  &=& \left(\frac{-\pi}{\log w}\right) (1-w^{\frac{\theta}{\pi}}) 
%w^{-\frac{\theta}{2\pi} \left(1-\frac{\theta}{\pi}\right)} 
%\prod_{n=1}^{\infty} \frac{(1-w^{n+\theta/\pi})
%(1-w^{n-\theta/\pi})}{(1-w^{n})^2}
%\eea
%thus
%\bea
%\ln \psi &=& \ln  \left(\frac{-\pi}{\log w}\right) +
%\ln(1-w^{\frac{\theta}{\pi}}) -\frac{\theta}{2\pi} 
%\left(1-\frac{\theta}{\pi}\right) \ln w + \sum_{n=1}^{\infty} 
%\ln \frac{(1-w^{n+\theta/\pi})(1-w^{n-\theta/\pi})}{(1-w^{n})^2}
%\eea
%Performing a re-summation in the last term gives
%\bea
%\ln \psi &=& \ln  \left(\frac{-\pi}{\log w}\right) +
%\ln(1-w^{\frac{\theta}{\pi}}) -\frac{\theta}{2\pi} 
%\left(1-\frac{\theta}{\pi}\right) \ln w + 2\sum_{m=1}^{\infty} 
%\frac{1}{m}\frac{w^m}{1-w^m}(1-\cosh m\theta/\pi)
%\eea
The approximations for the exponentials inside the square brackets near the 
critical surface are given in \eqref{expapprox2}. From there, we also see  that the integration over 
$x$ is well approximated by a gaussian in the $\ap s\to-\infty$ limit. The integration 
over $q$ is dominated by the end-point $q=0$ which demands 
that we expand the rest of the integrand as a power series in $q$. As we 
will see below, we need to expand the integrand beyond leading order in $q$ 
in order to extract the correct leading behavior. 
The expansions we need are:
\bea
P_+&=&q^{-1}(1-w^{1/2})^{10-D-S}(1+8q+\mathcal{O}(q^2))\\
\langle \hat{\mathcal{P}}_1 \hat{\mathcal{P}}_2 \hat{\mathcal{P}}_3 
\hat{\mathcal{P}}_4 
\rangle^+&=&a_0+ a_1 q +\mathcal{O}(q^2)\\
B^+(q)&=&b_1 q +\mathcal{O}(q^2)
\eea
where $a_0=\langle \hat{\mathcal{C}}_1 \hat{\mathcal{C}}_2 
\hat{\mathcal{C}}_3 \hat{\mathcal{C}}_4 \rangle$, 
which, in terms of the original $\theta_k$ variables is given by
\bea
16 a_0=\csc^2\theta_{32}\csc^2\theta_4(1+\ap t)^2+
\csc\theta_4\csc\theta_{32}
[(\ap s)^2\csc\theta_2\csc\theta_{43}-(\ap u)^2\csc\theta_3
\csc\theta_{42}]
\eea
and with a similar (but more cumbersome) expression for $a_1$. 
With these expansions, and integrating over the new variables $(\theta,r,x)$ 
we have
\bea
\mathcal{M}_{4,\rm ren}^+\simeq && 2\left(\frac{1}{8 \pi \ap }\right)^
{D/2}\pi^{20-2D-2S} e^{-\alpha^{\prime} |s| f(\lambda)} \int_R d\theta 
\, dr \, dx \,|J| 
e^{-\ap s\frac{(1-\lambda)^3}{2\lambda}(x-x_c)^2}\nn\\
&&\int_0^{\epsilon} \frac{dq}{q^2}\left(\frac{-1}{\ln q}\right)^{\gamma}
\left[a_0(e^{-2\ap s q^2(S_1-\lambda T_1)}-1)+q (a_1+8a_0) 
e^{-2\ap s q^2(S_1-\lambda T_1)}+b_1 q + \cdots\right]\nn\\
\label{amplitudeapprox1}
\eea
where $\gamma \equiv 15-3D/2-S$ and $|J|$ is the Jacobian for the transformation 
$d\theta_3 d\theta_2 =|J| dr dx$ which reads
\bea
|J| &=&  x^{-2} r \, [\sin\theta_4]^2 (r^2+2r\cos\theta_4+1)^{-1} 
\left(\frac{r^2}{x^2}+\frac{2r}{x}\cos\theta_4+1\right)^{-1}
\eea
The integration region $R$ in \eqref{amplitudeapprox1} is 
$0<\theta<\pi$,$0<r<\infty$, $0<x<1$ but avoiding the places where 
$S_n-\lambda T_n$ gets arbitrarily close to zero for all $\lambda$. By 
inspection, these regions correspond to the four vertices and four of 
the six edges in figure \eqref{fig:simplex}. We already mentioned that they 
correspond to tadpole diagrams and to loop insertions in external legs, 
which are also boundary regions of the moduli we are integrating over. 
According to the discussion in \cite{GrossManes}, we should also study these regions and 
extract their contributions. We shall do this at the end of this section 
and show that they produce subleading contributions in the hard scattering 
limit, thus, they can be neglected. Note also that the $B^+$ 
counterterm in \eqref{MpwithC2} is not being multiplied by an exponential factor with 
dependence in $q$ as is the case for $\langle \hat{\mathcal{P}}_1 
\hat{\mathcal{P}}_2 \hat{\mathcal{P}}_3 \hat{\mathcal{P}}_4 \rangle^+$. 
This implies that in the large $\ap s$ limit it is exponentially 
suppressed so we can neglect it\footnote{This is also true for the 
$\langle \hat{\mathcal{C}}_1 \hat{\mathcal{C}}_2 \hat{\mathcal{C}}_3 
\hat{\mathcal{C}}_4 \rangle^+$ term in \eqref{MpwithC2}, but we need 
to keep this term to ensure convergence of the integral at $q=0$}\!\!. 
Thus, in order to extract the leading contributions from the boundary 
region defined by $q= 0$, we now need to estimate the  integrals
\bea
I_1 &=& \int_0^{\epsilon} \frac{dq}{q^2}\left(\frac{-1}{\ln q}\right)^{\gamma}
(e^{-\beta q^2}-1)\label{I1}\\
I_2 &=& \int_0^{\epsilon} \frac{dq}{q}\left(\frac{-1}{\ln q}\right)^{\gamma}
e^{-\beta q^2}\label{I2}
\eea 
as $\beta \to \infty$ limit. Note that the $-1$ term inside the 
parentheses in $I_1$ is a result of the inclusion of the Neveu-Scherk 
counterterm. After the change $y=\beta q^2$,  for $I_1$ we have
\bea
I_1 &=& \frac{1}{2}\beta^{1/2}\left(\frac{2}{\ln \beta}\right)^{\gamma}
\int_0^{\beta \epsilon^2} dy\, y^{-3/2}(e^y-1)\left(1-\frac{\ln y}
{\ln \beta}\right)^{-p}\nn\\
&\simeq& \frac{1}{2}\beta^{1/2}\left(\frac{2}{\ln \beta}\right)^{\gamma}
\int_0^{\beta \epsilon^2} dy\, y^{-3/2}(e^y-1) \simeq -\sqrt{\pi\beta} 
\left(\frac{2}{\ln \beta}\right)^{\gamma}
\eea 
which is the leading term of $I_1$ as an expansion in powers of 
$(\ln \beta)^{-1}$. Similarly for $I_2$, we make the change $u=-\ln q$ 
yielding
\bea
I_2 &=& \int_{-\ln \epsilon}^{\infty} du \, u^{-\gamma} 
\exp[-\beta \exp[-2u]]\nn\\
&=& (\ln \beta)^{1-\gamma}\int_{-\ln \epsilon/\ln \beta}^{\infty} d\xi\, 
\xi^{-\gamma}\exp[-\exp[(1-2\xi)\ln \beta]] 
\eea
As $\beta \to \infty$ we see that the exponential factor 
$\exp[-\exp[(1-2\xi)\ln \beta]]$ effectively cuts the integration range to 
$1/2<\xi<\infty$ therefore, for small but fixed $\epsilon$ we have
\bea
I_2 &\simeq& (\ln \beta)^{1-\gamma}\int_{1/2}^{\infty} d\xi\, \xi^{-\gamma} = 
\frac{1}{\gamma-1}\left(\frac{2}{\ln \beta}\right)^{\gamma-1}
\eea
With these approximations for the $q$ integration, the amplitude in 
\eqref{amplitudeapprox1} now becomes
\beq\bal
\mathcal{M}_{4,\rm ren}^+ & \simeq 2\left(\frac{1}
{8 \pi \ap }\right)^{D/2}\pi^{20-2D-2S} e^{-\alpha^{\prime} |s| f(\lambda)} 
\int_R d\theta\,dr\,dx\, e^{-\ap s\frac{(1-\lambda)^3}{2\lambda}(x-x_c)^2}
|J|\nn\\
&\hspace{20pt}\left[-\sqrt{2\pi\ap s}\left(\frac{2}{\ln \ap |s|}\right)^{\gamma}
(S_1-\lambda T_1)^{1/2}a_0+\frac{1}{p-1}\left(\frac{2}{\ln\ap |s|}\right)
^{\gamma-1} (a_1+8a_0)\right]
\label{MpwithC3}
\eal\eeq
As mentioned above, the $x$ integral is very well approximated by a 
gaussian 
in the $\ap s\to-\infty$ limit. Thus, at leading order, we have
\bea
\int_0^1 dx\, e^{-\ap s\frac{(1-\lambda)^3}{2\lambda}(x-x_c)^2} 
h(x)&\simeq& h(x_c) \int_{-\infty}^{\infty} dx\, e^{-\ap s
\frac{(1-\lambda)^3}{2\lambda}(x-x_c)^2}\nn\\
&\simeq & h(x_c) \sqrt{\frac{2\pi\lambda}{\ap s (1-\lambda)^3}}
\label{gaussian}
\eea
where $h(x)$ simply tracks the complete dependence on the original 
$\theta_k$ variables 
of the rest of the integrand in \eqref{MpwithC2}.
Therefore, we now have
\beq\bal
\mathcal{M}_{4,\rm ren}^+ &\simeq  2\left(\frac{1}
{8 \pi \ap }\right)^{D/2}\pi^{20-2D-2S} 
e^{-\alpha^{\prime} |s| f(\lambda)}\sqrt{\frac{2\pi\lambda}{\ap s 
(1-\lambda)^3}}
\,\times \\ 
&\hspace{40pt}\left[-\sqrt{2\pi\ap s}
\left(\frac{2}{\ln \ap |s|}\right)^{\gamma} \int_R d\theta\,dr|J|
(S_1-\lambda T_1)^{1/2} a_0\right.\\
& \hspace{45pt}+\left.\frac{1}{\gamma-1}\left(\frac{2}{\ln\ap |s|}\right)^{\gamma-1} 
\int_0^{\pi}d\theta\int_0^{\infty}dr\,|J|(a_1+8a_0)\right]
\label{MpwithC4}
\eal\eeq
The integrals over $r$ and $\theta$ can not be evaluated in closed form, 
but we can simplify the expression above a bit further by inspecting the 
leading terms in the large $\ap s$ limit with $\lambda=-t/s$ held fixed.
We first notice that both functions $a_0$ and $a_1$ contain $(\ap s)^2$  
terms, therefore it would seem that the first of the integrals in 
\eqref{MpwithC2} would dominate in the large $\ap s$ limit. This is, 
however, not true. From \eqref{gaussian} we see that the integrands in 
\eqref{MpwithC2} need to be evaluated at $x=\frac{\sin\theta_{43} 
\sin\theta_2}{\sin\theta_{42}\sin\theta_3}=x_c=(1-\lambda)^{-1}$. The full 
expression for the factor $|J|a_0$ in terms of the new variables 
$\theta,r,x$ that enters in both integrands  is given by
\bea
16 |J| a_0
&=& r^{-1}[\sin\theta]^{-2}x^{-2} \left[(1+\ap t)^2
\frac{x^2}{(1-x)^2}+(\ap  s)^2
\frac{x}{1-x}-\alpha'^2(s+t)^2\frac{x^2}{1-x}\right]
\label{ctcritical}
\eea
From here, we can readily see that the coefficient of $(\ap s)^2$ 
inside the square brackets above is 
\bea
\frac{x}{(1-x)^2}\left(1-x(1-\lambda)\right)^2\label{critcond}
\eea
which vanishes precisely at the value $x=x_c=(1-\lambda)^{-1}$. Therefore, 
$a_0$ really contributes linearly in $s$ in the hard scattering limit, not 
quadratically. On the other hand, the factor $|J|a_1$ which enters in the  
second integral in \eqref{MpwithC2}, when evaluated at $x=x_c$, becomes
\beq\bal
|J|a_1=x^{-2}\left(r^2+2r\cos\theta+1\right)^{-1} \left(\frac{r^2}{x^2}+
\frac{2r}{x}\cos\theta+1\right)^{-1}\left[(\ap s)^2(1-\lambda)
r\sin^2\theta+\frac{\ap s}{2r\lambda}g(r,\theta)\right]
\eal\eeq
where
\bea
g(r,\theta)&\equiv& 1+2r^2\left(2-2\lambda+\lambda^2\right)+r^4\left(1-2\lambda\right)\nn\\
&& + \, 2r(2-\lambda)(1+r^2-\lambda r^2)\cos\theta+2r^2(1-\lambda)
\cos2\theta
\eea
thus, the contribution from $a_1$ does goes as $a_1 \sim (-\ap s)^2$ in 
the hard scattering limit and dominates over the one from $a_0$. Therefore, 
the leading behavior of the renormalized $\mathcal{M}_{4,\rm ren}^+$ 
amplitude is
\beq\bal
\mathcal{M}_{4,\rm ren}^+ &\simeq  2\left(\frac{1}{8 \pi \ap }
\right)^{D/2}\frac{\pi^{20-2D-2S}}{\gamma-1} e^{-\alpha^{\prime} |s| f(\lambda)}
\sqrt{\frac{2\pi\lambda}
{\ap s (1-\lambda)^3}}\left(\frac{2}{\ln\ap |s|}
\right)^{\gamma-1} \int_0^{\pi}d\theta\int_0^{\infty}dr\,|J|a_1\\
&\simeq 2\left(\frac{1}{8 \pi \ap }\right)^{D/2}\frac{\pi^{20-2D-2S}}{\gamma-1} 
e^{-\alpha^{\prime} |s| f(\lambda)} \sqrt{-2\pi\lambda}
\left(\frac{2}
{\ln\ap |s|}\right)^{\gamma-1}(1-\lambda)^{3/2} (-\ap s)^{3/2} F(\lambda)
\eal\eeq
where
\bea
F(\lambda)\equiv \int_0^{\pi}d\theta\int_0^{\infty}dr \frac{r\sin^2\theta (r^2+2r\cos\theta+1)^{-1}}
{(r^2(1-\lambda)^2+2r(1-\lambda)\cos\theta+1)}
\eea
We can now write a more succinct expression for the final behavior of the 
renormalized $\mathcal{M}^+$ part of amplitude in the hard scattering limit 
as
\bea
\mathcal{M}_{4,\rm ren}^+ \!\!&\simeq&\!\! G(\lambda)\, e^{-\alpha^{\prime} 
|s| f(\lambda)} 
\left(\frac{1}{\ln\ap |s|}\right)^{\gamma-1}(-\ap s)^{3/2}
\label{Mpfinal}
\eea 
with
\bea
G(\lambda)\equiv 2\left(\frac{1}{8 \pi \ap }\right)^{D/2}\frac{2^{\gamma-1}
\pi^{20-2D-2S}}{\gamma-1}(-2\pi\lambda)^{1/2}(1-\lambda)^{3/2}F(\lambda)
\eea
Note that, since in the hard scattering limit both $s$ and $t$ are large 
compared to $\ap{}^{-1}$, we have $\ln(-\ap s)=\ln(-\ap t) 
\left(1+\mathcal{O}(\frac{1}{\ln(-\ap t)})\right)$, thus at leading order 
we can write \eqref{Mpfinal} also as
\bea
\mathcal{M}_{4,\rm ren}^+ \!\!&\simeq&\!\! G(\lambda)\, e^{-\alpha^{\prime} 
|s| f(\lambda)} 
\left(\frac{1}{\ln\ap |t|}\right)^{\gamma-1}(-\ap s)^{3/2}
\label{Mpfinal2}
\eea
This form will be useful when we compare these results with the Regge 
behavior of the amplitude which is done in section \ref{hardtoregge}.

We now repeat the analysis of the $q\sim 0$ region for the NS$-$ spin structure, i.e., 
the $\mathcal{M}^-$ part of the amplitude. This one reads 
\bea
\mathcal{M}_{4,\rm ren}^- &=& 2\left(\frac{1}{8 \pi \ap }\right)^{D/2} 
\int_0^1 \frac{dq}{q} 
\left(\frac{-\pi}{\ln q}\right)^{(10-D)/2} P_-(q)\nn\\
&&\int \prod_{k=2}^4 d\theta_k \, \left[e^{-\ap s V_{\lambda}}\langle 
\hat{\mathcal{P}}_1 
\hat{\mathcal{P}}_2 \hat{\mathcal{P}}_3 \hat{\mathcal{P}}_4 \rangle^- 
-e^{-\ap s V_{\lambda}^0}
\langle \hat{\mathcal{C}}_1 \hat{\mathcal{C}}_2 \hat{\mathcal{C}}_3 
\hat{\mathcal{C}}_4 \rangle-B^-\right]
\label{MmwithC}
\eea
From here we see that the only differences with respect to the 
$\mathcal{M}^+$ case lie on the partition function $P_-(q)$ and the 
correlator $\langle \hat{\mathcal{P}}_1 
\hat{\mathcal{P}}_2 \hat{\mathcal{P}}_3 \hat{\mathcal{P}}_4 \rangle^-$. The 
exponential factors are the same as before. From equations \eqref{PP} and 
\eqref{HHm} we have
\bea
P_-(q)&=&2^4+\mathcal{O}(q^2)\\
\langle \hat{\mathcal{P}}_1 
\hat{\mathcal{P}}_2 \hat{\mathcal{P}}_3 \hat{\mathcal{P}}_4 \rangle^-&=& 
\langle \hat{\mathcal{P}}_1\hat{\mathcal{P}}_2 \hat{\mathcal{P}}_3 
\hat{\mathcal{P}}_4 \rangle^-_{q=0} + \mathcal{O}(q^2)
\eea
Expanding about the critical surface $(x,q)=(x_c,0)$ again, the amplitude 
\eqref{MmwithC} becomes
\bea
\mathcal{M}_{4,\rm ren} ^- &\simeq& 2\left(\frac{1}{8 \pi \ap }\right)^{D/2} 
e^{-\alpha^{\prime} |s| f(\lambda)}2^4\int \prod_{k=2}^4 d\theta_k 
\, e^{-\ap s\frac{(1-\lambda)^3}{2\lambda}
(x-x_c)^2}\int_0^{\epsilon} \frac{dq}{q} 
\left(\frac{-\pi}{\ln q}\right)^{(10-D)/2} \,
\times \nn\\
&&\left[e^{-2\ap s \, q^2(S_1-\lambda T_1)}\langle \hat{\mathcal{P}}_1 
\hat{\mathcal{P}}_2 \hat{\mathcal{P}}_3 \hat{\mathcal{P}}_4 \rangle^-_{q=0}
- \langle \hat{\mathcal{C}}_1 \hat{\mathcal{C}}_2 \hat{\mathcal{C}}_3 
\hat{\mathcal{C}}_4 \rangle+e^{-2\ap s \, 
q^2(S_1-\lambda T_1)}\mathcal{O}(q^2)\right]
\label{MmwithC2}
\eea
We again recall that the integral over the $\theta_k$ variables is dominated 
by the two dimensional surface
\bea
x=\frac{\sin\theta_{43}\sin \theta_{2}}
{\sin\theta_{42}\sin\theta_{3}}=(1-\lambda)^{-1}=x_c.
\eea
The integral over the 
cross ratio $x$ then becomes a gaussian which, at leading order, demands 
that we evaluate the expression inside the square brackets above at 
$\frac{\sin\theta_{43}\sin \theta_{2}}{\sin\theta_{42}\sin\theta_{3}}=
(1-\lambda)^{-1}$. It will be again 
convenient to separate the $s^2$ part of $\langle \hat{\mathcal{P}}_1 
\hat{\mathcal{P}}_2 \hat{\mathcal{P}}_3 \hat{\mathcal{P}}_4 \rangle^-_{q=0}$ 
as 
\bea
\langle \hat{\mathcal{P}}_1 
\hat{\mathcal{P}}_2 \hat{\mathcal{P}}_3 \hat{\mathcal{P}}_4 \rangle^-_{q=0} 
= As^2 + Bs + C
\eea
and, from equations \eqref{PP} and \eqref{HHm}, we obtain
\bea
A &\equiv& \frac{1}{16}\cot\theta_{4}\cot\theta_{32}
\left[\lambda^2\cot\theta_4\cot\theta_{32}-\cot\theta_2\cot\theta_{43}
-(1-\lambda)^2\cot\theta_3\cot\theta_{42}
\right]
\eea
Evaluating this expression on the critical surface implies that we make the  
replacement 
$\lambda=-\frac{\sin\theta_4\sin\theta_{32}}{\sin\theta_{43}\sin\theta_2}$. 
Remarkably, one can see that $A$ vanishes in this case, yielding
\bea
\langle \hat{\mathcal{P}}_1 
\hat{\mathcal{P}}_2 \hat{\mathcal{P}}_3 \hat{\mathcal{P}}_4 \rangle^-_{q=0} 
\to  Bs + C \label{Corrmcrit}
\eea 
on the critical surface. The coefficient of $s^2$ of the counterterm 
$\langle \hat{\mathcal{C}}_1 \hat{\mathcal{C}}_2 \hat{\mathcal{C}}_3 
\hat{\mathcal{C}}_4 \rangle$ also vanishes on this surface as derived in 
equations \eqref{ctcritical} and \eqref{critcond}. Given these facts, we 
can now estimate the contributions from the rest of the terms in 
\eqref{MmwithC2} as follows. The integration over the first term inside 
the square brackets in \eqref{MmwithC2} has the same form as \eqref{I2}, 
thus together with the $s$ factor coming from the correlator 
\eqref{Corrmcrit} it behaves as $s\,(\log\ap |s|)^{D/2-4}$. Due to the 
lack of the exponential factor in front it, the contribution from the 
counterterm 
$\langle \hat{\mathcal{C}}_1 \hat{\mathcal{C}}_2 \hat{\mathcal{C}}_3 
\hat{\mathcal{C}}_4 \rangle$ is exponentially suppressed. This was expected 
here since this counterterm is not necessary to make the behavior of the 
$\mathcal{M}^-$ amplitude convergent near the $q=0$ region\footnote{This 
counterterm is however necessary to cancel spurious divergences from 
certain regions in the $\theta_k$ integrals. The contributions from these 
regions will be analyzed separately at the end of this section.}. We can 
also estimate the contribution from all the rest of terms in the expansion 
in powers of $q$ by recalling that the exponential factor 
$\exp\{-2\ap s \, q^2(S_1-\lambda T_1)\}$ in \eqref{MmwithC2} cuts off 
the effective range of the $q$ integral to $\epsilon\sim s^{-1/2}$. Thus, 
since we have an expansion in even powers of $q$, the integral will produce 
a contribution $\sim s^{-1/2}\times (s^{-1/2})^{2n-1}
(\log\ap |s|)^{D/2-4}=s^{-n}(\log\ap |s|)^{D/2-4}$ with $n\geq 1$. 
The maximum power of $s$ that could come from the correlator 
$\langle \hat{\mathcal{P}}_1 \hat{\mathcal{P}}_2 \hat{\mathcal{P}}_3 
\hat{\mathcal{P}}_4 \rangle^-$ is $s^2$. Thus, even if there are no 
cancellations of these terms on the critical surface, the leading 
behavior coming from $\mathcal{O}(q^2)$ terms in \eqref{MmwithC2} is 
$s(\log\ap |s|)^{D/2-4}$. Finally, from \eqref{gaussian}, we already 
know that the integral over the cross ratio $x$ produces an overall factor 
of $s^{-1/2}$. Thus, putting everything together, we have that the leading 
behavior of $\mathcal{M}_4^--\mathcal{C}_4$ is
\bea
\mathcal{M}_{4,\rm ren}^-\sim e^{-\alpha^{\prime} |s| f(\lambda)}
\left(\frac{1}{\log\ap |s|}\right)^{D/2-4}(-\ap s)^{1/2}
\eea
which is definitely subleading with respect to 
$\mathcal{M}_{4,\rm ren}^+$ in \eqref{Mpfinal}.\\

We now turn to the study of the contributions from other regions that we 
have not analyzed yet. 
As mentioned before, the asymptotic behavior is governed by critical points 
of the second kind, i.e., the boundary regions of the integrated moduli. Thus, we also need 
to examine the region where $q\to 1$. To this end it is convenient to 
perform 
the Jacobi imaginary transformation $q=\exp\{2\pi^2/\ln w\}$, which maps the 
$q\sim 1$ region to $w\sim 0$. Using the corresponding transformations on 
the $\theta_1(\nu|\tau)$ function, we have
\bea
\theta_1\left({i\theta\ln w\over 2\pi},\sqrt{w}\right)
&=&-i\left({-2\pi\over\ln w}\right)^{1/2}
\exp\left\{{-\theta^2\ln w\over2\pi^2}
\right\}\theta_1(\theta,q)\label{jacobitheta1}\\
\theta_1^\prime(0,\sqrt{w})&=& 
\left({-2\pi\over\ln w}\right)^{3/2}\theta_1^\prime(0,q)
\label{jacobitheta1prime}
\eea
gives
\bea
\psi(\theta,q)&=&{\theta_1(\theta,q)\over\theta^\prime(0)}
=i{-2\pi\over\ln w }\exp\left\{{\theta^2\ln w\over2\pi^2}
\right\}{\theta_1\left({i\theta\ln w/ 2\pi},\sqrt{w}\right)\over
\theta_1^\prime(0,\sqrt{w})}\nonumber\\
&=&{\pi\over-\ln w }\exp\left\{-{\theta(\pi-\theta)\ln w\over2\pi^2}
\right\}(1-w^{\theta/\pi})\prod_{n=1}^\infty{(1-w^{n+\theta/\pi})
(1-w^{n-\theta/\pi})\over(1-w^n)^2}\label{jacobipsi}
\eea
We are thus interested in the small $w$ behavior of $\ln \psi$, therefore
\bea
\ln \psi &=& \ln \left({\pi\over-\ln w }\right) - 
{\theta(\pi-\theta)\ln w\over2\pi^2} + 
\ln(1-w^{\theta/\pi}) + \sum_{n=1}^{\infty} \ln {(1-w^{n+\theta/\pi})
(1-w^{n-\theta/\pi})\over(1-w^n)^2}\nn\\
&=& \ln \left({\pi\over-\ln w }\right) -{\theta(\pi-\theta)\ln w\over2\pi^2} 
+\mathcal{O}(w)
\label{logpsiw}
\eea
Keeping the first two terms is a good approximation as long as $\theta$ is 
not too close to zero or $\pi$. Using the approximation \eqref{logpsiw} we 
have
\bea
|V_{\lambda}| \approx  \Big|\frac{\ln w}{s\pi^2 }\sum_{i<j}  
\theta_{ji}(\pi-\theta_{ji})k_i \dt k_j + \mathcal{O}(w)\Big|
\eea
where the first term in \eqref{logpsiw} has vanished due to momentum 
conservation. We can readily see that at 
$w=0$ the function $V_{\lambda}$ increases logarithmically with $w$. Since 
the overall sign of $V_{\lambda}$ is negative, we see that the contribution 
from this region will be exponentially suppressed with respect to the 
one from $q\sim 0$ already computed. We have 
thus analyzed both boundaries, $q=0$ and $q=1$, and found that the 
first one dominates. 

The last pending task in this section is to estimate the contribution 
from the regions of integration we have avoided until now. As mentioned 
earlier, the regions where $S_n-\lambda T_n$ get arbitrarily close to 
zero invalidate the power series expansion in $q$ and the full 
integral over $q$ must be performed in order to obtain the correct 
asymptotic behavior coming from these places. This is a very complicated 
problem since the analytic approximations turn out to be difficult to 
analyze, however, we can estimate their contributions  and show that they 
are subleading with respect to the one from $q\sim 0$.  A crucial 
point is that, following \cite{GrossManes}, all the stationary points for 
the planar amplitude lie on the boundary of the integration region. Since 
we are now away from either $q=0$ and $q=1$ and focusing on all possible 
stationary regions that could come from the $\int d\theta_k$ integral, all 
we need to analyze are the boundary regions in the $\theta_k$ variables. These are 
the faces, edges and vertices of the 3-simplex shown in figure \ref{fig:simplex}. 

Recall that in the hard scattering limit the important factor is the one given 
in equation \eqref{psiprod2} which we write again here
\bea
\prod_{i<j} \psi (\theta_{ji})^{2\ap  k_i \cdot k_j} =  
e^{\ap  |s| \, V_{\lambda}}
\label{exponent1}
\eea
where
\bea
V_{\lambda} = \ln x -\lambda \ln(1-x) + 2\sum_{n=1}^{\infty} \frac{1}{n} 
\frac{q^{2n}}{1-q^{2n}}(S_n-\lambda T_n)
\label{exponent2}
\eea
From the expression for $V_{\lambda}$ we see that a maximum can also occurs 
for any value of $q$ provided that $x\sim x_c=\frac{s}{s+t}$ and 
$S_n-\lambda T_n \sim 0$. Notice that it is not possible to have an 
end-point-like contribution from the $\ln x-\lambda \ln(1-x)$ term in 
\eqref{exponent2} for fixed $\lambda$ because, since $\lambda<0$, this term 
does not get arbitrarily close to zero in the integration range $0<x<1$.
Therefore, this term will again provide with a stationary surface 
only from $x=x_c\equiv (1-\lambda)^{-1}$. Thus, now we need to analyze all 
possible \textbf{boundary} regions that could make $S_n-\lambda T_n$ 
vanish. After careful examinations, this will occur in the regions 
where all or all but one of the vertex 
operators coincide. The regions where all four vertex operator coincide 
correspond to the four vertices in figure \eqref{fig:simplex}. 
These are:
\bea
&&\theta_2=\theta_3=\theta_4=0 \label{vertex1}\\
&&\theta_2=\theta_3=0\,,\,\theta_4=\pi\label{vertex2}\\
&&\theta_2=0\,,\,\theta_3=\theta_4=\pi\,\, , \,\mbox{and}\label{vertex3}\\
&&\theta_2=\theta_3=\theta_4=\pi \label{vertex4}
\eea
Let us analyze one of the regions where all four vertex operators collapse, 
say, the vertex \eqref{vertex1}. It is convenient here to make the changes 
$\theta_4=\epsilon$, $\theta_3=\epsilon \eta_3$, 
$\theta_2=\epsilon \eta_2$ with $\epsilon$ small and expand everything in 
powers of $\epsilon$. In \cite{GrossManes} the authors also analyze these 
regions and point out that the asymptotic behavior of the amplitude does not 
depend on $\epsilon$ only for the superstring. The reason for this is that 
the regions in moduli space where $\epsilon\sim 0$ produce divergences that 
are due to the presence of tachyons which are absent in the superstring. \\
\begin{comment}
We can see this by using the GNS regulator, this is, we suspend total 
momentum conservation by an amount $p$ 
and then we take $p\to0$ at the end. Using this regulator we have
\bea
\prod_{i<j} \psi(\theta_{ji})^{2\ap  k_i \cdot k_j} \approx \prod_{i<j}
(\theta_{ji})^{2\ap  k_i \cdot k_j} 
\approx \epsilon^{\sum_{i<j}2\ap  k_i \cdot k_j}=\epsilon^
\eea
The regions where all four vertex operators coincide correspond to tachyons 
propagating with zero momentum while the regions where all but one vertex 
operator coincide are due to on-shell tachyons. In the NS+ model we are 
considering in this article we have projected out all odd G-parity states 
from the open string sector, therefore there are no open string 
tachyons.\\[5pt]
\end{comment}
Due to the form of $S_m-\lambda T_m$ there will only be even powers in 
$\epsilon$. In this case we have
\bea
S_m-\lambda T_m = \frac{8 m^2}{s} \left((s+t)\eta_2-
(s+t \eta_2)\eta_3\right) 
\, \epsilon^2 +\mathcal{O}(\epsilon^4)
\label{4coinexp}
\eea
The important point is that, since we have to evaluate this expression at 
the critical surface $x=x_c$, we have
\bea
x=\frac{\sin\theta_{43}\sin \theta_2}{\sin\theta_{42}\sin\theta_3} = 
\frac{(1-\eta_3)\eta_2}{\eta_3(1-\eta_2)} + \mathcal{O}(\epsilon^2)
=\frac{1}{1+t/s} 
\qquad \to \quad \eta_2 = \frac{s\eta_3}{s+t-t\eta_3} + 
\mathcal{O}(\epsilon^2)
\label{crosscondition}
\eea
Plugging this into \eqref{4coinexp} makes the entire coefficient multiplying
$\epsilon^{-2}$ in \eqref{4coinexp} to vanish! Therefore, on the critical 
surface we have $S_m-\lambda T_m \sim\mathcal{O}(\epsilon^4)$. The 
exponential factor \eqref{exponent2} then has its largest contribution to 
the integral for $s\,\epsilon^4 \sim 1$, which implies that the effective 
range for each variable $\theta_i$ is $\epsilon \sim s^{-1/4}$. Because we 
have a triple integral over these angles, the total contribution from 
the measure is $\sim s^{-3/4}$. Since $x$ is $\mathcal{O}(1)$ in 
$\epsilon$, the small corner studied here still contains the two 
dimensional plane $x=x_c=\frac{1}{1+t/s}$, which we already know contributes with a factor of 
$s^{-1/2} e^{-\alpha^{\prime} |s| f(\lambda)}$. 
Recall 
that the $s^{-1/2}$  factor comes from the gaussian approximation of 
the cross-ratio along this plane. The rest of the integrand only 
involves the contractions $\VEV{\hat{\mathcal{P}}_1\hat{\mathcal{P}}_2
\hat{\mathcal{P}}_3\hat{\mathcal{P}}_4}$. Note that our expression for 
the gluon amplitude includes counterterms that eliminate all possible 
divergences in the $\theta$ integrals. In particular, the corner of the 
integration region we are considering here is precisely one of the 
places that originally produced divergences. These were taken care by 
the counter-term\footnote{This counter-term has a two-fold purpose since 
it also cancels the divergence $\int \frac{dq}{q^2}$ for small $q$ which 
is re-interpreted as a renormalization of the coupling. This is the only 
divergence in the $q$ integration as long as the Dp-brane has $p<7$} that 
involves taking the $\VEV{\hat{\mathcal{P}}_1\hat{\mathcal{P}}_2
\hat{\mathcal{P}}_3\hat{\mathcal{P}}_4}$ correlator at $q=0$. Thus, 
starting from equation \eqref{MpwithC}, we see that we need to estimate 
the contribution of
\bea
e^{-2\ap s\sum_{n=1}^{\infty}\frac{1}{n}\frac{q^{2n}}{1-q^{2n}}
(S_n-\lambda T_n)}\VEV{\hat{\mathcal{P}}_1\hat{\mathcal{P}}_2
\hat{\mathcal{P}}_3\hat{\mathcal{P}}_4}-\VEV{\hat{\mathcal{C}}_1
\hat{\mathcal{C}}_2\hat{\mathcal{C}}_3\hat{\mathcal{C}}_4}
\eea
from the region in consideration. In this region we have that 
$s(S_n-\lambda T_n)\sim\mathcal{O}(1)$, 
therefore the prefactor of  the first term above is a number of order one. Now, 
expanding $\VEV{\hat{\mathcal{P}}_1
\hat{\mathcal{P}}_2\hat{\mathcal{P}}_3\hat{\mathcal{P}}_4}$ in powers of $\epsilon$ gives
\bea
\VEV{\hat{\mathcal{P}}_1\hat{\mathcal{P}}_2\hat{\mathcal{P}}_3
\hat{\mathcal{P}}_4}
=\alpha_1 \epsilon^{-4}+\alpha_2 \epsilon^{-2}+\mathcal{O}(1)
\eea
where the coefficient $\alpha_1$ above turns out to be
\bea
\alpha_1=\frac{s^2}{16 \eta_2 (\eta_3-1)(\eta_2-\eta_3)}-
\frac{s^2 (\lambda -1)^2}{16 
(\eta_2-1)(\eta_2-\eta_3) \eta_3}
+\frac{(s \lambda -1)^2}
{16 (\eta_2-\eta_3)^2}
\eea
Notice that this coefficient lacks $q$ dependence, which means that the 
$\VEV{\hat{\mathcal{C}}_1\hat{\mathcal{C}}_2\hat{\mathcal{C}}_3
\hat{\mathcal{C}}_4}$ term will have the exact same coefficient in its 
expansion in powers of $\epsilon$. Now, since we must demand the 
expression above to satisfy the condition \eqref{crosscondition} in order 
to lay on the critical surface, it is somewhat remarkable that the $s^2$ 
term in both coefficients $\beta_1$ and $\beta_2$ vanishes. As we will now 
show, this makes this contribution be smaller than the one computed 
from the $q\sim 0$ region, therefore it is subleading! Had not this been the case, 
this region would have dominated in 
the hard scattering regime and the entire leading behavior would 
have been much harder to obtain. Moreover, it is the $q\sim 0$ region 
the one that provides the correct asymptotic behavior for $s\gg t$ which 
matches with the Regge limit at high $t$ as we will show in the next 
section.

All in all, the total contribution from this corner has the following 
structure: (i) the cross-ratio $x$ contributes with a factor of 
$s^{-1/2}e^{-\alpha^{\prime} |s| f(\lambda)}$ as seen above; (ii) since 
the relevant range 
for each $\theta_i$ variable is $\Delta \theta \sim s^{-1/4}$, the triple 
integral provides a factor of $s^{-3/4}$; (iii) the rest of the integrand, 
namely the correlators 
$\VEV{\hat{\mathcal{P}}_1\hat{\mathcal{P}}_2\hat{\mathcal{P}}_3
\hat{\mathcal{P}}_4}$ and 
$\VEV{\hat{\mathcal{C}}_1\hat{\mathcal{C}}_2\hat{\mathcal{C}}_3
\hat{\mathcal{C}}_4}$, behave as $s \epsilon^{-4}\sim s^2$ on the plane 
$x=x_c$. Therefore, the total estimate is $s^{-1/2}e^{-\alpha^{\prime} 
|s| f(\lambda)} 
\times s^{-3/4} \times s^2=s^{3/4}e^{-\alpha^{\prime} |s| f(\lambda)}$ 
which definitely 
smaller than the one obtained in \eqref{Mpfinal} which came from the 
$q\sim 0$ region. It also straightforward to show, 
after suitable changes of variables, that the other three vertices 
\eqref{vertex2}, \eqref{vertex3}, and \eqref{vertex4} give an identical 
contribution.\\
A final estimation we need to obtain is the one from the regions that 
correspond to a 3-particle coincidence, that is, the regions where all 
but one of the vertex operators coincide in the moduli space. There are 
four of these regions and they correspond to four of the edges in the 
integration domain depicted in figure \ref{fig:simplex}. All of the edges 
corresponding to 
a loop insertion on an eternal leg will produce an 
important contribution in the hard scattering limit. These are  : 
$\theta_2=\theta_3=0$, $\theta_2=\theta_3=\theta_4$, 
$\theta_2=\pi-\theta_4=0$ 
and 
$\theta_3=\theta_4=\pi$. Let us focus on the first one. In this case 
it is again convenient to define $\theta_2=\eta_2\epsilon$ 
and $\theta_3=\epsilon$ and expand for small values of $\epsilon$. 
In this case we have
\bea
x=\eta_2 +\mathcal{O}({\epsilon})
\eea
and the analogous condition to \eqref{crosscondition} here is 
$\eta_2=(1-\lambda)^{-1}$ at leading order. Expanding $S_n-\lambda T_n$ 
in powers of $\epsilon$ yields
\bea
S_n-\lambda T_n = 2 n (\eta_2(1-\lambda)-1) 
\sin (2n \theta_4) \epsilon +\mathcal{O}(\epsilon^2)
\label{3coinexp}
\eea
from where see that the $\mathcal{O}(\epsilon)$ term vanishes on the 
critical surface. The $\mathcal{O}(\epsilon^2)$ does not vanish there, 
consequently, $S_m-\lambda T_m = \mathcal{O}(\epsilon^2)$. This implies 
that now the effective range of each $\theta_i$ variable in this corner 
of the integration
region is $\Delta \theta_i \sim \epsilon \sim s^{-1/2}$.
Expanding again $\VEV{\hat{\mathcal{P}}_1\hat{\mathcal{P}}_2
\hat{\mathcal{P}}_3 \hat{\mathcal{P}}_4}$ gives
\bea
\VEV{\hat{\mathcal{P}}_1\hat{\mathcal{P}}_2\hat{\mathcal{P}}_3
\hat{\mathcal{P}}_4}=\beta_1 \epsilon^{-2}+\mathcal{O}(1)
\eea
It will be again convenient to write the leading coefficient $\beta_1$ as 
a polynomial in $s$ as
\bea
\beta_1=b_0+b_1 s+b_2 s^2
\eea
As before, we single out the coefficient of the $s^2$ term $b_2$ for 
which we obtain
\bea
b_2=\left(1+\eta_2 (\lambda -1)\right)^2 \frac{\theta_2(0,q)^4 
\theta_3(0,q)^2 \theta_3(\theta_4,q)^2 \theta_4(0,q)^4}{16 (\eta_2-1)^2 
\eta_2 \theta_1(\theta_4,q)^2 \theta_1'(0,q)^2}
\eea
where $\theta_l(\theta_i,q)$ denotes the $l$-th Jacobi Theta function evaluated at
$(\theta_i,q)$.
From here we see immediately that this coefficient vanishes if 
$\eta_2=(1-\lambda)^{-1}$, which is precisely the value that $\eta_2$ 
acquires on the critical surface $x=x_c=(1-\lambda)^{-1}$. Therefore, when
we evaluate $\VEV{\hat{\mathcal{P}}_1\hat{\mathcal{P}}_2
\hat{\mathcal{P}}_3
\hat{\mathcal{P}}_4}$ on that surface we have
\bea
\beta_1=b_0+b_1 s
\eea
and likewise for 
$\VEV{\hat{\mathcal{C}}_1\hat{\mathcal{C}}_2\hat{\mathcal{C}}_3
\hat{\mathcal{C}}_4}$ since in that case the only difference is that 
the Jacobi Theta functions need to be evaluated at $q=0$, giving the result
\bea
\csc^2\theta_4
\frac{(1+\eta_2 (\lambda -1))^2 }{16 (\eta_2-1)^2 \eta_2}
\eea  
which also vanishes for the same value $\eta_2=(1-\lambda)^{-1}$. 
With these two results, we see that the biggest contribution from the 
correlators goes like $\sim \epsilon^{-2}s\sim s^2 $. Since 
$x=\mathcal{O}(1)$, the critical plane is again contained in the corner 
we are analyzing, producing again a factor of 
$s^{-1/2} e^{-\alpha^{\prime} |s| f(\lambda)}$. 
The since $\theta_2<\theta_3$, in this corner we also have that 
$d\theta_2 d\theta_2$ is $\mathcal{O}(\epsilon^2)$ 
thus giving a factor of $s^{-1}$. Putting everything together, we have that 
the corner $\theta_3\sim \theta_2\sim 0$ produces a total contribution 
$\sim s^2 \times s^{-1/2} e^{-\alpha^{\prime} |s| f(\lambda)} \times s^{-1}= 
s^{1/2}e^{-\alpha^{\prime} |s| f(\lambda)}$. It is also straightforward to 
check that the 
other two remaining corners produce the same answer. Therefore, we see again 
that these regions produce subleading behavior with respect to 
\eqref{Mpfinal}. \\

Summarizing, we have analyzed all the regions that produce dominant
contributions in the high energy regime at fixed angle.  These contributions correspond to the regions 
comprised of all possible stationary points of $V_{\lambda}$ (see equations 
\eqref{psiprod2} and \eqref{VsVt}). Our analysis yields that the leading contribution, among all the 
dominant regions, comes from the boundary 
\bea
q=0 \quad \mbox{with} \quad x=(1-\lambda)^{-1}
\eea
where $x$ was defined in \eqref{NSx}. 
Therefore, quoting the result from \eqref{Mpfinal2}, the behavior of the 
renormalized 4-point amplitude in the hard 
scattering regime is
\bea
\mathcal{M}_4^{\rm Hard}\!\!&\simeq&\!\! \, G(\lambda) \, 
e^{-\ap |s| f(\lambda)} 
\left(\log\ap|t|\right)^{1-\gamma}(\ap|s|)^{3/2} 
\label{Mfinal}
\eea 
$G(\lambda)$ only depends on the scattering angle as a function of $\lambda$ and is given by 
\bea
G(\lambda)\equiv 2g^2\left(\frac{1}{8 \pi \ap}\right)^{D/2}
\frac{2^{\gamma-1}\pi^{20-2D-2S}}{\gamma-1}
(-2\pi\lambda)^{1/2}(1-\lambda)^{3/2}F(\lambda)
\label{Glambda}
\eea 
with 
\bea
F(\lambda)\equiv \int_0^{\pi}d\theta\int_0^{\infty}dr 
\frac{r\sin^2\theta (r^2+2r\cos\theta+1)^{-1}}{
(r^2(1-\lambda)^2+2r(1-\lambda)\cos\theta+1)}
\label{Flambda}
\eea
where this last expression is the same one found in \cite{Rojas:2011pb}.
The expression for $F(\lambda)$ given in \eqref{Flambda} is convergent 
in the entire range $-\infty<\lambda<0$ and it only diverges when 
$\lambda$ approaches zero, in which case, it diverges logarithmically 
in $\lambda$. We will see that this is precisely what is needed in 
order to recover the Regge behavior from the hard scattering limit.

In conclusion, the amplitude is exponentially suppresed at high energies as 
expected for stringy amplitudes in this regime, but we also have an extra 
logarithmic falloff product of the presence of the D-branes. The full 
dependence in $\lambda$ contained in the function $G(\lambda)$ will be 
crucial in order to make contact with the results in \cite{ThornRojas} 
because taking the $-t/s=\lambda\to 0$ limit in \eqref{Mfinal} should reproduce the high $t$ 
limit of the Regge behavior. We will show in section \ref{hardtoregge} 
that this limit is indeed recovered.
\subsection{Comparison with tree amplitude}\label{Sec:tree}
At arbitrary energies, the tree amplitude for this polarization is
\bea
\mathcal{M}_4^{tree} = -g^2 \ee{1}{4}\ee{2}{3} 
\frac{\Gamma(1-\alpha' s)\Gamma(-\alpha' t)}
{\Gamma(-\alpha' s -\alpha' t)}
\label{tree}
\eea
where we have only omitted numerical factors for simplicity.
Using Stirling's approximation $\Gamma(1+x)\simeq x^x e^{-x}
(2\pi x)^{1/2}$, the $\alpha' s\to -\infty$ limit with $-t/s\equiv \lambda$ held fixed is
\bea
\mathcal{M}_4^{\rm Tree} &\sim&  -g^2 \ee{1}{4}\ee{2}{3} \sqrt{2\pi} (-\ap s)^{1+\ap t}
(-\ap t)^{-1/2-\ap t}(1+t/s)^{1/2+\ap s+\ap t } \nn\\
&\sim&  -g^2 \ee{1}{4}\ee{2}{3} \sqrt{2\pi} (-\ap s)^{1/2}(-\lambda)^{-1/2}(1-\lambda)^{1/2} 
e^{-\ap|s|f(\lambda)}
\label{hardlimit} 
\eea
where $f(\lambda)\equiv \lambda \ln(-\lambda)+(1-\lambda)
\ln(1-\lambda)\geq 0$. The ratio of the one-loop amplitude to the tree 
one in this regime is
\bea
\frac{\mathcal{M}_{\rm 1-loop}}{\mathcal{M}_{\rm tree}}  \sim -
\ap s\left(\frac{2}{\ln|\ap t|}\right)^{15-3D/2-S}
\label{ratio}
\eea 
Therefore, having computed the exact leading power of $s$ multiplying the 
exponential falloff allows us to assert that the planar one-loop amplitude 
dominates over the tree amplitude.
%This result is new in a sense because usually open string one-loop planar 
%amplitudes have the same exponential behavior as the tree level result 
%multiplied by the same power of $s$. In the case studied here the 
%exponential 
%decay is also the same as the tree level one, but the overall power of $s$ 
%is larger making the planar loop amplitude to dominate at high energies.
%-------------------------------------------------------------------------
%
%
\subsection{Recovery of the Regge behavior at high $t$}
\label{hardtoregge}
Recall that the Regge limit is 
obtained by taking $s$ to be large compared to $\alpha'^{-1}$ while 
keeping $t$ fixed, whereas the hard scattering regime is obtained 
by taking both $s$ and $t$ large compared to $\alpha'^{-1}$ while 
maintaining the ratio $\lambda=-t/s$ fixed. Therefore, we expect that when 
$s$ is large compared to $t$ in the hard scattering limit \eqref{Mfinal},   
this matches with the Regge limit when $t\gg \alpha'^{-1}$. The Regge limit 
in the type 0 model was obtained in \cite{ThornRojas} with the result
\bea
\mathcal{M}_4^{\rm Regge} \sim -g^2 \ee{1}{4}\ee{2}{3} (-\ap s)^{1+\ap t } \Gamma(-\ap t)
\log(-\ap s) \Sigma(t)
\label{MRegge}
\eea
where $\Sigma(t)$ is given by 
\bea 
\Sigma(t) \!\!\!&=& \!\!\! C\, g^2 \int_0^1 
\frac{dq}{q} \left(\frac{-\pi}{\ln q}\right)^{(10-D)/2} \!\!\!
\int_0^{\pi} \!\!\!d\theta 
\left((-\psi^2
[\ln \psi]^{\prime\prime})^{\ap t}
\frac{\ap t}{[\ln \psi]^{\prime\prime}}(P_+X^+-P_-X^-)\right.\nonumber\\
&& \left.-\frac{1}{4}(P_+-P_-)\left[(-\psi^2(\theta)
[\ln \psi]^{\prime\prime})^{\ap t }-1\right] 
[-\ln \psi]^{\prime\prime} \right)
\eea
giving the one-loop correction to the Regge trajectory. The functions $P_{\pm}$ and $\psi$  
are given in equations \eqref{P+} through \eqref{qpsi}. $X^{\pm}$ are defined in terms 
of the Jacobi Theta functions $\theta_i(\theta,q)$
\bea
X^+(q)&=&{1\over4}\theta_4(0)^4\theta_3(0)^4-{{\bfs E}\over\pi}
\theta_4(0)^4\theta_3(0)^2+{{\bfs E}^2\over\pi^2}\theta_3(0)^4\label{X+}\\
X^-(q)&=&-{1\over4}\theta_4(0)^4\theta_3(0)^4
+{{\bfs E}^2\over\pi^2}\theta_3(0)^4\label{X-}\\
{\bfs E}&=&{\pi\over6\theta_3(0)^2}\left({\theta_3(0)^4+\theta_4(0)^4}
-{\theta_1^{\prime\prime\prime}(0)\over\theta_1^\prime(0)}\right)
\eea
where we denote $\theta_i(0,q)\equiv\theta_i(0)$.
Using the infinite product representations 
\bea
\theta_3(0)&=&\prod_n(1-q^{2n})\prod_r(1+q^{2r})^2\\
\theta_4(0)&=&\prod_n(1-q^{2n})\prod_r(1-q^{2r})^2\\
{\theta_1^{\prime\prime\prime}(0)\over\theta_1^\prime(0)}
&=&-1+24\sum_n {q^{2n}\over(1-q^{2n})^2}
\eea
one can write $X^{\pm}$ explicitly in terms of $q$. The sums over $n$ are over positive integers and those over $r$
are over half odd integers.
As mentioned above, we need to take the limit $\ap t \gg 1$ in 
\eqref{MRegge}. Using Stirling's approximation   
$\Gamma(-\ap t) \sim \sqrt{2\pi} (-\ap t )^{-1/2-\ap t }e^{\ap t }$ 
and the fact that for large $\ap t$ the one-loop trajectory function 
becomes \cite{ThornRojas} 
\bea
\Sigma(t) \sim \ap t  \, [\log (-\ap t )]^{1-\gamma}
\eea
we have that 
\bea
\mathcal{M}_4^{\rm Regge} \underset{\alpha't \gg 1}{\sim} g^2 \ee{1}{4}\ee{2}{3}
(-\ap s)^{1+\ap t }(-\ap t )^{1/2-\ap t }e^{\ap t }\log(-\ap s)  
\, [\log (-\ap t )]^{1-\gamma}
\label{MRegge2}
\eea
We now expect to recover this result by taking the $s\gg t$ limit 
in \eqref{Mfinal}. This amounts to take the $\lambda \to 0$ limit of 
$F(\lambda)$ defined in \eqref{Flambda} and then putting this back into 
\eqref{Mfinal}.
For convenience, we write this integral here again
\bea
F(\lambda) = \int_0^{\infty}dr \int_0^{\pi} d\theta \, \frac{r\sin^2\theta(r^2+2r\cos\theta+1)^{-1}}
{(r^2(1-\lambda)^2+2r(1-\lambda)\cos\theta+1)}
\label{Flambda2}
\eea
This integral converges in the whole range $-\infty<\lambda<0$ but it 
gets larger and larger as $\lambda$ approaches zero. Recall that 
$\lambda = -t/s$ so this is precisely the limit we want to study. 
By putting $\lambda=0$ in the integrand of \eqref{Flambda2}, we see that 
the only singular region is the one given by $\theta \sim \pi$ and 
$r\sim 1$. There is an alternative way to note that this is the relevant 
region in the Regge limit. Recall that the saddle point which dominates 
in the high energy limit is given by $x=(1-\lambda)^{-1}$. From the 
definitions \eqref{nsvariables} we note that the region $\theta \sim \pi$ 
and $r\sim 1$ corresponds\footnote{Recall that the original $\theta_4$ 
variable was renamed $\theta$ here.} to 
$x=\frac{r\sin\theta_2}{\sin\theta_{42}}\sim 1$ which is precisely the 
location of the dominant saddle as $\lambda\to0$.  This perfectly matches 
with the fact that the Regge behavior of the amplitude is obtained from 
the region $\theta_2\sim\theta_3$, $\theta_4\sim \pi$ for which we have 
$x \sim 1-\theta_{32}(\pi-\theta_4)\csc^2\theta_3$. Thus, in the integrand  
above, let us replace $(1-\lambda)$ by $x^{-1}$ for notational convenience. 
Therefore, since the relevant region for integral above is given by 
$r \to x \to 1$, we focus on the corner $r\sim x$, $\theta\sim \pi$, thus
\bea
F(\lambda) &\sim& \int_{x-\delta}^{x+\delta}dr \int_{\pi-\epsilon}^{\pi} 
d\theta \, \frac{(\pi-\theta)^2}{((x-1)^2+x(\pi-\theta)^2)
((r/x-1)^2+(\pi-\theta)^2)}\nn\\
&\sim& 2 \int_0^{\epsilon} \frac{\theta}{(x-1)^2+x\theta^2}=
-2\ln|1-x|+\ln((1-x)^2+\epsilon^2)
\label{Flambda3}
\eea
Therefore, as $\lambda\to0$  for fixed $\epsilon$, we have
\bea
F(\lambda) &\sim&  -2 \left(\ln(-\lambda)-\ln(1-\lambda)\right) 
\sim 2 \ln (-\alpha's)
\eea
Putting this result back into \eqref{Glambda} gives
\bea
G(\lambda)&\simeq& 4g^2\left(\frac{1}{8 \pi \ap}\right)^{D/2}
\frac{2^{\gamma-1}\pi^{20-2D-2S}}{\gamma-1}
(-2\pi\lambda)^{1/2}\ln (-\alpha's)
\label{Glambda2}
\eea 
The exponential factor $e^{-\ap |s| f(\lambda)}$ in \eqref{Mfinal} can also 
be written as
\bea
e^{-\ap |s| f(\lambda)}&=&(-\lambda)^{\lambda \alpha' s}
(1-\lambda)^{\alpha's(1-\lambda)}\nn\\
&=&(-\lambda)^{-\alpha't}(1+t/s)^{\alpha's}(1-\lambda)^{\alpha't}
\eea
which in the $s\gg t$ ($\lambda\to 0$) limit then becomes
\bea
e^{-\ap |s| f(\lambda)}&\to & (-\lambda)^{-\alpha't}e^{\alpha't}
\eea
Plugging all these approximations back into the full hard scattering 
amplitude 
in \eqref{Mfinal} yields
\bea
\mathcal{M}_4^{\rm Hard}\!\!&\underset{s \gg t}{\sim}&\!\! 
g^2\ee{1}{4}\ee{2}{3}\left(\frac{1}{8 \pi \ap}\right)^{D/2}
\frac{2^{\gamma+1}\pi^{20-2D-2S}}{\gamma-1}
(-2\pi\lambda)^{1/2}\ln (-\alpha's) \, 
(-\lambda)^{-\alpha't}e^{\alpha't} \nn\\
&&\left(\log\ap|t|\right)^{1-\gamma}(\ap|s|)^{3/2} 
\label{MfinalRegge}
\eea 
Finally, replacing $\lambda=-t/s$ here one obtains
\bea
\mathcal{M}_4^{\rm Hard}\!\!&\underset{s \gg t}{\sim}&\!\! 
%\left(\frac{1}{8 \pi \ap}\right)^{D/2}
%\frac{2^{p+1}\pi^{20-2D-2S}}{p-1}(2\pi)^{1/2}
g^2\ee{1}{4}\ee{2}{3}\ln (-\alpha's) \, 
(-\alpha's)^{1+\alpha't}(-\alpha't)^{1/2-\alpha't}e^{\alpha't} 
\left(\log\ap|t|\right)^{1-\gamma}
\label{MfinalRegge2}
\eea 
which matches exactly with the expected result \eqref{MRegge2}. 

\section{Discussion and Conclusions}

As pointed out in the introductory section, considering only the planar diagrams in the multi-loop 
summation UV divergences in the 
open string channel
do not cancel among string diagrams (as it happens between the planar and Moebius strip diagrams for 
example) 
and a renormalization scheme is necessary. Moreover, since 
the one-loop expression for the amplitude is given in terms of an integral representation over the moduli, 
spurious divergences arise due to fact that the original integrals run over regions outside of their domain of convergence. 
For the case in study here, we show that all these spurious divergences and the UV ones can be regulated altogether by means 
of a single counterterm built out of suspending total 
momentum conservation before evaluating the integrals over the moduli.  Namely, for $p\equiv \sum_i k_i\neq0$, we first isolate the 
divergent parts, introduce the necessary counterterms, and we analytically continue the integrals to $p=0$ at the 
very end. As a result, we provide a novel expression for the $n$-gluon planar loop amplitude in type 0 theories which are completely 
free of all the spurious and UV divergences. If one is interested in the low energy limit, this new expression is now ready to give the correct field theory limit without having to 
be worried of the artifacts introduced by the spurious singularities originally present in the string loop amplitude.

We also studied in detail the high energy at fixed-angle limit (hard scattering) of the 4-gluon 
planar one-loop amplitude in these models using the renormalization procedure described above.
Since all the Mandelstam variables come multiplied with a factor of $\alpha'$, the hard scattering regime 
is equivalent to taking the tensionless limit ($\alpha'\to \infty$) with the external states held
at fixed momenta.

To extract the complete leading behavior 
of the amplitude and provide its full dependence on the kinematic invariants, it was necessary to carefully analyze 
all dominant regions.  Apart from the usual 
exponential drop-off, we also obtained the exact dependence on the scattering angle that multiplies the exponentially 
decaying factor which shows the existence of a smooth connection between the Regge and hard scattering regimes. 
Although we focus on the polarization structure 
that dominates in the Regge limit in order to correlate our results with those of \cite{ThornRojas}, our answers here 
are fully general and can be easily extended to all the other polarization structures.  Note that, contrary to the case of superstring amplitudes where the entire polarization structure can 
be factored out of the integration over the moduli (at least through one-loop), 
`gluon' amplitudes in type 0 theories are more convoluted since this 
factorization is, in general, not possible. 

It would be interesting to see if 
the smooth connection between the Regge and hard scattering regimes found here is also present for non-planar amplitudes.  In this case  
the amplitude is dominated by a saddle point which is located away from the 
boundaries at $q=0$ and  $q=1$, i.e., in the interior of the moduli. The saddle is given by the equation 
$\theta_4(0,q)/\theta_4(\pi,q)=(1-\lambda)^{1/4}$ where $\theta_4(\theta,q)$ is the fourth Jacobi theta function. As 
$\lambda \to 0$ the only solution for the saddle equation above is $q=0$, thus moving the saddle to the 
boundary. Also, the high energy ($\alpha'|t|\gg 1$) limit  of the Regge regime is again dominated by the 
$q\sim 0$ region. Therefore,  we should also expect a smooth transition between the hard and Regge behaviors 
for the 1-loop nonplanar diagram, although it would be nice to obtain this explicitly.

As pointed out in \cite{Thorn:Subcritical}, summing the planar open string diagrams to all loops by keeping the closed string 
tachyon (i.e., using type 0 strings)  causes a natural instability that could potentially explain confinement in gauge theories. 
Other indications of this phenomenon were also suggested in type 0 models
in the context of the AdS/CFT correspondence 
\cite{Klebanov:1998yya,Polyakov:Wall,Bergman:1999km,Klebanov:1998yy,Klebanov:1999ch,Minahan:1999yr}. Therefore, strings theories with tachyons in their closed string sector is a desireable feature. In a recent paper \cite{ADRV}, the scattering of closed strings off D-branes was studied in the high-energy Regge regime. At the one-loop level for planar diagrams they found that the dominant region is also the one we found in this work, namely the region where the inner boundary of the annulus shrinks to a point. Moreover, they were able to perform the sum of the leading contributions in this regime to all loops by means of an eikonal summation, yielding a non-zero result in terms of the vacuum expectation value of closed string vertex operators. Since each term in the sum comes from the region for the propagation of closed strings in the IR limit, we believe that a similar analysis can be performed in string theories with tachyons in their closed string sector (for instance, for the type 0 model studied here). Performing this sum could capture some of the effects of 
the closed string tachyons.

Finally, regarding the connections between higher spin theories \cite{fradkvas,vas,ss} and the tensionless limit of string theory \cite{Sagnotti,Bonelli:2003kh}, 
it would also be interesting to see if our results could be relevant for the construction of higher point 
vertices in 
higher spin theories using the methods of cutting loop amplitudes.

\begin{comment}
\begin{itemize}
\item It would be interesting to check the answer we obtained here with 
the extensions to the NS+ model of group theoretical approach by 
\cite{Moeller:2005ez}.
\item Multiloop analysis of the NS+ using Gross-Manes methods?
%\item Is there a constraint to the polarization of all string states 
%at high energy? I.e. do they try to lay on one particular direction or 
%hyperplane smaller than a 10 dimensional one? If so, could this imply 
%that at very high energies the open strings effectively reduce the number 
%of dimensions they live in? (at least perturbatively?) 
%\item Some numerics to compare leading with subleading contributions 
%(3 and 4-coincidence saddle?)
%\item Closed string sector? (Type-0 string)
\item 
\end{itemize}
\end{comment}

\vspace{20pt}
\noindent\underline{Acknowledgments}: I would like to thank Charles Thorn for guidance and very useful 
comments on the manuscript. I also thank Ido Adam for many suggestions and Horatiu Nastase and Mikhail Vasiliev for discussions.
Finally, I would like to acknowledge the hospitality of the University of Florida during the early stages of this work 
under the support of the Department of Energy under Grant No. DE-FG02-97ER-41029. 
This research was supported in part by FAPESP grant 2012/05451-8.

\appendix

\section{Orbifold Projection}\label{app:orb}
We discuss very briefly the alternative procedure for eliminating the 
massless scalars circulating the loop by projecting them out 
using an orbifold projection. It basically consists in demanding that the we keep only the states that are even under 
$a_n^I,b_r^I\to-a_n^I,-b_r^I$ for the components $I=D+S,D+S+1,\cdots, 10$ of the world-sheet 
oscillators. 
Thus, for the case when one has pure Yang-Mills theory in the $\alpha^{\prime}\to 0$ limit, i.e. $S=0$ (no adjoint massless scalars), we demand this 
condition for all the transverse components to the D-brane. This implies 
that in the partition functions in equations \eqref{P+} and \eqref{P-} now get modified as follows:
\bea
P_+ &\to& q^{-1}\frac{1}{2}\left[{\prod_r(1+q^{2r})^8\over\prod_n(1-q^{2n})^8}\right.\\
&&\left.\qquad\qquad + q^{(10-D-S)/4}\left(\frac{-\pi}{4\ln q}\right)^{(D+S-10)/2}{\prod_r(1+q^{2r})^{D+S-2}\prod_n(1+q^{2n})^{10-D-S}\over\prod_n(1-q^{2n})^{D+S-2}\prod_r(1-q^{2r})^{10-D-S}}
\right]\nn\\
P_- &\to& 2^{4}\frac{1}{2}\left[{\prod_n(1+q^{2n})^8\over\prod_n(1-q^{2n})^8}\right.\\
&&\left.\qquad\qquad +\left(\frac{-\pi}{\ln q}\right)^{(D+S-10)/2}{\prod_n(1+q^{2n})^{D+S-2}\prod_r(1+q^{2r})^{10-D-S}\over\prod_n(1-q^{2n})^{D+S-2}\prod_r(1-q^{2r})^{10-D-S}}
\right]
\eea
It is worth noticing that in the case of the maximal number of scalars circulating the loop, i.e. $D+S=10$, the modified partition functions become
\bea
P_+ &\to& q^{-1}{\prod_r(1+q^{2r})^8\over\prod_n(1-q^{2n})^8}\\
P_- &\to& 2^{4}{\prod_n(1+q^{2n})^8\over\prod_n(1-q^{2n})^8}
\eea
which are identical to the partition functions in the case without 
orbifold projections\footnote{Which in turn coincides with the  non-abelian D-brane projections in the $D+S=10$ case as well}. 

In \cite{ThornRojas} we computed the one-loop to the leading Regge trajectory using the projection 
procedure suggested in \cite{Thorn:Nonabelian}. If we use the new partition functions for the orbifold 
projection, the new Regge trajectory is given by
\bea \label{sigma}
\Sigma(t) \eqe -\frac{4g^2\alpha^{\prime 2-D/2}}{(8\pi^2)^{D/2}} \int_0^1 
\frac{dq}{q} \left(\frac{-\pi}{\ln q}\right)^{(10-D)/2} 
\int_0^{\pi} d\theta 
\left((-\psi^2(\theta)
[\ln \psi]^{\prime\prime})^{\alpha^{\prime} t}
\frac{\alpha^{\prime} t}{[\ln \psi]^{\prime\prime}}(P_+X^+-P_-X^-)\right.\nonumber\\
&& \left.-\frac{1}{4}(P_+-P_-)\left[(-\psi^2(\theta)
[\ln \psi]^{\prime\prime})^{\alpha^{\prime} t}-1\right] 
[-\ln \psi]^{\prime\prime} \right)
\eea
with the $P_{\pm}$ functions defined above and the rest is the same as before. 

The low energy (field theory) limit of \eqref{sigma} is governed by the contributions from the $q\sim 1$ 
region. Thus, it is more convenient to go back to the original $w$ variable where $w=e^{2\pi^2/\log q}$ and expand in 
powers of $w\sim 0$. Performing the Jacobi transform to write the new partition functions as functions 
of $w$ gives  
\beq\bal
P_+^{\rm orb}\!\!&=&\!\!\frac{1}{2w^{1/2}} \left(\frac{-2\pi}{\ln w}\right)^{4}
\left[\frac{\prod_r(1+w^r)^8}{\prod_n(1-w^n)^8}+\frac{\prod_r(1+w^r)^{D+S-2}\prod_r(1-w^r)^{10-D-S}}
{\prod_n(1-w^n)^{D+S-2}\prod_n(1+w^n)^{10-D-S}}\right]\\
P_-^{\rm orb}\!\!&=&\!\!\frac{1}{2w^{1/2}} \left(\frac{-2\pi}{\ln w}\right)^{4}
\left[\frac{\prod_r(1-w^r)^8}{\prod_n(1-w^n)^8}+\frac{\prod_r(1-w^r)^{D+S-2}\prod_r(1+w^r)^{10-D-S}}
{\prod_n(1-w^n)^{D+S-2}\prod_n(1+w^n)^{10-D-S}}\right]
\eal\eeq
we see that the low energy limit $\alpha^{\prime}\to 0$ is not modified 
since this regime is governed by the $w\sim 0$ behavior which does not 
change as we can see by expanding the new partition functions in this limit, 
where 
\bea
P_{\pm}^{\rm orb} &\sim& \frac{1}{w^{1/2}} \left(\frac{-2\pi}{\ln w}\right)^{4}
\left[1\pm(D+S-2)w^{1/2}+\mathcal{O}(w)\right]\nn
\eea
which is the same asymptotic behavior that the nonabelian D-brane 
construction provides.
\section{Counterterms for logarithmic divergences}\label{app:Bc}
The expression for the $B^{\pm}$ counterterm is more cumbersome because it 
is the sum of four terms which correspond to the four different edges that 
contribute with logarithmic divergences in the $\theta$ integrals. We list 
them here:
\bea
B_1^{\pm} \!\!\!\!&=& \!\!\!\!\frac{1}{4} \,\theta_{42}^{\ap (s+t)} \,
\theta_{43}^{-\ap  s} \theta_{32}^{-\ap  t-1} \times\nn \\
&\times & \!\!\!\left[\left(\mathcal{P}(\theta_4)-
\mathcal{P}(\theta_4)_C\right)
(1+\ap t)\theta_{32}^{-1} + \right.\nn\\
&&+\left.\left(\chi_+^2(\theta_4)-\chi_+^2(\theta_4)_C\right) 
\left(\ap t(1+\ap t) \theta_{32}^{-1} + 
(\ap s)^2\theta_{43}^{-1}-\ap{}^2(s+t)^2
\theta_{42}^{-1}\right) \right]\nn\\
B_2^{\pm} \!\!\!\!&=& \!\!\!\!\frac{1}{4} \,(\pi-\theta_3)^{\ap (s+t)} \,
\theta_{43}^{-\ap  s} (\pi-\theta_{4})^{-\ap  t-1} \times\nn \\
&\times & \!\!\!\left[\left(\mathcal{P}(\theta_2)-
\mathcal{P}(\theta_2)_C\right)
(1+\ap t)(\pi-\theta_{4})^{-1} + \right.\nn\\
&&+\left.\left(\chi_+^2(\theta_2)-\chi_+^2(\theta_2)_C\right) 
\left(\ap t(1+\ap t) (\pi-\theta_{4})^{-1} + 
(\alpha'  s)^2\theta_{43}^{-1}-\alpha'^2(s+t)^2
(\pi-\theta_3)^{-1}\right) \right]\nn\\
B_3^{\pm} \!\!\!\!&=& \!\!\!\!\frac{1}{4} \,(\pi-\theta_{42})^{\ap (s+t)} 
\,\theta_{2}^{-\ap  s} (\pi-\theta_{4})^{-\ap  t-1} \times\nn \\
&\times & \!\!\!\left[\left(\mathcal{P}(\theta_3)-
\mathcal{P}(\theta_3)_C\right)
(1+\ap t)(\pi-\theta_{4})^{-1} + \right.\nn\\
&&+\left.\left(\chi_+^2(\theta_3)-\chi_+^2(\theta_3)_C\right) 
\left(\ap t(1+\ap t) (\pi-\theta_{4})^{-1} + 
(\alpha's)^2\theta_{2}^{-1}-\alpha'^2(s+t)^2(\pi-\theta_{42})^{-1}\right) 
\right]\nn\\
B_4^{\pm} \!\!\!\!&=& \!\!\!\!\frac{1}{4} \,\theta_3^{\ap (s+t)} \,
\theta_2^{-\ap  s} \theta_{32}^{-\ap  t-1} \times\nn \\
&\times & \!\!\!\left[\left(\mathcal{P}(\theta_4)-
\mathcal{P}(\theta_4)_C\right)(1+\ap t)\theta_{32}^{-1} + \right.\nn\\
&&+\left.\left(\chi_+^2(\theta_4)-\chi_+^2(\theta_4)_C\right) \left(\ap t
(1+\ap t) \theta_{32}^{-1} + 
(\alpha's)^2\theta_{2}^{-1}-\alpha'^2(s+t)^2\theta_3^{-1}\right) \right]
\label{Bct}
\eea 
therefore, with these definitions, $B^{\pm}=\sum_{i=1}^4 B_i^{\pm}$.
%Notice that both $B^+$ and $B^-$ are the same which is a nice thing 
%because it 
%allows to see explicitly the cancellation of the open string tachyon 
%($q\sim 1$) in these expressions.
Note that, because of the form of these counter-term integrands, none of 
them is singular in the $\theta_4\sim\pi$, $\theta_2\sim\theta_3$ region 
which 
is the dominant region in the large $-s$ fixed $t$ limit, therefore they 
will 
not contribute to the one-loop correction to the Regge trajectory. This 
is why 
it was not necessary to include them in \cite{ThornRojas}. The fact 
that they are also non-singular in the remaining egde, namely 
$\theta_2\sim0$, $\theta_3 \sim \theta_4$ suggests that they do not 
contribute 
to the regime where $t$ is large and $s$ is held fixed either.\\


\begin{thebibliography}{}

\bibitem{'tHooft:1973jz} 
  G.~'t Hooft,
  ``A Planar Diagram Theory for Strong Interactions,''
  Nucl.\ Phys.\ B {\bf 72}, 461 (1974).
  %%CITATION = NUPHA,B72,461;%%
  %3450 citations counted in INSPIRE as of 18 Jul 2013

\bibitem{Maldacena:1997re} 
  J.~M.~Maldacena,
  ``The Large N limit of superconformal field theories and supergravity,''
  Adv.\ Theor.\ Math.\ Phys.\  {\bf 2}, 231 (1998)
  [hep-th/9711200].
  %%CITATION = HEP-TH/9711200;%%
  %9126 citations counted in INSPIRE as of 24 Jul 2013

\bibitem{Witten:1998qj} 
  E.~Witten,
  ``Anti-de Sitter space and holography,''
  Adv.\ Theor.\ Math.\ Phys.\  {\bf 2}, 253 (1998)
  [hep-th/9802150].
  %%CITATION = HEP-TH/9802150;%%
  %6123 citations counted in INSPIRE as of 24 Jul 2013

\bibitem{Gubser:1998bc} 
  S.~S.~Gubser, I.~R.~Klebanov and A.~M.~Polyakov,
  ``Gauge theory correlators from noncritical string theory,''
  Phys.\ Lett.\ B {\bf 428}, 105 (1998)
  [hep-th/9802109].
  %%CITATION = HEP-TH/9802109;%%
  %5364 citations counted in INSPIRE as of 24 Jul 2013
\bibitem{Thorn:Summing}
  C.~B.~Thorn,
  ``Summing Planar Open String Loops on a Worldsheet Lattice with Dirichlet and Neumann Boundaries,''
  Phys.\ Rev.\  {\bf D80}, 086010 (2009).
  [arXiv:0906.3742 [hep-th]].  

\bibitem{Thorn:Digital}
  C.~B.~Thorn,
  ``Digitizing the Neveu-Schwarz Model on the Lightcone Worldsheet,''
  Phys.\ Rev.\  {\bf D82}, 065009 (2010).
  [arXiv:1005.2924 [hep-th]]. 

\bibitem{Papathanasiou:2012mi} 
  G.~Papathanasiou and C.~B.~Thorn,
  ``Closed String Self-energy on the Lightcone Worldsheet Lattice,''
  Phys.\ Rev.\ D {\bf 86}, 066002 (2012)
  [arXiv:1206.5554 [hep-th]].
  %%CITATION = ARXIV:1206.5554;%%
  %3 citations counted in INSPIRE as of 17 Oct 2013

%\cite{Papathanasiou:2012fn}
\bibitem{Papathanasiou:2012fn} 
  G.~Papathanasiou and C.~B.~Thorn,
  ``Worldsheet Propagator on the Lightcone Worldsheet Lattice,''
  Phys.\ Rev.\ D {\bf 87}, 066005 (2013)
  [arXiv:1212.2900 [hep-th]].
  %%CITATION = ARXIV:1212.2900;%%
  %2 citations counted in INSPIRE as of 17 Oct 2013

%\cite{Papathanasiou:2013nta}
\bibitem{Papathanasiou:2013nta} 
  G.~Papathanasiou and C.~B.~Thorn,
  ``Open String Self-energy on the Lightcone Worldsheet Lattice,''
  Phys.\ Rev.\ D {\bf 88}, 026014 (2013)
  [arXiv:1305.5850 [hep-th]].
  %%CITATION = ARXIV:1305.5850;%%
  %1 citations counted in INSPIRE as of 17 Oct 2013

%\cite{Magnea:2013lna}
\bibitem{Magnea:2013lna} 
  L.~Magnea, S.~Playle, R.~Russo and S.~Sciuto,
  %``Multi-loop open string amplitudes and their field theory limit,''
  JHEP {\bf 1309}, 081 (2013)
  [arXiv:1305.6631 [hep-th]].
  %%CITATION = ARXIV:1305.6631;%%
  %3 citations counted in INSPIRE as of 14 Nov 2013

\bibitem{Klebanov:1998yya} 
  I.~R.~Klebanov and A.~A.~Tseytlin,
  ``D-branes and dual gauge theories in type 0 strings,''
  Nucl.\ Phys.\ B {\bf 546}, 155 (1999)
  [hep-th/9811035].
  %%CITATION = HEP-TH/9811035;%%
  %193 citations counted in INSPIRE as of 14 Nov 2013

\bibitem{Klebanov:1998yy} 
  I.~R.~Klebanov and A.~A.~Tseytlin,
  ``Asymptotic freedom and infrared behavior in the type 0 string approach to gauge theory,''
  Nucl.\ Phys.\ B {\bf 547}, 143 (1999)
  [hep-th/9812089].
  %%CITATION = HEP-TH/9812089;%%
  %118 citations counted in INSPIRE as of 14 Nov 2013

\bibitem{Klebanov:1999ch} 
  I.~R.~Klebanov and A.~A.~Tseytlin,
  ``A Nonsupersymmetric large N CFT from type 0 string theory,''
  JHEP {\bf 9903}, 015 (1999)
  [hep-th/9901101].
  %%CITATION = HEP-TH/9901101;%%
  %101 citations counted in INSPIRE as of 14 Nov 2013

\bibitem{Minahan:1999yr} 
  J.~A.~Minahan,
  ``Asymptotic freedom and confinement from type 0 string theory,''
  JHEP {\bf 9904}, 007 (1999)
  [hep-th/9902074].
  %%CITATION = HEP-TH/9902074;%%
  %79 citations counted in INSPIRE as of 14 Nov 2013

\bibitem{Thorn:Subcritical}
  C.~B.~Thorn,
  ``Subcritical String and Large N QCD,''
  Phys.\ Rev.\  {\bf D78}, 085022 (2008).
  [arXiv:0808.0458 [hep-th]].  
%\cite{Thorn:Nonabelian}

\bibitem{Polchinski:1998rq} 
  J.~Polchinski,
  ``String theory. Vol. 1: An introduction to the bosonic string,''
  Cambridge, UK: Univ. Pr. (1998) 402 p

\bibitem{ThornRojas}
  F.~Rojas, C.~B.~Thorn,
  ``The Open String Regge Trajectory and Its Field Theory Limit,''
  Phys.\ Rev.\  {\bf D84}, 026006 (2011).
  [arXiv:1105.3967 [hep-th]].

 \bibitem{kunsztst}
  Z.~Kunszt, A.~Signer and Z.~Trocsanyi,
  ``One loop helicity amplitudes for all 2 $\to$ 2 processes in QCD and N=1
  supersymmetric Yang-Mills theory,''
  Nucl.\ Phys.\ B {\bf 411} (1994) 397
  [arXiv:hep-ph/9305239];
  %%CITATION = HEP-PH 9305239;%%
%\bibitem{ellissexton}
For earlier calculations, see R.~K.~Ellis and J.~C.~Sexton,
  Nucl.\ Phys.\ B {\bf 269}, 445 (1986);
%\bibitem{bernk4gluons}
Z.~Bern and D.~A.~Kosower,
    Nucl.\ Phys.\ B {\bf 379}, 451 (1992).

\bibitem{chakrabartiqt}
  D.~Chakrabarti, J.~Qiu and C.~B.~Thorn,
  ``Scattering of glue by glue on the light-cone worldsheet. I: Helicity
  non-conserving amplitudes,''
Phys.\ Rev.\ D {\bf 72} (2005) 065022, arXiv:hep-th/0507280.
  %%CITATION = HEP-TH 0507280;%%

\bibitem{chakrabartiqt2}
  D.~Chakrabarti, J.~Qiu and C.~B.~Thorn,
``Scattering of glue by glue on the light-cone worldsheet. II: Helicity
conserving amplitudes,''
  Phys.\ Rev.\  D {\bf 74} (2006) 045018
  [Erratum-ibid.\  D {\bf 76} (2007) 089901]
  [arXiv:hep-th/0602026].

\bibitem{thornresir}
  C.~B.~Thorn,
  ``Resolution of Infrared Divergences in 
  Gluon-Gluon Scattering Regulated on a Lightcone Worldsheet Lattice,''
  Phys.\ Rev.\  {\bf D82 } (2010)  125021.
  [arXiv:1010.5998 [hep-th]].

\bibitem{Alessandrini:1972jy} 
  V.~Alessandrini, D.~Amati and B.~Morel,
  ``The asymptotic behaviour of the dual pomeron amplitude,''
  Nuovo Cim.\ A {\bf 7}, 797 (1972).
  %%CITATION = NUCIA,A7,797;%%

\bibitem{Dorn:1974er} 
  H.~Dorn, D.~Ebert and H.~-J.~Otto,
  ``High-Energy Behavior of Nonplanar and Planar Dual Multiloop Amplitudes,''
  Acta Phys.\ Polon.\ B {\bf 6}, 599 (1975).
  %%CITATION = APPOA,B6,599;%%

\bibitem{DornKaiser} 
  H.~Dorn, H.~J.~Kaiser,
  ``Asymptotic Behavior of the Planar One-Loop Correction to the Regge Trajectory in the Dual Model,''
  Acta Phys.\ Polon.\ B {\bf 6}, 17 (1975).

\bibitem{Otto:1976pu} 
  H.~J.~Otto, V.~N.~Pervushin and D.~Ebert,
  ``On Renormalization of Regge Trajectories in Dual Models,''
  Theor.\ Math.\ Phys.\  {\bf 35}, 308 (1978)
  [Teor.\ Mat.\ Fiz.\  {\bf 35}, 48 (1978)].
  %%CITATION = TMPHA,35,308;%%

\bibitem{Moeller:2005ez}
  N.~Moeller, P.~C.~West,
  ``Arbitrary four string scattering at high energy and fixed angle,''
  Nucl.\ Phys.\  {\bf B729 } (2005)  1-48.
  [hep-th/0507152].    
\bibitem{GrossManes}
  D.~J.~Gross, J.~L.~Manes,
  ``The High-energy Behavior Of Open String Scattering,''
  Nucl.\ Phys.\  {\bf B326}, 73 (1989).
\bibitem{GrossMende1}
  D.~J.~Gross, P.~F.~Mende,
  ``The High-Energy Behavior of String Scattering Amplitudes,''
  Phys.\ Lett.\  {\bf B197}, 129 (1987) 

\bibitem{thornsantafe} C. B. Thorn, unpublished comments, 
 Santa Fe Institute workshop, November 8-10, 1985.

\bibitem{Dixon:1986iz} 
  L.~J.~Dixon and J.~A.~Harvey,
  ``String Theories in Ten-Dimensions Without Space-Time Supersymmetry,''
  Nucl.\ Phys.\ B {\bf 274}, 93 (1986).
  %%CITATION = NUPHA,B274,93;%%
  %302 citations counted in INSPIRE as of 14 Nov 2013  

\bibitem{Seiberg:1986by} 
  N.~Seiberg and E.~Witten,
  ``Spin Structures in String Theory,''
  Nucl.\ Phys.\ B {\bf 276}, 272 (1986).
  %%CITATION = NUPHA,B276,272;%%
  %358 citations counted in INSPIRE as of 14 Nov 2013

\bibitem{Thorn:Nonabelian} 
  C.~B.~Thorn,
  ``Nonabelian D-branes, Open Strings, and Gauge Theory,''
  Phys.\ Rev.\ D {\bf 78}, 106008 (2008)
  [arXiv:0809.1085 [hep-th]].
  %%CITATION = ARXIV:0809.1085;%%  
%\cite{Thorn:Summing}
 %\cite{Goddard:1972ky}

\bibitem{Bergman:1997rf} 
  O.~Bergman and M.~R.~Gaberdiel,
  ``A Nonsupersymmetric open string theory and S duality,''
  Nucl.\ Phys.\ B {\bf 499}, 183 (1997)
  [hep-th/9701137].
  %%CITATION = HEP-TH/9701137;%%
  %126 citations counted in INSPIRE as of 14 Nov 2013

\bibitem{neveuscherkrenorm} 
  A.~Neveu and J.~Scherk,
  ``Gauge invariance and uniqueness of the renormalisation of dual models with unit intercept,''
  Nucl.\ Phys.\ B {\bf 36}, 317 (1972).
  %%CITATION = NUPHA,B36,317;%%  

\bibitem{Minahan:1987ha} 
  J.~A.~Minahan,
  ``One Loop Amplitudes on Orbifolds and the Renormalization of Coupling Constants,''
  Nucl.\ Phys.\ B {\bf 298}, 36 (1988).
  %%CITATION = NUPHA,B298,36;%%
  %106 citations counted in INSPIRE as of 15 Oct 2013

\bibitem{goddardreg}
  P.~Goddard,
  ``Analytic renormalization of dual one-loop amplitudes,''
  Nuovo Cim.\  A {\bf 4} (1971) 349.
  %%CITATION = NUCIA,A4,349;%%  

\bibitem{Neveu:1970iq}
  A.~Neveu, J.~Scherk,
  ``Parameter-free regularization of one-loop unitary dual diagram,''
  Phys.\ Rev.\  {\bf D1}, 2355-2359 (1970).  

\bibitem{nsetal} 
D.J. Gross, A. Neveu, J. Scherk and J.H. Schwarz, Phys. Rev. D2 (1970) 697;\\
C.S. Hsue, B. Sakita and M.A. Virasoro, Phya. Rev. D2 (1970) 2857

\bibitem{GSW2} 
  M.~B.~Green, J.~H.~Schwarz and E.~Witten,
  ``Superstring Theory. Vol. 2: Loop Amplitudes, Anomalies And Phenomenology,''
  Cambridge, Uk: Univ. Pr. (1987) 596 P. (Cambridge Monographs On Mathematical Physics)

\bibitem{Rojas:2011pb} 
  F.~Rojas,
  ``A Note on High-Energy Scattering of Open Superstrings,''
  Phys.\ Rev.\ D {\bf 89}, 086002 (2014)
  [arXiv:1111.7319 [hep-th]].
  %%CITATION = ARXIV:1111.7319;%%
  %1 citations counted in INSPIRE as of 01 May 2014

\bibitem{Polyakov:Wall} 
  A.~M.~Polyakov,
  %``The Wall of the cave,''
  Int.\ J.\ Mod.\ Phys.\ A {\bf 14}, 645 (1999)
  [hep-th/9809057].
  %%CITATION = HEP-TH/9809057;%%
  %263 citations counted in INSPIRE as of 14 Nov 2013

\bibitem{Bergman:1999km}
  O.~Bergman and M.~R.~Gaberdiel,
  %``Dualities of type 0 strings,''
  JHEP {\bf 9907} (1999) 022
  [hep-th/9906055].
  %%CITATION = HEP-TH/9906055;%%
  %91 citations counted in INSPIRE as of 14 Nov 2013

\bibitem{ADRV} 
  G.~D'Appollonio, P.~Di Vecchia, R.~Russo and G.~Veneziano,
  ``High-energy string-brane scattering: Leading eikonal and beyond,''
  JHEP\ {\bf 1011}, 100  (2010)
  [arXiv:1008.4773 [hep-th]].
  %%CITATION = JHEPA,1011,100;%%   

\bibitem{fradkvas}
E.~S.~Fradkin and M.~A.~Vasiliev,
%``Cubic Interaction In Extended Theories Of Massless Higher Spin Fields,''
Nucl.\ Phys.\ B291 (1987) 141,
%%CITATION = NUPHA,B291,141;%%
%``Candidate To The Role Of Higher Spin Symmetry,''
Annals Phys.\  177 (1987) 63,
%%CITATION = APNYA,177,63;%%
%``On The Gravitational Interaction Of Massless Higher Spin Fields,''
Phys.\ Lett.\ B189 (1987) 89,
%%CITATION = PHLTA,B189,89;%%
%``Gravitational Interaction Of Massless High Spin (S > 2) Fields,''
JETP Lett.\  44 (1986) 622 [Pisma Zh.\ Eksp.\ Teor.\ Fiz.\  44
(1986) 484],
%%CITATION = JTPLA,44,622;%%
%``Superalgebra Of Higher Spins And Auxiliary Fields,''
Int.\ J.\ Mod.\ Phys.\ A3 (1988) 2983.
%%CITATION = IMPAE,A3,2983;%%

\bibitem{vas}
M.~A.~Vasiliev,
%``Consistent Equation For Interacting Gauge Fields Of All Spins In (3+1)-Dimensions,''
Phys.\ Lett.\ B243 (1990) 378,
%%CITATION = PHLTA,B243,378;%%
%``Properties Of Equations Of Motion Of Interacting Gauge Fields Of All Spins In (3+1)-Dimensions,''
Class.\ Quant.\ Grav.\  8 (1991) 1387,
%%CITATION = CQGRD,8,1387;%%
%``Algebraic Aspects Of The Higher Spin Problem,''
Phys.\ Lett.\ B257 (1991) 111,
%%CITATION = PHLTA,B257,111;%%
%``More on equations of motion for interacting massless fields of all spins in (3+1)-dimensions,''
Phys.\ Lett.\ B285 (1992) 225;
%%CITATION = PHLTA,B285,225;%%
M.~A.~Vasiliev,
%``Nonlinear equations for symmetric massless higher spin fields in  (A)dS(d),''
Phys.\ Lett.\ B {\bf 567} (2003) 139 [arXiv:hep-th/0304049].
%%CITATION = HEP-TH 0304049;%%
For reviews see: M.~A.~Vasiliev,
%``Higher-spin gauge theories in four, three and two dimensions,''
Int.\ J.\ Mod.\ Phys.\ D5 (1996) 763 [arXiv:hep-th/9611024],
%%CITATION = HEP-TH 9611024;%%
%``Higher spin gauge theories: Star-product and AdS space,''
arXiv:hep-th/9910096;
%%CITATION = HEP-TH 9910096;%%
%``Progress in higher spin gauge theories,''
arXiv:hep-th/0104246.
%%CITATION = HEP-TH 0104246;%%

\bibitem{ss}
E.~Sezgin and P.~Sundell,
%``Higher spin N = 8 supergravity,''
JHEP {\bf 9811} (1998) 016 [arXiv:hep-th/9805125].
%%CITATION = HEP-TH 9805125;%%
%``Doubletons and 5D higher spin gauge theory,''
JHEP {\bf 0109} (2001) 036 [arXiv:hep-th/0105001],
%%CITATION = HEP-TH 0105001;%%
%``Towards massless higher spin extension of D = 5, N = 8 gauged supergravity,''
JHEP 0109 (2001) 025 [arXiv:hep-th/0107186],
%%CITATION = HEP-TH 0107186;%%
%``7D bosonic higher spin theory: Symmetry algebra and linearized  constraints,''
Nucl.\ Phys.\ B {\bf 634} (2002) 120 [arXiv:hep-th/0112100],
%%CITATION = HEP-TH 0112100;%%
%``Massless higher spins and holography,''
Nucl.\ Phys.\ B {\bf 644} (2002) 303 [Erratum-ibid.\ B {\bf 660}
(2003) 403] [arXiv:hep-th/0205131],
%%CITATION = HEP-TH 0205131;%%
%``Analysis of higher spin field equations in four dimensions,''
JHEP {\bf 0207} (2002) 055 [arXiv:hep-th/0205132],
%%CITATION = HEP-TH 0205132;%%
%``Holography in 4D (super) higher spin theories and a test via cubic  scalar couplings,''
arXiv:hep-th/0305040.
%%CITATION = HEP-TH 0305040;%%
J.~Engquist, E.~Sezgin and P.~Sundell,
%``On N = 1,2,4 higher spin gauge theories in four dimensions,''
Class.\ Quant.\ Grav.\  {\bf 19} (2002) 6175
[arXiv:hep-th/0207101],
%%CITATION = HEP-TH 0207101;%%
%``Superspace formulation of 4D higher spin gauge theory,''
Nucl.\ Phys.\ B {\bf 664} (2003) 439 [arXiv:hep-th/0211113].
%%CITATION = HEP-TH 0211113;%%

\bibitem{Sagnotti} 
  A.~Sagnotti and M.~Tsulaia,
  %``On higher spins and the tensionless limit of string theory,''
  Nucl.\ Phys.\ B {\bf 682}, 83 (2004)
  [hep-th/0311257].
  %%CITATION = HEP-TH/0311257;%%
  %151 citations counted in INSPIRE as of 14 Nov 2013
%\cite{Bonelli:2003kh}

\bibitem{Bonelli:2003kh} 
  G.~Bonelli,
  %``On the tensionless limit of bosonic strings, infinite symmetries and higher spins,''
  Nucl.\ Phys.\ B {\bf 669}, 159 (2003)
  [hep-th/0305155].
  %%CITATION = HEP-TH/0305155;%%
  %77 citations counted in INSPIRE as of 14 Nov 2013

\end{thebibliography}
\end{document}